\documentclass[a4paper,twoside,10pt]{report}

\usepackage[USenglish]{babel} 
\usepackage[ansinew]{inputenc}

\usepackage[Bjornstrup]{fncychap}
\usepackage[light,math]{iwona}
\usepackage[T1]{fontenc}

\usepackage{amsmath}%
\usepackage{amsfonts}%
\usepackage{amssymb}%
\usepackage{graphicx}
\usepackage{feyn}
\usepackage{slashed}
\usepackage{cite}
\usepackage{bbm}
\usepackage{array}
\usepackage{multirow}

\def\spa#1.#2{\langle#1\,#2\rangle}
\def\spb#1.#2{[#1\,#2]}

\def\spab#1.#2.#3{\langle\mskip-1mu{#1}
                  | #2 | {#3}]}

\def\spba#1.#2.#3{[\mskip-1mu{#1}
                  | #2 | {#3}\rangle}

\def\spbb#1.#2.#3.#4{[\mskip-1mu{#1}
                     | {#2} \ {#3} | {#4}]}

\def\spaa#1.#2.#3.#4{\langle\mskip-1mu{#1}
                     | {#2} \ {#3} | {#4}\rangle}

\newtheorem{prop}{Proposition}[section]
\newtheorem{theo}{Theorem}[section]
\newcommand{\trule}{\rule[-1.5mm]{0mm}{6mm}}
\newcommand{\bea}{\begin{eqnarray}}
\newcommand{\eea}{\end{eqnarray}}

\begin{document}

\pagestyle{empty}



\begin{titlepage}

\begin{center}

\vspace{1cm}
\Large
\textsc{\LARGE{Scattering Amplitudes in Gauge Theories}}\\

\vspace{3cm}

\textsc{Diplomarbeit\\[0.5\baselineskip]
of\\[0.5\baselineskip]
Ulrich Schubert}\\

\vspace{2.5cm}
\textsc{June 6, 2013}\\ 

\vspace{1cm}
\textsc{Supervisor:\\
Prof. Dr. Wolfgang Hollik\\
\vspace{0.25in}
Co-Supervisor:\\
Dr. Pierpaolo Mastrolia}

\vspace{2cm}
\textsc{Technische Universist\"at M\"unchen	\\
Physik-Department}\\

\end{center}
\vspace{2.2cm}
\begin{minipage}[t]{0.25\textwidth}
\begin{center}
\includegraphics[width=0.9\textwidth]{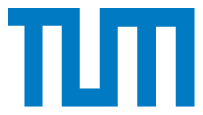}
\end{center}
\end{minipage}
\begin{minipage}[t]{0.25\textwidth}
\begin{center}
\includegraphics[width=0.5\textwidth]{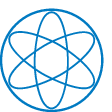}
\end{center}
\end{minipage}
\begin{minipage}[t]{0.25\textwidth}
\begin{center}
\includegraphics[width=0.5\textwidth]{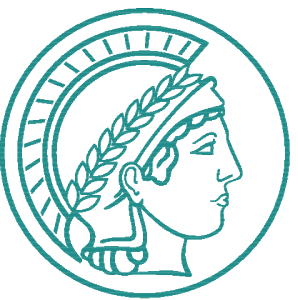}
\end{center}
\end{minipage}
\begin{minipage}[t]{0.25\textwidth}
\begin{center}
\includegraphics[width=0.9\textwidth]{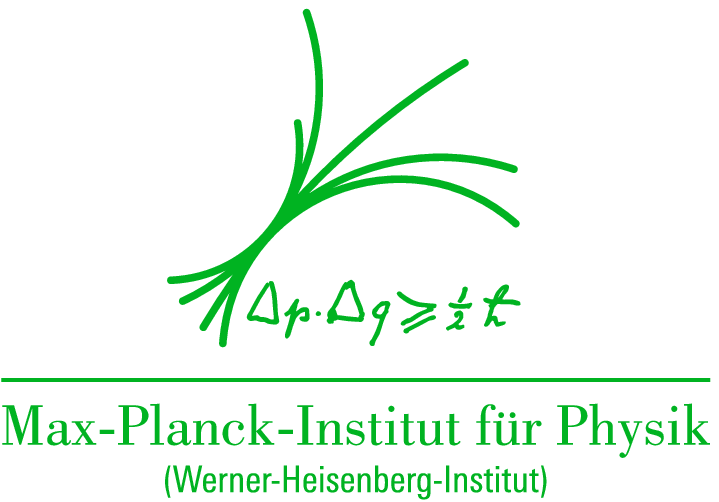}
\end{center}
\end{minipage}

\end{titlepage}

\begin{center}
\section*{Abstract}
\end{center}
\label{sec:Abstract}
This thesis is focused on the development of new mathematical methods
for computing multi-loop scattering amplitudes
in gauge theories.
In this work we combine, for the first time, the unitarity-based
construction for integrands, and the recently introduced
integrand-reduction through multivariate polynomial division.
After discussing the generic features of this novel reduction
algorithm, we will apply it to the one- and two-loop five-point
amplitudes in ${\cal N}=4$ sYM.
The integrands of the multiple-cuts are generated from products of
tree-level amplitudes within the super-amplitudes formalism. The
corresponding expressions will be used for the analytic reconstruction
of the polynomial residues.
Their parametric form is known a priori, as derived by means of
successive polynomial divisions using the Gr\"obner basis associated
to the on-shell denominators.
The integrand reduction method will be exploited to investigate the
color-kinematic duality for multi-loop ${\cal N}=4$ sYM scattering
amplitudes. Our analysis yields a suggestive, systematic way to
generate graphs which automatically satisfy the color-kinematic
dualities.
Finally, we will extract the leading ultra-violet divergences of
five-point one- and two-loop amplitudes in ${\cal N}=4$ sYM, which
represent a paradigmatic example for studying the UV behavior of
supersymmetric amplitudes.

\newpage
\tableofcontents 

\cleardoublepage 

\pagestyle{plain}

\newpage

\chapter{Introduction}

Particle Physics is the study of the subatomic constituents of matter that can rarely be held in one place, either because they move at the speed of light, or because their lifetime is too tiny. At the same time, the measurement apparatus cannot be so small to explore a fundamental particle without being a particle itself.
Hence, studying the interactions between colliding particles and identifying the consequences of this collision is the only key to access the subatomic world. 

The physical quantity that describes the interaction process is called {\it scattering cross section}. It specifies the area where the particles do collide, times the quantum mechanical probability of the collision to take place. Quantum mechanical probability densities are defined as the absolute squared value of a quantum mechanical amplitude, called {\it scattering amplitude}, which can be seen as the element of the scattering matrix (the $S$-matrix). 
Any scattering amplitude receives contributions from different processes which are indistinguishable at the quantum level, namely leaving the same traces in the measurement apparatus. To form a scattering cross section, all the processes contributing to the same scattering amplitude have to be summed before being squared.

The improving of the experimental precision in particle physics demands for more accurate predictions from the theoretical side. All computations of observables related to interacting particles rely on the postulate that the properties of final states can be described by partonic reactions. Perturbation Theory is a powerful tool for describing the quantum behavior of particles: at the leading order (LO), particle scattering is depicted in terms of {\it tree} graphs; and higher accuracy is reached by including terms which, beyond LO, are represented by diagrams containing {\it loops}. On the one hand loop diagrams provide access to physics beyond the range of sensitivity of the current experiments, because of the room heavier particles have to circulate around the loop and on the other hand they represent the quantum corrections to tree graphs, which are purely classical.
The canonical formulation of Quantum Field Theory (QFT) is built on the two pillars of {\it locality} and {\it unitarity}. The standard apparatus of Lagrangians and Feynman integrals allows us to make these two fundamental principles manifest. This approach, however, requires the introduction of unphysical redundancy in our description of physics. Accordingly, the computation of scattering amplitudes requires the evaluation of highly non-trivial integrals, which can be understood as the generalization of averaging on non-observable degrees of freedom.
The study of scattering amplitudes is fundamental to our understanding of QFT, although the canonical formalism turns out to be blind to astonishingly simple properties of the gauge-invariant physical observables of the theory. There are powerful new mathematical structures underlying the extraordinary properties of scattering amplitudes in gauge theories, and studying them lead us into direct contact with a very active area of current research in mathematics, 
like Algebraic Geometry \cite{Mastrolia:2012an,Zhang:2012ce,ArkaniHamed:2012nw}.

Scattering amplitudes lay in the heart of perturbative quantum field theories.
The recent advances in their calculational tools have been initiated by a more profound  understanding 
of the multi-channel factorization properties which arise under complex deformations of the kinematics 
imposed by on-shell \cite{Cachazo:2004kj,Britto:2004ap} and 
generalized unitarity-cut conditions \cite{Bern:1994zx,Britto:2004nc}.

Moreover analyticity and unitarity of scattering amplitudes \cite{OldUnitarity} have then been strengthened through the revealing of the mathematical structures present at the singularities.
Their study exposed a quadratic recurrence relation for tree-level amplitudes, the BCFW-recursion \cite{Britto:2004ap}, its link to the leading singularity of one-loop amplitudes \cite{Britto:2004nc} and a relation between 
numerator and denominators 
of one-loop Feynman integr{\it als}, yielding the multipole decomposition of 
Feynman integr{\it ands}, characterizing the by now known as OPP method
\cite{Ossola:2006us}. 

These new insights, which originate from a reinterpretation 
of tree-level scattering within the twistor string theory \cite{Witten:2003nn},
have inspired the study of novel mathematical frameworks in the more 
supersymmetric sectors of quantum field 
theories, such as dual conformal symmetries 
\cite{Drummond:2006rz}, 
grassmanians 
\cite{ArkaniHamed:2009dn}, 
Wilson-loops/gluon-amplitudes duality 
\cite{Alday:2007hr}, 
color/kinematic 
and gravity/gauge dualities  \cite{Bern:2008qj,Bern:2010ue},
as well as on-shell \cite{Cachazo:2004kj,Britto:2004ap,Badger:2005zh,Bern:2005cq} 
and generalized unitarity-based methods 
\cite{Bern:1994zx,Britto:2004nc,GenUn,Ellis:2007br},
and more generally the breakthrough advances in automating 
the evaluation of multi-particle scattering one-loop amplitudes, as demanded by the
experimental programs at hadron colliders.

In most cases the direct integration of Feynman integrals is prohibitive, hence 
the evaluation of scattering amplitudes beyond 
the leading order is addressed in two stages: 
First the amplitude is expressed in terms of an universal integral basis, and secondly the evaluation of the elements of the basis, called {\it master integrals} (MI's).

At one-loop, 
the advantage of knowing {\it a priori} that the basis of MI's is formed by scalar 
one-loop functions \cite{Passarino:1978jh},
as well as the availability of their analytic expression \cite{vanOldenborgh:1989wn},
allowed the community to focus on the development of efficient algorithms  
for extracting the coefficients multiplying each MI's. 
Improved tensor decomposition \cite{Denner:2005nn}, 
complex integration and contour deformation \cite{Nagy:2006xy},
on-shell and generalized unitarity-based methods, and 
integrand-reduction techniques \cite{Ossola:2006us,Ellis:2007br,Mastrolia:2010nb,Heinrich:2010ax} 
led to impressive results which only few years ago were considered inconceivable,
and to such a high level of automation \cite{Mastrolia:2010nb,Ossola:2007ax} 
that different scattering processes at the 
next-to-leading order accuracy can be handled by single, yet multipurpose, codes 
\cite{Hahn:2010zi,vanHameren:2009dr,Bevilacqua:2010mx,Hirschi:2011pa,GoSam}.

At higher-loop, and in particular at two-loop to begin with, the situation is different.
The basis of MI's is not known a priori. 
MI's are identified at the end of the reduction procedure, and afterwards the problem of their 
evaluation arises.
The most used multi-loop reduction technique is the well-known Laporta algorithm \cite{Laporta:2001dd},
based on the solution of algebraic systems of equations obtained through integration-by-parts 
identities \cite{Tkachov:1981wb}. 
The recent progress in calculating higher loop amplitudes  
has been enabled by the improvement of mathematical methods dedicated to Feynman 
integrals, such as 
difference \cite{Laporta:2001dd,Lee:2010ug} 
and 
differential \cite{Kotikov:1991pm} 
equations, 
Mellin-Barnes integration \cite{Smirnov:1999gc}, asymptotic expansions \cite{Beneke:1997zp}, sector decomposition
\cite{Binoth:2000ps}, complex integration and contour deformation 
\cite{Anastasiou:2007qb} -- to list few of them. \\

In this work we aim at extending the combined use of 
{\it unitarity-based methods} and 
{\it integrand-reduction techniques},
in order to accomplish 
the semi-analytic reduction of two-loop amplitudes to MI's.
The use of unitarity-cuts and complex momenta for on-shell internal particles 
turned unitarity-based methods into very efficient
tools for computing scattering amplitudes. These methods exploit two general properties of 
scattering amplitudes, such as analyticity and unitarity. The former guarantees that amplitudes can be reconstructed from 
the knowledge of their singularities, while the latter ensures that 
the residues at the 
singular points factorize into products of simpler amplitudes.
Unitarity-based methods are founded on the underlying representation 
of scattering amplitudes as a linear combination of MI's, and their principle is 
the extraction of the coefficients entering in such a linear combination 
by matching the cuts of the amplitudes onto the cuts of each MI.

Cutting rules as computational tools have been introduced at two-loop in the context 
of supersymmetric amplitudes \cite{Bern:1997nh} and later applied to the case of pure 
QCD amplitudes \cite{Bern:2000dn}.
The use of complex momenta for internal particles to fulfill the multiple 
cuts of two-loop amplitudes has been proposed for extending 
the one-loop quadruple-cut technique, 
to the octa-cut \cite{Buchbinder:2005wp, Cachazo:2008vp}, 
as well as the method of maximal cuts \cite{maximalcut}. \\

The multi-particle pole decomposition for the integrands 
of arbitrary scattering amplitudes emerges
from the combination of analyticity and unitarity 
with the idea of a reduction under the integral sign.

The principle of an integrand-reduction method is 
the underlying multi-particle pole expansion for the integrand of any scattering amplitude,
or, equivalently, the relation between numerator and denominators of the integrand:
a representation where the numerator of each Feynman integral is expressed as a combination 
of products of the corresponding denominators, with polynomial coefficients \cite{Ossola:2006us,Mastrolia:2011pr,Zhang:2012ce,Mastrolia:2012an}.

The key feature in the integrand-decomposition is represented by 
the shape of the residues on the multi-particle pole before integration:
each residue is a (multivariate) polynomial in the 
{\it irreducible scalar products} (ISP's) formed by the loop momenta and 
either external momenta or polarization vectors constructed out of them; 
the scalar products appearing in the residues are 
by definition {\it irreducible}, namely 
they cannot be expressed in terms of the denominators of the integrand \cite{Mastrolia:2011pr}. 

The polynomial structure of the multi-particle residues is a {\it qualitative} information
that turns into a {\it quantitative} algorithm for decomposing arbitrary 
amplitudes in terms of MI's at the integrand level. 
In the context of an integrand-reduction, any explicit integration procedure 
and/or any matching procedure between cuts of amplitudes and cuts of MI's 
is replaced by {\it polynomial fitting}, which is a simpler operation.

Decomposing the amplitudes in terms of MI's amounts to reconstructing the full polynomiality 
of the residues, {\it i.e.} it amounts to determining all the coefficients of each polynomial. \\

In this work we will meet the criteria to constrain the 
polynomial form of the residue on each multiple-cut of an arbitrary two-loop amplitude.
Unlike the one-loop case, where the residues of the multiple-cut have been 
systematized for all the cuts, in the two-loop case, their form 
is still unknown.
Their existence is a prerequisite for establishing a relation between numerator 
and denominators of any two-loop integrand.
Their implicit form can be given in terms of unknown coefficients, which are 
determined through polynomial fitting.
As in the one-loop case,
the full reconstruction of the polynomial residues requires only the knowledge of the 
integrand evaluated at explicit values of 
the loop momenta as many times as the number of the unknown coefficients.

Another feature of the integrand-reduction algorithm we are describing is 
that the determination of the polynomial form of 
the residues amounts to choose a basis of MI's, which 
does not necessarily need to be known a priori \cite{Mastrolia:2011pr}.
In fact, as we will see, each ISP appearing in the polynomial residues
is the numerator of a potential MI which may appear in the final result
(other than the scalar integrals). 

The parametric form of the polynomial residues is process-independent and it can be determined a priori, from the topology of the corresponding multiple cut, namely from the diagram identified by the denominators that go simultaneously on-shell. The actual value of the coefficients is clearly process-dependent, and its determination is indeed the goal of the integrand-reduction. Integrand-reduction methods determine the (unknown) coefficients by polynomial fitting, through the evaluation of the (known) integrand at values of the loop-momenta fulfilling the cut conditions.

The integrand-reduction methods have been originally developed at one loop by Ossola, Papadopoulos and Pittau (OPP) \cite{Ossola:2006us}. 
Extensions  beyond one-loop were first proposed by Mastrolia and Ossola in~\cite{Mastrolia:2011pr}, and later refined by Badger, Frellersvig and Zhang \cite{Badger:2012dp}.
Both the numerator and the denominators of any integrand are
multivariate polynomials in the components of the loop variables.
As recently shown in tandem by Zhang~\cite{Zhang:2012ce}, and Mastrolia, Ossola, Mirabella and Peraro \cite{Mastrolia:2012an},  
the  decomposition of the integrand can be obtained by performing the 
 {\it multivariate polynomial division} between the numerator and
the denominators, using  basic  principles of
algebraic geometry, like the division modulo a Gr\"obner basis.
Moreover,  the multivariate polynomial divisions gives a systematic classification 
of the polynomial structures of the residues,  leading to both the identification 
of the MI's and the determination  of their coefficients. 
By using multivariate polynomial division, 
a systematic classification of a four dimensional integral basis for two-loop integrands is doable \cite{Feng:2012bm}.

In \cite{Mastrolia:2012an}, the mathematical framework for the multi-loop integrand-reduction algorithm was developed. Accordingly,  the residues are
uniquely determined by the denominators involved in the corresponding multiple cut,
and the multi-particle pole decomposition for any scattering amplitude is achieved through a 
 simple {\it integrand recurrence relation}.
The algorithm is valid  for arbitrary amplitudes, irrespective of  
the number of loops, the particle content (massless or massive), 
and of the diagram topology (planar or non-planar). Interestingly, 
at one-loop our algorithm allows for a simple derivation of the one-loop OPP reduction formula. 
The spurious terms, when present, naturally 
arise from the structure of the denominators entering the generalized cuts.  
In the same work \cite{Mastrolia:2012an}, 
the so called {\it maximum-cut theorem} was proven. The theorem deals with cuts 
where the number of on-shell conditions is equal to the number of integration variables
and therefore the loop momenta are completely localized.
The theorem ensures that the number of independent
solutions of the maximum-cut is equal to the
number of coefficients parametrizing the corresponding residue.
The maximum cut theorem generalizes at any loop the simplicity of  
the one-loop quadruple-cut \cite{Britto:2004nc,Ossola:2006us},
where the two coefficients parametrizing the residue
are determined by the two solutions of the cut. \\

New approaches tackling the evaluation of one-loop multi-parton amplitudes
have recently been under intense development (see ~\cite{Ellis:2011cr}
and \cite{Alday:2008yw,Britto:2010xq,Henn:2011xk,Bern:2011qt,Carrasco:2011hw,Dixon:2011xs,Ita:2011hi}
for a comprehensive review). As previously said, various new structures have been uncovered in the amplitudes of gauge and gravity theories.

One such structure is the duality between color and kinematics found by Bern, Carrasco, and Johansson (BCJ) \cite{Bern:2008qj,Bern:2010ue}. This duality is conjectured to hold at all loop orders in Yang-Mills theory and its supersymmetric counterparts. Besides imposing strong constraints on gauge-theory amplitudes, whenever a gauge-theory loop integrand is obtained, where the duality is manifest, gravity integrands can be simply obtained from the {\it double-copy} relation, namely by replacing color factors with specified kinematic numerator factors.
The duality between color and kinematics has been confirmed in numerous tree-level studies \cite{Tye,StringBCJ,YMSquared,TreeBCJConf,OConnell,JJ2,TreeAllN}.
 At loop level, the duality remains a conjecture, but there is already significant nontrivial evidence in its favor for supersymmetric theories \cite{Bern:2010ue,ck4l,OneTwoLoopN4,SchnitzerBCJ,White,OneLoopN1Susy}
  and for special helicity configurations in non-supersymmetric pure Yang-Mills theory \cite{Bern:2010ue,OConnellRational,Bern:2013yya}. \\
Results obtained for gauge-theory amplitudes may be used 
to study the ultra-violet divergences of the corresponding gravity amplitudes. Recent years have seen a renaissance in the study of ultra-violet divergences in gravity theories, mainly due to the greatly improved ability to carry out explicit multiloop computations in gravity theories \cite{Bern:2010ue,ck4l,OneTwoLoopN4,SchnitzerBCJ,Bern:1997nh,LowerLoopSupergravity,ThreeloopHalfMax,TwoloopHalfMax}. The unitarity method also has revealed hints that multiloop supergravity theories may be better behaved in the ultra-violet than power counting arguments based on standard symmetries suggest \cite{Finite}. 
However the question of whether it is possible to construct a finite supergravity is still an open one.
The duality between color and kinematics and its associated double-copy formula offer a new angle on the ultra-violet divergences in supergravity theories \cite{Bern:2010ue,ck4l,ThreeloopHalfMax,TwoloopHalfMax,BoelsUV}.

In this thesis, I combine for the first time two of the most advanced techniques for the evaluation of scattering amplitudes in gauge theories: the color-kinematic duality and the integrand-reduction via polynomial division. 
I will apply them to the decomposition of one- and two-loop five-point amplitudes in ${\cal N}=4$ sYM theory in four dimension. In particular, I will use the super-amplitudes tree-level formalism for constructing the integrands of the multiple-cuts. Such integrands are the input for the determination of the polynomial residues within the integrand-reduction algorithm.
The color-kinematic equations are then used for imposing additional restrictions on the shape of the residues, which are needed for constructing graph numerators satisfying the duality. 
The decomposition of the amplitudes in terms of MI's is finally used to determine the UV behavior of the ${\cal N}=4$  sYM amplitudes.
The results contained in this thesis are already available in the literature \cite{Bern:2006vw,Carrasco:2011mn},
but they will be obtained in a new fashion. The methodology hereby presented is purely four-dimensional, nevertheless, the extension to dimensionally regulated amplitudes, as well as to non-supersymmetric theories, may follow exactly the same procedure. Therefore the technique hereby presented has the potential to become a standard technique for the reduction of multi-loop Feynman amplitudes.

This thesis is organized as follows: In the first chapter we will introduce the general notation used to describe the kinematic and gauge group part of scattering amplitudes. The next chapter discusses the conjectured color-kinematics duality at tree- and loop-level. We will then move to the calculation of one-loop amplitudes with the OPP method and the calculation of multi-loop amplitudes via the newly proposed integrand-reduction through multivariate polynomial division. Afterwards we will cover an efficient description of scattering amplitudes in $\mathcal{N}$=4 sYM. Chapters four and five will apply the former described techniques to derive the five-point one-loop and two-loop amplitudes, with the help of the color-kinematical duality. Chapter six will contain a derivation of the same one- and two-loop amplitude only from general properties of $\mathcal{N}$=4 sYM. The final chapter discusses the leading ultra-violet properties of the previously calculated five-point one- and two-loop amplitudes. \\
All numerical and semi-analytic computations presented in this thesis have been obtained using Mathematica with the Spinor-Helicity Formalism package S@M \cite{Maitre:2007jq}.  
\chapter{Scattering Amplitudes}

\section{Spinor Helicity Formalism}
Before we start our discussion on amplitudes it is important to determine some notation. We will work in the so called spinor helicity formalism see \cite{Dixon,peskin,srednicki,Elvang:2013cua} for reviews.\\
In the 1980s Berends and Wu used the fact that every lightlike momentum can be written in terms of massless spinors \cite{calkul} and that the polarization vectors also have a presentation as spinors \cite{polvectora,polvectorb} to develop a very compact formalism for amplitudes. \\
A spinor $\Psi$ is a four-vector which solves the Dirac equation
\begin{gather}
\left(i \slashed{\partial}-m \right)\Psi=0
\end{gather}
where we used the Feynman slash notation $\slashed{\partial}=\gamma^\mu \partial_\mu$. The four-vector $\Psi$ can be split into a part with positive and a part with negative helicity also called right-handed and left-handed respectively
$\Psi=\begin{pmatrix}
\psi_-\\
\psi_+ 
\end{pmatrix}
$. Using the Weyl representation of the gamma matrices
\begin{gather}
\gamma^\mu=\begin{pmatrix} 0 & \sigma^\mu \\ \overline{\sigma}^\mu & 0 \end{pmatrix}
\end{gather}
we can see that the mass term mixes both helicities
\begin{gather}
\begin{pmatrix}
-m & i\partial_\mu \sigma^\mu \\
i\partial_\mu \overline{\sigma}^\mu & -m 
\end{pmatrix}
\begin{pmatrix}
\psi_- \\
\psi_+ 
\end{pmatrix}=0
\end{gather} 
where $\sigma^\mu= \begin{pmatrix} \mathbbm{1} \\ \sigma^x \\ \sigma^y \\ \sigma^z \end{pmatrix}$ and $\overline{\sigma}^\mu= \begin{pmatrix} \mathbbm{1} \\ -\sigma^x \\ -\sigma^y \\ -\sigma^z \end{pmatrix}$ are vectors of Pauli matrices. In the massless limit the equations for positive and negative helicity will decouple and give us the Weyl equations
\begin{gather}
\partial_\mu \sigma^\mu \psi_+=0\\
\partial_\mu \overline{\sigma}^\mu \psi_-=0.
\end{gather}
The solutions to these equations are plane waves $\psi_+=\lambda e^{-ip\cdot x}$ and $\psi_-=\tilde{\lambda}e^{-ip\cdot x}$, where $\lambda$ and $\tilde{\lambda}$ are two component Weyl spinors.\footnote{Usually there would be a second set of spinors for plane waves with negative frequencies, but since we are working with massless spinors they are the same up to normalization conventions.} \\
Fixing the normalization in the usual way
\begin{gather}
\lambda^\dagger \lambda=2p_0 \\
\tilde{\lambda}^\dagger \tilde{\lambda}=2p_0 
\end{gather}
we find the following solutions for the spinors
\begin{gather}
\lambda = \begin{pmatrix} \sqrt{p^-}e^{-i\varphi_p} \\ -\sqrt{p^+} \end{pmatrix} \\
\tilde{\lambda} = \begin{pmatrix} -\sqrt{p^+}e^{-i\varphi_p} \\ \sqrt{p^-} \end{pmatrix}
\end{gather}
where we used a different basis for the momenta defined by
\begin{gather}
p^\pm = p^0 \pm p^z\\
e^{\pm i \varphi_p}= \frac{p^x \pm i p^y}{\sqrt{p^+} \sqrt{p^-}}.
\end{gather}
The latter definition is useful since $e^{ i \varphi_p}e^{- i \varphi_p}=1$.
In order to form scalar products between the spinor it is advantageous to go back to the full four-vector so we define
\begin{gather}
\Lambda=\begin{pmatrix} 0 \\ \lambda  \end{pmatrix} \\
\tilde{\Lambda}=\begin{pmatrix}  \tilde{\lambda} \\ 0  \end{pmatrix}
\end{gather}
and form the Lorentz invariant spinor products
\begin{gather}
\overline{\Lambda}_i \Lambda_j = \spa{i}.j \\
\overline{\tilde{\Lambda}}_i \tilde{\Lambda}_j = \spb{i}.j
\end{gather}
where i and j are two particles with momentum $p_i$ and $p_j$.
With the Gordon Identity and the spin-decoupled completeness relation we are able to connect a momentum to a spinor product
\begin{gather}
\spab{i}.{\gamma^\mu}.i = 2 p_i^\mu \\
|i\rangle [i|= \Lambda_i\overline{\tilde{\Lambda}}_i= \slashed{p}_i \left(\frac{1-\gamma^5}{2} \right) \\
|i] \langle i|= \tilde{\Lambda}_i\overline{\Lambda}_i= \slashed{p}_i\left(\frac{1+\gamma^5}{2} \right).
\end{gather}
From the definitions of the spinors we can derivate the first five identities:  \\
1. Antisymmetry
\begin{gather}
\spa{i}.j= - \spa{j}.i \ \ \ \ \ \ \ \ \ \spb{i}.j =- \spb{j}.i
\end{gather}
2. Fierz rearrangement
\begin{gather}
\spab{i}.\gamma^\mu.j \spab{k}.\gamma_\mu.l=2 \spa{i}.k \spb{l}.j
\end{gather}
3. Charge Conjugation of a current
\begin{gather}
\spab{i}.\gamma^\mu.j = \spba{i}.\gamma^\mu.j
\end{gather}
4. Schouten Identity
\begin{gather}
\spa{i}.j \spa{k}.l + \spa{i}.k \spa{l}.j + \spa{i}.l \spa{j}.k =0 \\
\spb{i}.j \spb{k}.l + \spb{i}.k \spb{l}.j + \spb{i}.l \spb{j}.k =0
\end{gather}
5. Momentum conservation
\begin{gather}
\sum_i^n{\spb{j}.i \spa{i}.k}=0.
\end{gather}
Furthermore to formulate an amplitude completely in terms of spinors we need a representation of the polarization vector in terms of spinors \cite{polvectora,polvectorb}
\begin{gather}
\epsilon^\mu(i^+,j)=+\frac{\spab{j}.\gamma^\mu.i }{\sqrt{2} \spa{j}.i } \\
\epsilon^\mu(i^-,j)=-\frac{\spba{j}.\gamma^\mu.i }{\sqrt{2} \spb{j}.i }
\label{polvec}
\end{gather}
where $j$ is light-like reference vector, which is not collinear with $i$. The difference between two choices of reference vectors corresponds to an on-shell gauge transformation
\begin{gather}
\epsilon^\mu(i^+,j')-\epsilon^\mu(i^+,j)=-\frac{ \spa{i}.j \spb{j}.{j'} \spab{j'}.\gamma^\mu.{i}  + \spa{i}.{j'} \spb{j'}.{j} \spab{j}.\gamma^\mu.i  } {\sqrt{2} \spa{j'}.i \spa{j}.i \spb{j'}.{j} }\\
=- \frac{ \spab{i}.{\slashed{j} \slashed{j}' \gamma^\mu }.i + \spab{i}.{\slashed{j}' \slashed{j} \gamma^\mu }.i  }{\sqrt{2} \spa{j'}.i \spa{j}.i \spb{j'}.{j} }\\
= - \frac{\sqrt{2} p_j \cdot p_{j'} \spab{i}.\gamma^\mu.i }{ \spa{j'}.i \spa{j}.i \spb{j'}.{j} } \\
= - \frac{ \sqrt{2} \spa{j}.{j'} \spb{j'}.j  p_i^\mu }{ \spa{j'}.i \spa{j}.i \spb{j'}.{j} } \\ 
=\frac{ \sqrt{2} \spa{j'}.j } {\spa{i}.{j'} \spa{i}.j }  p_i^\mu 
\end{gather}
where we used the completeness relation and the charge conjugation of a current. \\
Moreover we can check that the definition in (\ref{polvec}) satisfies the properties of a polarization vector. \\
First they have to be perpendicular to its momentum
\begin{gather}
\epsilon(i^+,j) \cdot p_i= +\frac{\spab{j}.\slashed{p}_i.i }{\sqrt{2} \spa{j}.i }=0 \\
\epsilon(i^-,j) \cdot p_i =-\frac{\spba{j}.\slashed{p}_i.i }{\sqrt{2} \spb{j}.i }=0
\end{gather}
where we used the Dirac Equation in the last step. Secondly they need to be properly normalized
\begin{gather}
\epsilon(i^+,j) \cdot \epsilon(i^-,j)=-\frac{\spab{j}.\gamma^\mu.i \spba{j}.\gamma_\mu.{i} }{2 \spa{j}.i \spb{j}.i } =1\\
\epsilon(i^+,j) \cdot \epsilon(i^+,j)=\frac{ \spab{j}.\gamma^\mu.i \spab{j}.\gamma_\mu.i } {2\spa{j}.i \spa{j}.i }=0
\end{gather}
where we used the Fierz Identity and the antisymmetry of the spinor product.
With this formalism it is possible to efficiently describe the kinematical factors of an amplitude. But there is a second part to the amplitude, namely, the color structure. It turns out that we can treat this part rather trivially as we will see in the next section.

\section{Color Decomposition}
Generally there are two parts in the Feynman rules of a gauge theory. One is the structure constant of the underlying gauge group and the other one is kinematical part of the external particles. Here we will present a way of solving the color structure beforehand. This will correspond to a reordering of the amplitude in terms of so called color-ordered amplitudes. This has two advantages. First, the number of independent color-ordered amplitudes is lower than the number Feynman diagrams and secondly each color-ordered amplitude is gauge invariant in contrast to a single Feynman diagram.\\
In our case the underlying gauge group is $SU(N)$ with structure constants $f^{abc}$. Normally the Feynman rules are written in terms of these structure constants but in order to strip off the color we will find it beneficial to transform them into the fundamental representation where we replace the structure constants with the corresponding traces over the generators of $SU(N)$
\begin{gather}
f^{abc}=-\frac{i}{\sqrt{2}}\left( (T^a)_i^j(T^b)_j^k(T^c)_k^i-(T^a)_i^j(T^c)_j^k(T^b)_k^i \right).
\label{trafo}
\end{gather}
The fundamental indices $i,j$ and $k$ represent the rows and columns in the generators and the adjoint indices $a,b$ and $c$ tell us which generator we have to use. \\
The next step is to remove all contracted adjoint indices by the repeated use of the Fierz-Identity
\begin{gather}
(T^a)_i^j(T^a)_k^l= \delta_i^l \delta_k^j - \frac{1}{N} \delta_i^j \delta_k^l .
\label{sun}
\end{gather}
Finally we are able to collect all independent traces of generators appearing in the amplitude and their corresponding kinematical parts. These kinematical parts are then called color-ordered amplitudes. This leads us to an alternative expansion of the tree-level amplitude
\begin{gather}
\mathcal{A}_n(1,2..,n)=g^{n-2}\sum_{\sigma \in S_n/Z_n}{Tr(T^{\sigma(1)}T^{\sigma(2)}..T^{\sigma(n)})A_n(\sigma(1),\sigma(2)..,\sigma(n))}
\label{colororderd}
\end{gather}
where $A_n(\sigma(1),\sigma(2)..,\sigma(n))$ are the color-ordered amplitudes and the sum runs over all permutations $S_n$ of n elements except over the cyclic ones denoted by the subgroup $Z_n$. \\
Later we will apply this color decomposition to an example but first we should discuss a graphical approach to convert the color factors.\\

\subsection{Color Algebra in Graphform}
\begin{figure}[h]
	\centering
		\includegraphics[width=0.50\textwidth]{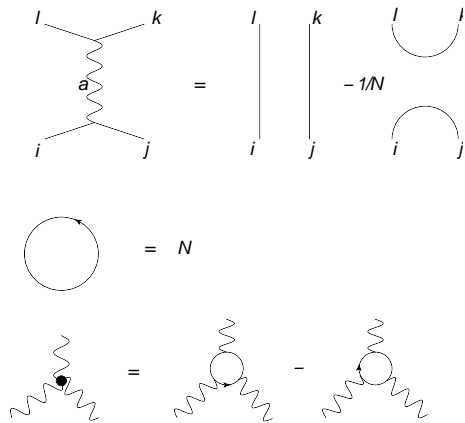}
	\caption{Ways to simplify a graph representation of the color algebra}
	\label{fig:sun id}
\end{figure}

\begin{figure}[h]
	\centering
		\includegraphics[width=0.20\textwidth]{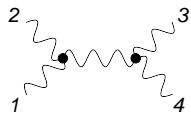}
	\caption{Graph representation of the color factor $f^{12a}f^{34a}$}
	\label{fig:schannel}
\end{figure}

An even faster way to transform the color factors is to represent them with graphs, do the color algebra there, and then read off the corresponding traces. To do that every fundamental index is represented by an 'quark' line and every adjoint index is represented by a 'gluon' line. This means the full generator $(T^a)_i^j$ is represented by an quark-gluon vertex. The three main operations for simplifying these graphical representations are displayed in figure \ref{fig:sun id} which are just graphical representations of the equations (\ref{trafo},\ref{sun}). \\
As an example we are going to show how a s-channel color factor of a four-point amplitude
\begin{gather}
c_s=f^{12a}f^{a34}
\end{gather}
is translated into traces over the generators. The color factor can be represented by the graph in figure \ref{fig:schannel}.
The next step is to use the identities displayed in figure \ref{fig:sun id} to simplify the graph.
\begin{figure}[h]
	\centering
		\includegraphics[width=0.70\textwidth]{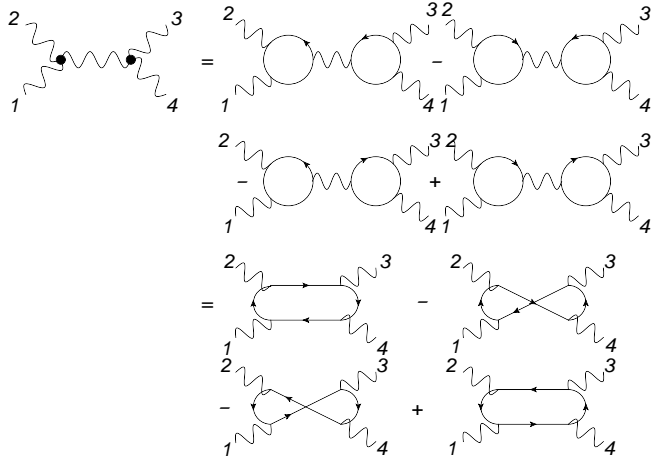}
		\caption{Using graphical methods to show eq. (\ref{schantrace})}
	\label{fig:schantrace}
\end{figure}
From the last line in figure \ref{fig:schantrace} we can simply read off the ordering of the generators in the trace by following the closed fermion line.
\begin{gather}
c_s=Tr(T^1T^2T^3T^4)-Tr(T^1T^3T^4T^2)-Tr(T^1T^2T^4T^3)+Tr(T^1T^4T^3T^2)
\label{schantrace}
\end{gather}

\subsection{Color Decomposition of a Four-Point Amplitude}
Going back to our color decomposition of tree-level amplitudes in (\ref{colororderd}) we will apply this algorithm now to a four-point amplitude.\\
In the normal Feynman diagram expansion we have four diagrams with their corresponding color factors
\begin{gather}
\mathcal{A}_4=c_sk_s+c_tk_t+c_uk_u+c_sk_{s,4}+c_tk_{t,4}+c_uk_{u,4}
\label{feynman4point}
\end{gather}
where $k$ denotes the kinematical part and $k_{x,4}$ is the four-point vertex part which has the color factor of channel $x$. The color factors of the channels are given by
\begin{gather}
c_s=f^{12a}f^{a34} \\
c_t=f^{23a}f^{a41} \\
c_u=f^{13a}f^{a24}. 
\end{gather}
We already translated the first color factor into traces over generators in (\ref{schantrace}). The other color factors can be obtained by a simple relabeling of the s-channel color factor $c_s$
\begin{gather}
c_t=Tr(T^1T^2T^3T^4)-Tr(T^1T^4T^2T^3)-Tr(T^1T^3T^2T^4)+Tr(T^1T^4T^3T^2)\\
c_u=Tr(T^1T^3T^4T^2)-Tr(T^1T^3T^2T^4)-Tr(T^1T^4T^2T^3)+Tr(T^1T^2T^4T^3).
\label{uchantrace}
\end{gather}
If we plug the transformed color factors into (\ref{feynman4point}) we find
\begin{gather}
\begin{split}
\mathcal{A}_4=\left(Tr(T^1T^2T^3T^4)-Tr(T^1T^3T^4T^2)-Tr(T^1T^2T^4T^3)+Tr(T^1T^4T^3T^2)\right)\\\left(k_s+k_{s,4}\right)\\
+\left(Tr(T^1T^2T^3T^4)-Tr(T^1T^4T^2T^3)-Tr(T^1T^3T^2T^4) +Tr(T^1T^4T^3T^2)\right)\\
\left(k_t+k_{t,4} \right)\\
+\left(Tr(T^1T^3T^4T^2)-Tr(T^1T^3T^2T^4)-Tr(T^1T^4T^2T^3)+Tr(T^1T^2T^4T^3)\right)\\
\left(k_u+k_{u,4} \right)
\end{split}\\
\begin{split}
=Tr(T^1T^2T^3T^4)\left(k_s+k_{s,4}+k_t+k_{t,4}\right)\\
+Tr(T^1T^3T^4T^2)\left(-k_s-k_{s,4}+k_u+k_{u,4}\right)\\
+Tr(T^1T^2T^4T^3)\left(-k_s-k_{s,4}+k_u+k_{u,4}\right)\\
+Tr(T^1T^4T^3T^2)\left(k_s+k_{s,4}+k_t+k_{t,4}\right)\\
+Tr(T^1T^4T^2T^3)\left(-k_t-k_{t,4}-k_u-k_{u,4}\right)\\
+Tr(T^1T^3T^2T^4)\left(-k_t-k_{t,4}-k_u-k_{u,4}\right).
\label{kinfactor}
\end{split}
\end{gather}
Now we can define the kinematical objects in front of the traces as the color-ordered amplitudes
\begin{gather}
\begin{split}
=Tr(T^1T^2T^3T^4)A(1234)+Tr(T^1T^3T^4T^2)A(1342)\\
+Tr(T^1T^2T^4T^3)A(1243)+Tr(T^1T^4T^3T^2)A(1432)\\
+Tr(T^1T^4T^2T^3)A(1423)+Tr(T^1T^3T^2T^4)A(1324).
\label{full4point}
\end{split}
\end{gather}
At first it looks like we introduced more objects then before in (\ref{feynman4point}) but from our definitions in (\ref{kinfactor}) we immediately find the first identities for the color-ordered amplitudes
\begin{gather}
\label{cyclic}
A(ijkl)=A(jkli)\\
A(ijkl)=A(lkji)
\label{reflection}
\end{gather}
which are based on the cyclicity of the trace and C-invariance. It turns out that there are even more identities for color-ordered amplitudes which greatly simplify the number of diagrams which we have to calculate.\cite{Kleiss-Kuijf,Bern:2008qj} \\
As we have seen in our simple example of a four-point amplitude. The color-ordered amplitude $A(1234)$ had only contributions coming from s-channel, the t-channel and from the contact term. This statement can be generalized to: Every color-ordered amplitude can only have contributions from Feynman graphs where adjacent particles become collinear and form a pole in the amplitude and the corresponding contact terms.\\
This will become important later on when we are talking about the color-kinematics duality. 

\subsection{Photon Decoupling Equation}
With some basic knowledge of color-ordered amplitudes we can derive another identity for tree amplitudes. If we have a look at the complete four-point amplitude again (\ref{full4point}) we find
\begin{gather}
\begin{split}
\mathcal{A}_4^{tree} = g^2 \left[ Tr(T^1T^2T^3T^4)A(1,2,3,4) + Tr(T^1T^2T^4T^3)A(1,2,4,3)\right. \\  Tr(T^1T^3T^2T^4)A(1,3,2,4) + Tr(T^1T^4T^2T^3)A(1,4,2,3) \\
 + \left. Tr(T^1T^3T^4T^2)A(1,3,4,2)+Tr(T^1T^4T^3T^2)A(1,4,3,2) \right]
\end{split} \\
\begin{split}
=  g^2 A(1,2,3,4) \left[ Tr(T^1T^2T^3T^4)+ Tr(T^1T^4T^3T^2) \right] \\
 + g^2 A(1,4,2,3)  \left[ Tr(T^1T^4T^2T^3) + Tr(T^1T^3T^2T^4) \right] \\
 +g^2 A(1,3,4,2) \left[Tr(T^1T^3T^4T^2) + Tr(T^1T^2T^4T^3) \right] 
\end{split}
\label{full color}
\end{gather}
where we used the cyclicity of the color-ordered amplitudes (\ref{cyclic}) and the reflection identity (\ref{reflection}).
We would find the same decomposition as in equation (\ref{full color})  for the gauge group $U(N)=SU(N)\times U(1)$ where the $U(1)$ corresponds to a 'photon'. The correspondence can be seen by having a look at the field strength tensor
\begin{equation}
F_{\mu \nu}^a= \partial_\mu A_\nu^a - \partial_\nu A_\mu^a - i g f^{a}_{bc} A_\mu^b A_\nu^c .
\end{equation}
Since the generator of the $U(1)$ is the unit matrix, the part proportional to the structure constant will drop out, reducing the non-abelian field strength tensor to an abelian one. 
\begin{gather}
F_{\mu \nu}= \partial_\mu A_\nu - \partial_\nu A_\mu
\end{gather}
But this means that the corresponding particle, the photon, will have no interactions with the particles associated with the $SU(N)$ group. Therefore every scattering amplitude containing a photon will vanish. \\
In order to derive the photon decoupling equation we will pick one of the generators to be the generator of $U(1)$ and than demand that our full amplitude must vanish. Here we choose the fourth generator to be the unit matrix $T^4= \mathbbm{1}$
\begin{gather}
\begin{split}
0 =  g^2 A(1,2,3,4) \left[ Tr(T^1T^2T^3)+ Tr(T^1T^3T^2) \right] \\
 + g^2 A(1,4,2,3)  \left[ Tr(T^1T^2T^3) + Tr(T^1T^3T^2) \right] \\
 +g^2 A(1,3,4,2) \left[Tr(T^1T^3T^2) + Tr(T^1T^2T^3) \right] 
\end{split} \\
\begin{split}
\Rightarrow 0 =  g^2 \left(A(1,2,3,4) +  A(1,4,2,3) + A(1,3,4,2) \right) \\
 \left[ Tr(T^1T^2T^3)+Tr(T^1T^3T^2) \right] .
\end{split} 
\end{gather}
There is at least one combination of generators which will make the second part non-vanishing, hence the linear combination of color-ordered amplitudes must be zero
\begin{gather}
  A(1,2,3,4) +  A(1,4,2,3) + A(1,3,4,2) =0 .
\end{gather}
This is the photon decoupling equation for a four-point amplitude. A second way of deriving this equation is to calculate the full matrix element and demand that there is no difference between the matrix elements of $U(N)$ and $SU(N)$. This condition then boils down to the photon decoupling identity. \\
Furthermore the derivation of the photon decoupling identity can be generalized to an arbitrary number of legs
\begin{gather}
A_n(1,2,3,..,n)+A_n(2,1,3,..,n)+A_n(2,3,1,..,n)+..+A_n(2,3,..,1,n)=0.
\end{gather}

\subsection{Color-Ordered Feynman Rules}
\begin{figure}[h]
\begin{minipage}[b]{0.45\linewidth}
	\centering
		\includegraphics[width=0.3\textwidth]{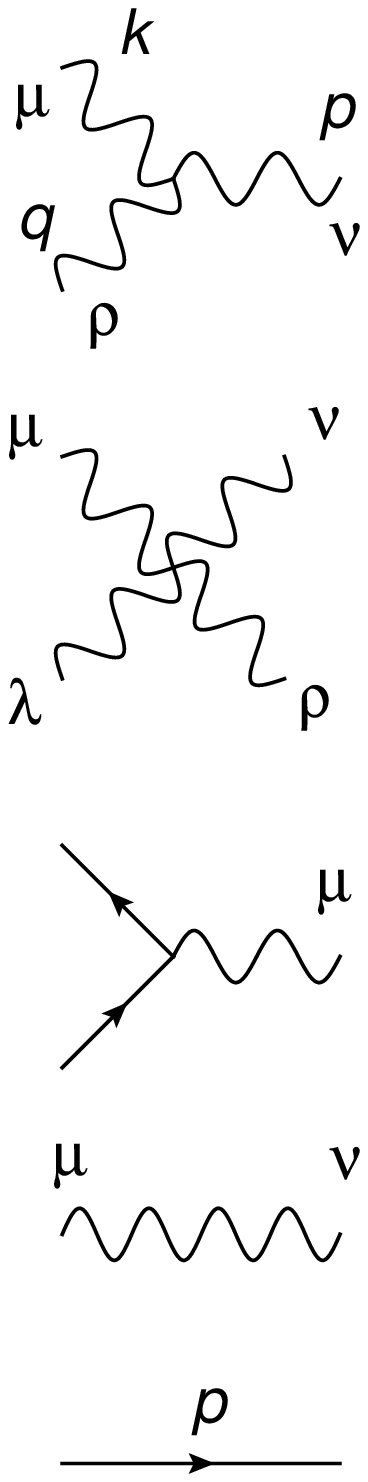}
		\label{asdn}
		\vspace{0.2cm}
	\end{minipage}
	\begin{minipage}[b]{0.45\linewidth}
	\begin{gather*}
	\frac{i}{\sqrt{2}} \left(\eta_{\nu\rho}(p-q)_\mu+\eta_{\rho \mu}(q-k)_\nu+\eta_{\mu \nu}(k-p)_\rho   \right) \\[0.7cm]
	\frac{i}{2}\left( 2\eta_{\mu \rho} \eta_{\nu \lambda} - (\eta_{\mu \nu}\eta_{\rho \lambda} + \eta_{\mu \lambda} \eta_{\nu \rho} ) \right) \\[0.7cm]
	\frac{i}{\sqrt{2}}\gamma_\mu \\[0.3cm]
	\frac{-i}{p^2} \eta_{\mu \nu} \\[0.1cm]
	\frac{i \slashed{p}}{p^2}  
	\end{gather*}
	
	\end{minipage}
\caption{Color-ordered Feynman rules for QCD}
\label{feynmanrulesqcd}
\end{figure}

The question raised should be if there is a direct way to obtain these color-ordered amplitudes. The answer is yes. We can find color-ordered Feynman rules which allow us to compute these amplitudes directly from their contributing Feynman diagrams. We can obtain these rules by transforming the structure constants in the normal Feynman rules into the fundamental representation and then collect all pieces which correspond to a specific ordering of external particles. Doing this we arrive at the color-ordered Feynman rules displayed in figure \ref{feynmanrulesqcd}.

\subsubsection{Obtaining $A(1234)$ through Color-Ordered Feynman Rules}
\begin{figure}[h]
	\centering
		\includegraphics[width=0.65\textwidth]{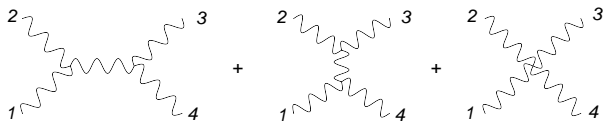}
	\caption{Feynman diagrams contributing to the color-ordered four-point amplitude $A(1,2,3,4)$}
	\label{fig:feynmanfourpoint}
\end{figure}

The color-ordered four-point amplitude $A(1^-2^-3^+4^+)$ has contributions from three Feynman diagrams: a s-channel, a t-channel and the four-point vertex which are displayed in figure \ref{fig:feynmanfourpoint}.
Using the color-ordered Feynman rules in figure \ref{feynmanrulesqcd} we find 
\begin{gather}
\begin{split}
A(1^-2^-3^+4^+)= \ \ \ \ \ \ \ \ \ \ \ \ \ \ \ \ \ \ \ \ \ \  \frac{ig^2}{2}\left[\frac{-i}{s_{12}} \left[ \epsilon(1^-,i) \cdot \epsilon(2^-,i) (p_1-p_2)^\mu \right. \right.\\
\left. + \epsilon^\mu(2^-,i)(2p_2+p_1)\cdot\epsilon(1^-,i)+\epsilon^\mu(2^-,i)(-2p_1-p_2)\cdot \epsilon(2^-,i)  \right] \\
\cdot \left[ \epsilon(3^+,j) \cdot \epsilon(4^+,j) (p_3-p_4)^\mu+ \epsilon^\mu(4^+,j)(2p_4+p_3)\cdot\epsilon(3^+,j) \right. \\
\left.+\epsilon^\mu(3^+,j)(-2p_3-p_4)\cdot \epsilon(4^+,j) \right]
+\frac{-i}{s_{14}}\left[ \epsilon(4^+,j) \cdot \epsilon(1^-,i) (p_4-p_1)^\mu \right. \\
\left. + \epsilon^\mu(1^-,i)(2p_1+p_4)\cdot\epsilon(4^+,j)+\epsilon^\mu(4^+,j)(-2p_4-p_1)\cdot \epsilon(1^-,i)  \right]\\
\cdot \left[ \epsilon(2^-,i) \cdot \epsilon(3^+,j) (p_2-p_3)^\mu+ \epsilon^\mu(3^+,j)(2p_3+p_2)\cdot\epsilon(2^-,i)+ \right.\\
\left. \epsilon^\mu(2^-,i)(-2p_2-p_3)\cdot \epsilon(3^+,j) \right] 
-i \left[2\epsilon(1^-,i)\cdot \epsilon(3^+,j)\epsilon(2^-,i)\cdot \epsilon(4^+,j) \right. \\
\left.  -\epsilon(1^-,i) \cdot \epsilon(2^-,i)\epsilon(3^+,j)\cdot \epsilon(4^+,j)-\epsilon(1^-,i)\cdot \epsilon(4^+,j)\epsilon(2^-,i)\cdot \epsilon(3^+,j) \right] \bigg] .
\end{split}
\end{gather}
The reference momenta in the polarization vectors correspond to different gauge choices. Since  color-ordered amplitudes are gauge invariant objects we can use any convenient gauge. We only have to be careful to stick to one gauge choice during the calculation.\\
A handy gauge is to choose a momentum of the opposite helicity as reference momentum. This will lead to the vanishing of most scalar products between the polarization vectors. In our example we will choose $p_4$ as a reference momentum for the polarization vectors with negative helicity and $p_1$ with positive helicity. Therefore all but the scalar product of $\epsilon(2^-,4)\cdot \epsilon(3^+,1)$ will vanish and we are only left with
\begin{gather}
A(1^-2^-3^+4^+)=\frac{-2g^2}{s_{12}} \epsilon(2^-,4)\cdot \epsilon(3^+,1) \: p_2 \cdot \epsilon(1^-,4) \: p_3 \cdot \epsilon(4^+,1) \\
=-g^2\frac{\spa1.2 ^2 \spb3.4 }{\spa3.4 \spb4.1 \spa4.1 }. 
\end{gather}
The next step is multiplying and dividing by $\spa1.2$ and applying momentum conservation which gives us the well known result
\begin{gather}
A(1^-2^-3^+4^+)=g^2\frac{\spa1.2 ^4}{\spa1.2 \spa2.3 \spa3.4 \spa4.1 }
\end{gather}

\section{Factorization and Recursion Relations}
Next we will turn to a basic property of any amplitude: factorization. This property controls the behavior of an amplitude in the limit where one of the internal propagators goes on-shell. By inverting this limit and with the help of complex analysis it was possible to derive the BCFW recursion relations \cite{Britto:2004ap} which enables us to build tree-level amplitudes from a product of two lower-point amplitudes.  \\

\subsection{Factorization}
Factorization tells us that an amplitude splits up into a product of two lower point amplitudes if one of the internal propagators goes on-shell
\begin{gather}
A(1,2,..,i,..,n)\overset{(\sum_{j=1}^i p_j)^2 \rightarrow m_j }{\rightarrow} \sum_\lambda{ A_L(1,2,..,i,l^\lambda)\frac{1}{(\sum_{j=1}^i p_j)^2}A_R(-l^{-\lambda},i+1,..,n)}
\end{gather}
where the sum runs over all possible intermediate particles and $m_j$ is the mass of the internal propagator.\\
Factorization is connected to the unitarity of the S-Matrix which makes it an extremely fundamental attribute of scattering amplitudes. Moreover it is also one of the reason why MHV and $\overline{MHV}$ amplitudes have such easy expressions. Because if we try to factorize an MHV amplitude we only have three gluons with negative helicity to distribute between the lower point amplitudes. However since all amplitudes of the type $A(1^\pm,2^+,3^+..,\text{n}^+)$ vanish the factorization channel must also vanish. \\ 
\subsection{A Recursive Relation for Tree-Level Amplitudes}
In \cite{Britto:2004ap} it was first described and proven that it is possible to reconstruct the whole color-ordered amplitude if we know all its factorization channels. If we consider a tree-level amplitude $A(1,2,..,i,..,a-1,a,..,j,..,n)$ this can be achieved by introducing a complex shift on two external legs $i$ and $j$
\begin{gather}
\hat{p}^\mu_i=p^\mu_i+z \epsilon^\mu_{ij} \\
\hat{p}^\mu_j=p^\mu_j-z \epsilon^\mu_{ij}
\end{gather}
where the complex shift is defined as 
\begin{gather}
\epsilon^\mu_{ij}= \frac{ \spab{i}.{\gamma^\mu}.j} {2} .
\end{gather}
The introduced shift leaves the square of the momenta and the total momentum conservation invariant
\begin{gather}
(\hat{p}_i)^2=(p_i)^2 \\
\hat{\slashed{p}}_i+ \hat{\slashed{p}}_j = p_i+p_j 
\label{invmandel}.
\end{gather}
Since we have a complex shift we can now consider a contour integral over the shifted amplitude
\begin{gather}
\oint{\frac{dz}{2\pi i} \frac{1}{z}A(z)}
\end{gather} 
where we manually introduced a pole at $z=0$. If $A(z)$ vanishes as $z$ approaches infinity then this integral vanishes. But Cauchys theorem also tells us that this contour integral is given by its residues.
\begin{gather}
\oint{\frac{dz}{2\pi i} \frac{1}{z}A(z)}= \sum_{i} Res\left(\frac{A(z)}{z},\zeta_i \right) .
\end{gather} 
The residues of a simple pole $\zeta$ is given by
\begin{gather}
Res\left(\frac{A(z)}{z},\zeta \right)=\underset {z \rightarrow \zeta}  {\text{lim}} (z-\zeta)\frac{A(z)}{z} .
\label{residue}
\end{gather} 
The only pole in $z$ of an amplitude comes from on-shell intermediate particles and therefore correspond to a factorization channels. But as we have already seen in eq. (\ref{invmandel}) the shifts we introduce cancel out when both shifted legs are on the same side of the factorization channel. So the only possibility to obtain a pole is from a configuration where $\hat{i}$ and $ \hat{j}$ are on different sides of the factorization channel. If we consider leg $\hat{j}$ to be on one side with the legs $a$ to $b$ the intermediate on-shell propagator is given by
\begin{gather}
s_{a...b}(z)= \left( \sum^b_{m=a}{p^\mu_m} - z \epsilon^\mu_{ij} \right)^2 \\
= \left( \sum^b_{m=a}{p^\mu_m} \right)^2 - z \spab{i}.{\sum^b_{m=a}{\slashed{p}_m}}.j \\
= s_{a..b}(0) - z \spab{i}.{\sum^b_{m=a}{\slashed{p}_m}}.j  . 
\end{gather}
Therefore we have the following pole
\begin{gather}
\zeta = \frac{ s_{a..b} (0) } { \spab{i}.{\sum^b_{m=a}{\slashed{p}_m} }.j } .
\end{gather}
Looking back at the whole expression for the residue (\ref{residue}) we find
\begin{gather}
Res\left( \frac{A(z)}{z},\zeta \right) = \underset {z \rightarrow \zeta}  {\text{lim}} (z-\zeta)\frac{A(z)}{z} \\
=\underset {z \rightarrow \zeta} {\text{lim}} \frac{1}{z}  A_L(b+1,..,\hat{i},..,a-1,l) \frac{(z-\zeta)}{s_{a..b}(0)- z \spab{i}.{\sum^b_{m=a} { \slashed{p}_m }}.j} A_R(-l,a,..,\hat{j},..,b) \\
= \underset {z \rightarrow \zeta} {\text{lim}} \frac{1}{z} A_L(b+1,..,\hat{i},..,a-1,l) \frac{(z-\zeta)}{s_{a..b}(0)(1 - \frac{z}{\zeta}) } A_R(-l,a,..,\hat{j},..,b) \\
= \underset {z \rightarrow \zeta} {\text{lim}}  \frac{1}{z} A_L(b+1,..,\hat{i},..,a-1,l)\frac{\zeta (z-\zeta)}{s_{a..b}(0)(\zeta - z) } A_R(-l,a,..,\hat{j},..,b) \\
= \underset {z \rightarrow \zeta} {\text{lim}} -\frac{1}{z} A_L(b+1,..,\hat{i},..,a-1,l)\frac{\zeta}{s_{a..b}(0)} A_R(-l,a,..,\hat{j},..,b) \\
=- A_L(b+1,..,\hat{i},..,a-1,l)\frac{1}{s_{a..b}(0)} A_R(-l,a,..,\hat{j},..,b).
\end{gather}
Taking into account that every possible subset of external legs, where the shifted legs $\hat{i}$ and $\hat{j}$ are in different sets, provide a pole $\zeta_i$ we get the following expression for the contour integral
\begin{gather}
\oint{\frac{dz}{2\pi i} \frac{1}{z}A(z)}= A(0) + \sum^{n}_{i=1} Res \left( \frac{A(z)}{z},\zeta_i \right) 
\end{gather}
where we used that the manually introduced pole at $\zeta_0=0$ is given by 
\begin{gather}
Res\left( \frac{A(z)}{z},\zeta_0 \right) = \underset {z \rightarrow 0}  {\text{lim}} (z)\frac{A(z)}{z}=A(0) .
\end{gather}
But since this contour integral vanishes we arrive at the following recursion relation
\begin{gather}
A(0)=\sum_{a,b}{ A_L(b+1..,\hat{i},..,a-1,l(\zeta))\frac{1}{s_{a..b}(0)}A_R(-l(\zeta),a,..,\hat{j},..,b)}
\end{gather}
which connects the real amplitude $A(0)$ to its factorization channels into lower-point amplitudes evaluated at complex momenta.
\chapter{Color-Kinematic Duality}
The color-kinematics duality discovered by Bern, Carrasco and Johansson (BCJ) \cite{Bern:2008qj}  provides a different way of dealing with the color factors of an amplitude. As the name suggest this approach treats the color and kinematic factors of gauge amplitudes on the same footing. Therefore we will introduce a new relation between numerators of an amplitude which is the kinematical analog to the well known Jacobi Identity of the structure constants
\begin{equation}
f^{abi}f^{icd} + f^{aci}f^{idb} - f^{adi}f^{ibc} =0 .
\end{equation}
We will split the discussion of the color-kinematic duality into two parts. First we will consider the duality at tree-level where it has been shown to exist \cite{bcjproove1,bcjproove2,bcjproove3,bcjproove4} and later at loop level where it is still a conjecture. However there is a vast variety of nontrivial evidence in the literature \cite{Bern:2010ue,ck4l,OneTwoLoopN4,SchnitzerBCJ,White,OneLoopN1Susy,OConnellRational,Bern:2013yya}.

\section{The Duality at Tree-Level}
At tree-level the duality enables us to expand the full amplitude in the following way
\begin{gather}
\mathcal{A}_n^{tree}(1,2,..n)=\sum_i{\frac{n_ic_i}{\prod_j{p^2_j}}}
\label{bcj amplitude}
\end{gather}
where $n_i$ and $c_i$ are the kinematical and the color factor respectively which both fulfill all possible Jacobi Identities. For example if we have a Jacobi Identity of the form
\begin{gather}
c_i+c_j+c_k=0
\end{gather}
then we will demand that the corresponding numerators satisfy the same identity
\begin{gather}
n_i+n_j+n_k=0 .
\end{gather}
 In the upcoming part we will outline how such a representation can be found for a four-point amplitude.\\

\subsection{Duality Representation of a Four-Point Amplitude}
\begin{figure}[h]
	\centering
		\includegraphics[width=0.60\textwidth]{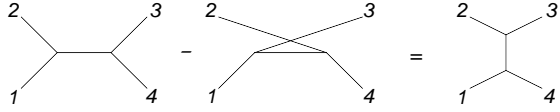}
	\caption{Jacobi Identity at four-points}
	\label{fig:jacobiid}
\end{figure}

A four-point amplitude can be expanded in its channels accordingly to equation (\ref{bcj amplitude})
\begin{gather}
\mathcal{A}_n^{tree}(1,2,3,4)=\frac{n_sc_s}{s}+\frac{n_tc_t}{t}+\frac{n_uc_u}{u}
\label{4bcj}
\end{gather}
where the Mandelstem variables are given by $s=(p_1+p_2)^2$, $t=(p_1+p_4)^2$ and $u=(p_1+p_3)^2$.
 At first glance equation (\ref{4bcj}) looks a lot like the normal expansion in Feynman diagrams but note that the contact term has already been absorbed into the three-point vertices and that the numerators are gauge invariant since they are linear combinations of color-ordered amplitudes as we will see later. \\
In order to find such an representation we should first consider the Jacobi Identity at four-points shown in figure \ref{fig:jacobiid}.

\begin{equation}
f^{12b}f^{b34} - f^{42b}f^{b31} = f^{41b}f^{b23}
\label{colorf1}
\end{equation}
or expressed in the three channels we have
\begin{gather}
c_s-c_u=c_t.
\label{colorf2}
\end{gather}
Now we will demand that the corresponding kinematical parts fulfill the same identity
\begin{gather}
n_s-n_u=n_t .
\label{numbcj4}
\end{gather}
The first step to find such numerators is to expand the well known color-ordered amplitudes in the channels they can factorize in
\begin{gather}
\begin{split}
 A(1,2,3,4)=\frac{n_s} {s} + \frac{n_t}{t} \\ 
 A(1,3,4,2)=-\frac{n_u} {u} - \frac{n_s}{s} \\
 A(1,4,2,3)=-\frac{n_t} {t} + \frac{n_u}{u} .
\end{split}
 \label{bcjpoles}
\end{gather}
The sign of these numerators are not arbitrarily chosen, they correspond to the right ordering of color factors we defined before (\ref{colorf1},\ref{colorf2}). In order to get an expression in terms of color-ordered amplitudes for these numerators we will write down functional equations, which demand that the numerators behave exactly like their color factors
\begin{gather}
\begin{split}
n_s(1234)=-n_s(1243)\\
n_s(1234)=-n_s(2134)\\
n_s(1234)=n_s(4321) .
\end{split}
\label{colorsym}
\end{gather}
To solve these equations we will choose an ansatz in terms of the two independent color-ordered amplitudes and the two independent Mandelstam variables. From dimensional analysis we conclude that the numerators consist of an color-ordered amplitudes multiplied by a Mandelstam variable, therefore we can form the following ansatz
\begin{gather}
n_s(1234)=\alpha_1 sA(1234)+\alpha_2 s A(1423)+\alpha_3 t A(1234)+\alpha_4 t A(1423) .
\end{gather}
The first equation in (\ref{colorsym}) is trivially satisfied since the color-ordered amplitudes and the Mandelstam variables are invariant under reflections. But the last two equations give us
\begin{gather}
\alpha_1=\alpha_3\\
\alpha_2=0\\
\alpha_4=0
\end{gather}
where we used the following identities for four-point amplitudes
\begin {gather}
tA(1,2,3,4)= u A(1,3,4,2) \\
sA(1,2,3,4)= u A(1,4,2,3) \\
tA(1,4,2,3)= s A(1,3,4,2).
\label{4pteq}
\end{gather}
Now we can repeat these steps for the other two numerators and arrive at
\begin{gather}
n_s=\alpha_1 \frac{sA(1234)}{u}(u-t)\\
n_t=\beta_1 \frac{tA(1234)}{u}(u-s)\\
n_u=\gamma_1 \frac{uA(1234)}{u}(s-t) .
\end{gather}
With the help of the expansion of the color-ordered amplitudes in their poles (\ref{bcjpoles}) we can fix all three parameters $\alpha_1,\beta_1$ and $\gamma_1$
\begin{gather}
n_s= \frac{sA(1234)}{3u}(u-t)\\
n_t= \frac{tA(1234)}{3u}(u-s)\\
n_u= \frac{uA(1234)}{3u}(s-t) .
\end{gather}
We can check that this representation satisfies the Jacobi Identity
\begin{gather}
n_s-n_t=\frac{A(1234)}{3u} \left(su-st -tu + st \right)= \frac{uA(1234)}{3u}(s-t)=n_u 
\end{gather}
and that the expansion in (\ref{4bcj}) is really valid
\begin{gather}
\mathcal{A}_n^{tree}(1,2,3,4)=\frac{n_sc_s}{s}+\frac{n_tc_t}{t}+\frac{n_uc_u}{u} \\
= \frac{A(1234)}{3u} \left[ (u-t)c_s + (u-s)c_t + (s-t)c_u  \right] \\
= \frac{A(1234)}{3u} \left[ u (c_s+c_t) + s (c_u-c_t) + t (-c_s-c_u)  \right] \\
= \frac{1}{3} \left[ A(1234) (c_s+c_t) + A(1423) (c_u-c_t) +  A(1342) (-c_s-c_u)  \right]
\end{gather}
where we used the identities in (\ref{4pteq}). Transforming the color factors into the adjoint representation, using the photon decoupling identity and collecting all terms corresponding to an independent trace we arrive at
\begin{gather}
\begin{split}
=Tr(T^1T^2T^3T^4)A(1234)+Tr(T^1T^3T^4T^2)A(1342)\\
+Tr(T^1T^2T^4T^3)A(1243)+Tr(T^1T^4T^3T^2)A(1432)\\
+Tr(T^1T^4T^2T^3)A(1423)+Tr(T^1T^3T^2T^4)A(1324)
\end{split}
\end{gather}  
which we already know as the color decomposition of a four-point amplitude (\ref{full4point}).
We might get the impression that these numerators are completely determined by the derivation we introduced above, but this is not the case as we will discuss in the next section.

\subsection{Gauge Freedom of Numerators}
If we apply a transformation on the numerators of the following form
\begin{gather}
n'_s=n_s+\alpha(k_i,\epsilon_i)s \\
n'_t=n_t-\alpha(k_i,\epsilon_i)t \\
n'_u=n_u-\alpha(k_i,\epsilon_i)u 
\end{gather}
we will keep the pole structure of the amplitude (\ref{4bcj}) intact since the prefactor of our transformation $\alpha(k_i,\epsilon_i)$ will vanish. Furthermore it also leaves the Jacobi Identity for the kinematical factors invariant
\begin{gather}
n'_s-n'_t-n'_u=n_s-n_t-n_u+\alpha(k_i,\epsilon_i)\left(s+t+u\right)=n_s-n_t-n_u
\end{gather}
since in the massless case we have $s+t+u=0$.
These transformation are called Generalized Gauge Freedom. 
We can also choose an expression
\begin{equation}
\alpha(k_i,\epsilon_i)=\frac{n_u}{u}
\end{equation}
which sets $n'_u=0$. But our u-pole is not gone it has just been absorbed into the numerators $n'_s$ and $n'_t$.
With this choice our amplitudes now have the following expansion
\begin{gather}
A(1,2,3,4)=\frac{n'_s}{s}+\frac{n'_t}{t} \\
A(1,3,4,2) = -\frac{n'_s} {s} \\
A(1,4,2,3) = -\frac{n'_t} {t} .
\end{gather}
From this we can already see that if we plug the last two equations into the first one we arrive at the photon decoupling equation. But if we take the last two on themselves we can derive
\begin{gather}
0=sA(1,3,4,2) +n'_s \\
0=tA(1,4,2,3) + n'_t \\
\Rightarrow sA(1,3,4,2) +n'_s= tA(1,4,2,3) + n'_t .
\end{gather}
Realizing that the kinematic Jacobi equation now reads $0=n'_s-n'_t$
we find
\begin{equation}
sA(1,3,4,2)= tA(1,4,2,3)
\end{equation}
which is one of the equations which were known at four-points (\ref{4pteq}). Similarly we can derive all the other equations by setting the other numerators to zero
\begin{gather}
tA(1,2,3,4)=uA(1,3,4,2)\\
sA(1,2,3,4)=uA(1,4,2,3) .
\end{gather}

\subsection{Duality Representation of a n-Point Amplitude}
From \cite{amplitude school} we can see that the way of finding BCJ conform numerators, we presented here, can be generalized.\\
\begin{enumerate}
	\item Form a basis of color-ordered amplitudes and expand them in their poles 
	\item Write down all Jacobi Relations and graph automorphisms 
	\item Create an ansatz out of the color-ordered amplitudes times Mandelstam variables with the right dimension of the numerators 
	\item Solve the functional equations of step two with the ansatz
\end{enumerate}
The obvious problem of this method is that the ans\"atze get unmanageable since the number of independent color-ordered amplitudes and the number of Mandelstem variables grow very fast.

\section{The Duality at Loop-level}
The expansion of a loop-level amplitude is very similar to the one at tree-level. 
\begin{gather}
\mathcal{A}^{L-loop} = \int{\left(\prod_m^L{\frac{d^Dl_m}{(2\pi)^D}}\right)\sum_{\text{graphs,perms}}\frac{1}{S_i}\frac{N_i C_i}{\prod_j{D_j}}}
\label{loopbcj}
\end{gather}
where $S_i$ is a symmetry factor which accounts for over counting in the sum.
As in the tree-level case we will demand that the numerators will behave accordingly to their color factors. This means if we have a Jacobi relation of the form
\begin{gather}
C_i+C_j+C_k=0
\end{gather} 
we demand that the numerators to follow the same equations
\begin{gather}
N_i+N_j+N_k=0 .
\end{gather}
It is important to mention that unlike in the tree-level case, the BCJ relations at loop-level have not been proven. This means if we construct an amplitude with the help of the BCJ equation we have to check if their unitarity cuts are correct. This has been done for several examples so far \cite{Carrasco:2011mn,4point4loop,4point5loop,4point6loop}. \\
The real strength in using BCJ relations lies in the fact that the number of graphs that we need to generate is drasticly reduced as we will see in the case of the two-loop five-point amplitude.  

\section{From Gauge Theory to Gravity}
Probably the most important feature of the BCJ relations is the provided connection to gravity by the double copy procedure. If we have a BCJ expansion of an gauge theory amplitude (\ref{bcj amplitude}) or (\ref{loopbcj}) we can simply replace the color factor with another copy of the corresponding kinematical factor to obtain a gravity amplitude. For example if we consider an L-loop amplitude in $\mathcal{N}$=4 sYM where the numerators satisfy all BCJ equations
\begin{gather}
\mathcal{A}^{\text{sYM}}= \int{\left(\prod_m^L{\frac{d^Dl_m}{(2\pi)^D}} \right) \sum_{\text{graphs,perms}} {\frac{1}{S_i}\frac{N_i C_i}{\prod_j{p_j^2}}} } ,
\end{gather}
then we can connect it to an $\mathcal{N}$=8 SUGRA amplitude by replacing the color factors $C_i$ with another copy of kinematical factor $N_i$
\begin{gather}
\Rightarrow \mathcal{A}^{\text{SUGRA}}= \int{\left(\prod_m^L{\frac{d^Dl_m}{(2\pi)^D}}\right) \sum_{\text{graphs,perms}} { \frac{1}{S_i}\frac{N_i \tilde{N}_i}{\prod_j{p_j^2}}} } 
\end{gather}
where the R-symmetry indices in the second numerator have been shifted
\begin{gather}
\tilde{N}_i= N_i |_{\eta^a_i \rightarrow \eta^{a+4}_i} .
\end{gather}
This shift might be better understood after we will discuss amplitudes in $\mathcal{N}$=4 sYM in chapter 5. \\
The details of which product of numerators is known to lead to which gravity theory are given in the table below.
\begin{center}
\begin{tabular}{c|c|c}
gauge numerator $n$ & gauge numerator $\tilde{n}$ & Gravity \\
\hline
$\mathcal{N}$=4 sYM & $\mathcal{N}$=4 sYM & $\mathcal{N}$=8 SUGRA  \\ 
$\mathcal{N}$=4 sYM & $\mathcal{N}$=0 sYM & $\mathcal{N}$=4 SUGRA  \\
$\mathcal{N}$=0 sYM & $\mathcal{N}$=0 sYM & $\mathcal{N}$=0 SUGRA  
\end{tabular}
\end{center}
Here $\mathcal{N}$=0 SUGRA consists out of a graviton, an antisymmetric tensor and a dilaton.\\
At tree-level this connection has the following form
\begin{gather}
\mathcal{A}^{sYM}= \sum_{\text{graphs}} { \frac{n_i c_i} {\prod_j{p_j}} } \\
\Rightarrow \mathcal{A}^{SUGRA} = \sum_{\text{graphs}}{  \frac{n_i \tilde{n}_i} {\prod_j{p_j}} }
\end{gather}
and has been known before in the form of the KLT relations \cite{KLT}.
 These relations, originally derived in string theory but valid in the field theory limit, connect gravity and gauge theories exactly through this double copy procedure. But if we have an amplitude in the BCJ representation it has been conjectured that the double copy procedure can be applied directly.\footnote{Similar to the BCJ equations there have been several non trivial examples where} \\
This connection is rather remarkable. At the Lagrangian level the two theories are quite different, while Yang-Mills gauge theories have a rather simple Lagrangian which only involves three- and four-point vertices, gravity has an infinite numer of interactions in the Lagrangian. Therefore it is actually much easier to calculate an amplitude in $\mathcal{N}$=8 SUGRA through the double copy procedure then directly obtaining it.
Due to the BCJ relations and the double copy procedure it has become possible to directly carry out multiloop calculations for gravity amplitudes, which stimulated a renaissance in the study of ultra-violet properties of gravity theories.

\newpage
\chapter{Calculating Loop Amplitudes}
If direct integration is prohibitive the general strategy of calculating a loop amplitude is to rewrite the Feynman integrals as a linear combination of master integrals
\begin{gather}
\int{\prod^l_{i=1} { \left( \frac{  d^D q_i} {(2 \pi)^D} \right) } \ \ I(q_1,..,q_l)   } = \sum_{j=1}^n{a_j M_j}, 
\label{pass-velt}
\end{gather}
where $M_j$ is our basis of master integrals and $a_j$ are their coefficients. \\
The basis of master integrals is theory independent and therefore we can solve the MI's for all theories and focus on the generation of the theory dependent coefficients. \\
One strategy to obtain the coefficients is to extend equation (\ref{pass-velt}) to the integrand level. In order to achieve this we will need to introduce additional terms; the so called spurious terms
\begin{gather}
I(q_1,..,q_l)= \sum_{j=1}^m a_j M'_j .
\label{binte}
\end{gather}
Here m is in general bigger then n because $M'_j$ denotes the integrands of the master integrals $M_j$, which also contain spurious integrands. If we integrate equation (\ref{binte}) the spurious integrands we will vanish and therefore we arrive back at equation (\ref{pass-velt}). Since we are now working at the integrand level we achieved that any explicit integration procedure and/or any matching procedure between cuts of amplitudes and cuts of master integrals is replaced by simple polynomial fit.  \\
In the following chapter we will discuss two ways of calculating loop amplitudes which are both based on a reduction algorithm which are valid before integration. First we will explain how an one-loop amplitude is calculated through the OPP method \cite{Ossola:2006us,Ellis:2007br} and then move to the calculation of arbitrary loop amplitudes via the integrand-reduction through multivariate polynomial division \cite{Mastrolia:2012an}. The coefficients appearing in the linear combination (\ref{binte}) will then be fitted by evaluating product of tree amplitudes at the solution of the corresponding cut system.

\section{Integrand-Reduction for One-Loop Amplitudes}
For simplicity we will write all upcoming formulas for massless particles in four dimensions. 

But it is important to note that these methods are valid for massive particles as well and  have been extended to arbitrary dimensions \cite{Ellis:2007br,Mastrolia:2010nb}. \\
At one-loop it was realized that one can choose a basis of master integrals which only has scalar integrals \cite{Passarino:1978jh}. As the name suggests a scalar integral is of the form 
\begin{gather}
\int{\frac{d^4 q}{(2\pi)^4} \frac{1}{\prod_i{D_i}}}
\end{gather}
where $i=1,2,3$ or $4$. Each of these integrals has a theory dependent coefficient which we want to determine.

\subsection{The OPP Algorithm}
Instead of determining the coefficients directly we will pull them under the integral sign and delay the integration till these coefficients are determined. Hence we reduce the problem of determining the coefficients to a simple polynomial fit. Using a polynomial fit comes at a price namely the introduction of additional spurious terms which will vanish after integration.\\
With this idea the OPP parametrization of the numerator takes the following form
\begin{gather}
\begin{split}
N(q)= \sum_{i_0 < i_1 < i_2 < i_3}^{m-1} {  \left[d_{i_0i_1i_2i_3}+\tilde{d}_{i_0i_1i_2i_3}(q)\right] \prod_{i\neq i_0,i_1,i_2,i_3}^{m-1}{D_i} } \\
+ \sum_{i_0 < i_1 < i_2 }^{m-1} { \left[c_{i_0i_1i_2}+\tilde{c}_{i_0i_1i_2}(q)\right] \prod_{i\neq i_0,i_1,i_2}^{m-1}{D_i}} \\
+ \sum_{i_0 < i_1  }^{m-1} { \left[b_{i_0i_1}+\tilde{b}_{i_0i_1}(q)\right] \prod_{i\neq i_0,i_1}^{m-1}{D_i}} \\
+ \sum_{i_0 }^{m-1} { \left[a_{i_0}+\tilde{a}_{i_0}(q)\right] \prod_{i\neq i_0}^{m-1}{D_i} } 
\end{split}
\label{oppnumerator}
\end{gather}
where $d_{i_0i_1i_2i_3}$, $c_{i_0i_1i_2}$, $b_{i_0i_1}$ and $a_{i_0}$ are constants and $\tilde{d}_{i_0i_1i_2i_3}(q)$, $\tilde{c}_{i_0i_1i_2}(q)$, $\tilde{b}_{i_0i_1}(q)$ and $\tilde{a}_{i_0}(q)$ are polynomials in q.
If we consider an integrand instead of the numerator, cancel all propagators and combine the constant and spurious terms in each channel into one residue $\Delta$ we see the driving principle of this reduction
\begin{gather}
\begin{split}
\frac{N(q)} {D_0..D_{n-1}} =\sum_{i_0 < i_1 < i_2 < i_3}^{m-1} \left( \frac{ \Delta_{i_0i_1i_2i_3}(q) }{D_{i_0}D_{i_1}D_{i_2}D_{i_3}} +\frac{ \Delta_{i_0i_1i_2}(q) }{D_{i_0}D_{i_1}D_{i_2}} \right.  \\ \left.
+\frac{\Delta_{i_0i_1}(q) } {D_{i_0}D_{i_1}} +\frac{ \Delta_{i_0}(q) }{D_{i_0}} \right),
\end{split}
\label{oppintegrand}
\end{gather}
 namely the multi-pole nature of any amplitude. This means if we fix the loop momenta in a way that it sets certain internal propagators on-shell then we expect the amplitude to have a pole in this channel. The OPP reduction can be understood as separating these poles into different terms and therefore making this feature explicit. \\
As mentioned earlier if we integrate expression (\ref{oppintegrand}) the spurious terms will vanish and we will arrive at the one-loop basis of master integrals
\begin{gather}
\begin{split}
\int{\frac{d^4 q}{ (2 \pi)^4} \frac{N(q)} {D_0..D_{n-1}}} = \sum_{i_0 < i_1 < i_2 < i_3}^{m-1} \left( d_{i_0i_1i_2i_3} \int{\frac{d^4 q}{ (2 \pi)^4} \frac{ 1 }{D_{i_0}D_{i_1}D_{i_2}D_{i_3}} } \right.  \\ 
+ c_{i_0i_1i_2} \int{\frac{d^4 q}{ (2 \pi)^4} \frac{ 1 }{D_{i_0}D_{i_1}D_{i_2}}}  \left.
+ b_{i_0i_1} \int{\frac{d^4 q}{ (2 \pi)^4} \frac{1 } {D_{i_0}D_{i_1}} } + a_{i_0} \int{\frac{d^4 q}{ (2 \pi)^4} \frac{ 1 }{D_{i_0}} } \right) .
\end{split}
\end{gather}  

In order to use the OPP method we have to identify all spurious terms which can appear in the residues. These spurious terms are scalar products between the loop momentum and auxiliary vectors which are independent of the theory. Thus if we determine them once we can reuse them in other calculations.

\subsubsection{Obtaining the Spurious Terms}
To see how we can determine the spurious terms we will consider a four-point one-loop amplitude
\cite{amplitudeschool}. \\
In four dimensions any Lorentz vector can be expressed in terms of a four dimensional basis e.g. for our loop momentum we find 
\begin{gather}
q^\mu=x_1p_1^\mu + x_2 p_2^\mu + x_3 v^\mu + x_4 v^\mu_\perp
\end{gather}
where the auxiliary vectors $v$ and $v_\perp$ are given by
\begin{gather}
v^\mu = p_3 \cdot \epsilon_{41}  \epsilon_{14}^\mu + p_3 \cdot \epsilon_{14} \epsilon^\mu_{41} \\
v^\mu_\perp =p_3 \cdot \epsilon_{41}  \epsilon_{14}^\mu - p_3 \cdot \epsilon_{14} \epsilon^\mu_{41}  .
\end{gather}
Here $\epsilon_{ij}^\mu = \frac{\spab{i}.\gamma^\mu.j}{2}$ denotes the polarization vector $\epsilon^\mu (i^+,j)$ without the normalization. The auxiliary vector $v_\perp$ is perpendicular to all external momenta
\begin{gather}
v_\perp^\mu \cdot (p_i)=0 .
\end{gather}
At one-loop it is always possible to express a scalar product between the loop momentum and the momentum of an external particle as a linear combination of propagators. But in order for OPP expansion (\ref{oppnumerator}) to be meaningful we need the residues $\Delta$ to be independent of the propagators $D_i$. Otherwise we would be able to absorb terms which are proportional to the propagators into lower-point residues and the expansion would be arbitrary. \\
Therefore our residues must be independent of scalar products between the loop momenta and the momenta of external particles. To see which other scalar products we can form we have to consider the loop momentum basis for each residue.

\subsubsection*{Quadruple-Cut Residue}
\begin{figure}[h]
	\centering
		\includegraphics[width=0.15\textwidth]{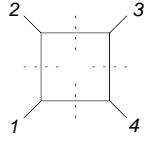}
	\caption{Quadruple-cut of a four-point one-loop amplitude}
	\label{fig:boxcut}
\end{figure}
For the quadruple-cut displayed in figure \ref{fig:boxcut} the overall momentum conservation reduces our independent external momentums down to three. But since $v_\perp$ is perpendicular to all these momenta we can use this vector as the fourth base element for the Lorentz vectors
\begin{gather}
B=(p_1,p_2,p_3,v_\perp) .
\end{gather}
As we have mentioned earlier we can not have any scalar products with external momenta. So the only possibility is to have scalar products of the form $q \cdot v_\perp$ in the residue.
From renormalizability we know that the power of loop momentum can not exceed the number of propagators in the denominator. Therefore the quadruple-cut residue takes the following form
\begin{gather}
\begin{split}
\Delta_{1234}(q)=c_{0;1234}+c_{1;1234} (q \cdot v_\perp)+c_{2;1234} (q \cdot v_\perp)^2\\
+c_{3;1234} (q \cdot v_\perp)^3+ c_{4;1234} (q \cdot v_\perp)^4 .
\end{split}
\end{gather}
From the corresponding quadruple-cut equations 
\begin{gather}
D_1=q^2 \equiv 0 \\
D_2=(q+p_1)^2 \equiv 0 \\
D_3=(q+p_1+p_2)^2 \equiv 0 \\
D_4=(q+p_1+p_2+p_3)^2 \equiv 0
\end{gather}
we can see that as soon as we solve the quadratic equation for $D_1$ all other quadratic equations reduce to linear ones. Hence we actually have one quadratic and three linear equations, so we have only two solutions to a quadruple-cut. But this is bad news since we are not able to fit the coefficients of the residue at the corresponding cut. \\
Fortunately it is possible to reduce the number of coefficients down to two. This can be seen if we expand the metric tensor in our Lorentz basis
\begin{gather}
\eta^{\mu \nu}=a_1 p_1^\mu p_2^\nu+a_2 p_1^\mu p_3^\nu +a_3 p_2^\mu p_2^\nu + a_4 \frac{v_\perp^\mu v_\perp^\nu}{v_\perp^2}.
\end{gather}
and use this to calculate the square of the loop momentum
\begin{gather}
\begin{split}
q^2 = q_\mu \eta^{\mu \nu} q_\nu\\
= a_1 q \cdot p_1 q \cdot p_2+a_2 q \cdot p_1 q \cdot p_3+a_3 q \cdot p_2 q \cdot p_3+a_4 \frac{(q \cdot v_\perp)^2}{v_\perp^2} .
\end{split}
\label{qsquare}
\end{gather}
But since we are on the quadruple-cut the square of the loop momentum is set to zero. This means we can use equation (\ref{qsquare}) to connect the square of our auxiliary vector $v_\perp$ to scalar products including only external momenta. But these scalar products between external momenta and the loop momentum are always reducible. In detail this means we can absorb the scalar products of the form $(q \cdot v_\perp)^i$ with $i \geq 2$ into either the linear term for odd $i$s or into the constant term for even $i$s. Therefore our residue includes only terms which are maximal linear in the loop momentum
\begin{gather}
\Delta_{1234}(q) = c_{0;1234}+c_{1;1234} (q \cdot v_\perp) .
\label{4res}
\end{gather}

\subsubsection*{The Triple-Cut Residue}
\begin{figure}[h]
	\centering
		\includegraphics[width=0.15\textwidth]{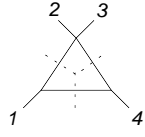}
	\caption{Triple-cut of a four-point one-loop amplitude}
	\label{fig:trianglecut}
\end{figure}

For a triple-cut displayed in figure \ref{fig:trianglecut} we only have two independent external momenta. This means we need two auxiliary vectors to form the basis
\begin{gather}
B(p_1,p_4,\epsilon_{14},\epsilon_{41}) .
\end{gather}
By the same argument as we used in the quadruple-cut we can only have scalar products of the form
$q \cdot \epsilon_{14}$ and $ q \cdot \epsilon_{41}$ with a maximum of three powers in $q$.
Therefore the residue takes the following form
\begin{gather}
\begin{split}
\Delta_{013}(q)=c_{0;013}+c_{1;013} q \cdot \epsilon_{14} +c_{2;013} q \cdot \epsilon_{41}
+c_{3;013} (q \cdot \epsilon_{14})^2 \\
 +c_{4;013} (q \cdot \epsilon_{41})^2  +c_{5;013} (q \cdot \epsilon_{41})^3 +c_{6;013} (q \cdot \epsilon_{14})^3 +c_{7;013} q \cdot \epsilon_{14} q \cdot \epsilon_{41} \\
  +c_{8;013} (q \cdot \epsilon_{14})^2 (q \cdot \epsilon_{41}) +c_{9;013} (q \cdot \epsilon_{41})^2 q \cdot \epsilon_{14}.   
\end{split}
\end{gather}
But if we expand the loop momenta in terms of our triple-cut basis
\begin{gather}
q^\mu=a_1 p_1 + a_2 p_4 + a_3 \epsilon_{14}+a_4 \epsilon_{41}
\end{gather} 
and have a look at the three cut conditions  
\begin{gather}
D_1=q^2 \equiv 0 \rightarrow a_1 a_2=a_3 a_4  \\
D_2=(q+p_1)^2 \equiv 0 \rightarrow a_2=0 \\
D_3=(q-p_4)^2 \equiv 0 \rightarrow a_1 =0 
\end{gather}
we obtain the constraint that $a_3a_4=0$.
Therefore all terms with both scalar products $q \cdot \epsilon_{41}$ and $q \cdot \epsilon_{14}$ in our residues will vanish and we only have
\begin{gather}
\begin{split}
\Delta_{013}(q)=c_{0;013}+c_{1;013} q \cdot \epsilon_{14} +c_{2;013} q \cdot \epsilon_{41} 
+c_{3;013} (q \cdot \epsilon_{14})^2 \\
+c_{4;013} (q \cdot \epsilon_{41})^2  +c_{5;013} (q \cdot \epsilon_{41})^3 +c_{6;013} (q \cdot \epsilon_{14})^3 .
\end{split}
\label{3res}
\end{gather}

\subsubsection*{The Double-Cut Residue}
\begin{figure}[h]
	\centering
		\includegraphics[width=0.20\textwidth]{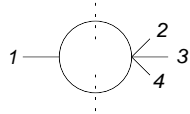}
	\caption{Double-cut of a four-point one-loop amplitude}
	\label{fig:bubblecut}
\end{figure}

For a double-cut displayed in figure \ref{fig:bubblecut} we only have one independent external momenta and we have to use three auxiliary vectors to form the basis
\begin{gather}
B=(p_1,p_4,\epsilon_{14},\epsilon_{41}) .
\end{gather}
Therefore we can form three different scalar products $q \cdot p_1$,$q \cdot \epsilon_{14}$ and $q \cdot \epsilon_{41}$. From the renormalizability condition we know that we can only have up to quadratic terms in the residue. Putting this together we find the following parametric form of the residue
\begin{gather}
\begin{split}
\Delta_{01}(q)=c_{0;01}+c_{1;01} q \cdot p_1+c_{2;01} q \cdot \epsilon_{14} 
+c_{3;01} q \cdot \epsilon_{41} +c_{4;01} (q \cdot p_1)^2 \\
 +c_{5;01} (q \cdot \epsilon_{14})^2
+c_{6;01} (q \cdot \epsilon_{41} )^2+ c_{7;01} q \cdot p_1  q \cdot \epsilon_{14} + c_{8;01} q \cdot p_1  q \cdot \epsilon_{41} \\
 +_{9;01} q \cdot \epsilon_{14}  q \cdot \epsilon_{41} .
\end{split}
\end{gather}
By the same argument we had before the scalar products belonging to the coefficient $c_{9;01}$ vanish and therefore the residue for the bubble reduces to
\begin{gather}
\begin{split}
\Delta_{01}(q)=c_{0;01}+c_{1;01} q \cdot p_1+c_{2;01} q \cdot \epsilon_{14} 
+c_{3;01} q \cdot \epsilon_{41} +c_{4;01} (q \cdot p_1)^2 \\
 +c_{5;01} (q \cdot \epsilon_{14})^2
+c_{6;01} (q \cdot \epsilon_{41} )^2+ c_{7;01} q \cdot p_1  q \cdot \epsilon_{14} + c_{8;01} q \cdot p_1  q \cdot \epsilon_{41} .
\end{split}
\label{2res}
\end{gather}

\subsubsection*{The Single-Cut Residue}

\begin{figure}[h]
	\centering
		\includegraphics[width=0.15\textwidth]{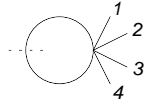}
	\caption{Single-cut of a four-point one-loop amplitude}
	\label{fig:tadpolecut}
\end{figure}

For a single-cut displayed in figure \ref{fig:tadpolecut} we have to use four auxiliary vectors as the basis
\begin{gather}
B=(p_1,p_4,\epsilon_{14},\epsilon_{41}) .
\end{gather}
Therefore we can form four scalar products between the loop momentum and auxiliary vectors. If we take into account the renomalizability condition we can find the following form of the residue
\begin{gather}
\Delta_{0}(q)=c_{0;0}+ c_{1;0} q \cdot p_1 + c_{2;0} q \cdot p_4 + c_{3;0} q \cdot \epsilon_{14} + c_{4;0} q \cdot \epsilon_{41} .
\label{1res}
\end{gather} \\[0.1cm]
With the knowledge of the parametric form of the residue it is now possible to simply evaluate the numerator at certain values of the loop momenta in order to determine the theory dependent coefficients $c$. Reminding ourselves of the OPP parametrized numerator
\begin{gather}
\begin{split}
N(q)= \sum_{i_0 < i_1 < i_2 < i_3}^{m-1} \left( \Delta_{i_0i_1i_2i_3}(q) \prod_{i\neq i_0,i_1,i_2,i_3}^{m-1}{D_i} 
+\Delta_{i_0i_1i_2}(q) \prod_{i\neq i_0,i_1,i_2}^{m-1}{D_i} \right. \\
+\left. \Delta_{i_0i_1}(q) \prod_{i\neq i_0,i_1}^{m-1}{D_i} 
+ \Delta_{i_0}(q)  \prod_{i\neq i_0}^{m-1}{D_i} \right)
\end{split}
\label{oppnum}
\end{gather}
we see that we have a large number of unknowns. In order to determine the unknowns efficiently one triangularizes the system of equations. This can be done by evaluating the integrand at loop momenta where several propagators go on-shell. This reduces the number of contributing residues as we can see from equation (\ref{oppnum}). \\
A second advantage from this way of fitting the coefficients comes from the possibility to use the unitarity method before integration. This method connects amplitudes where internal propagators are set on-shell to product of simpler amplitudes. Hence we can skip the step of calculating the full numerator and instead we are able to directly fit the coefficients from products of simpler amplitudes.

\subsection{Residues from Unitarity Cuts}
In the situation where are a set of certain internal propagators go on-shell
\begin{gather}
D_i=..=D_j=0.
\end{gather}
the unitarity of the S-Matrix will ensure that the amplitude factorizes into a product of lower-loop amplitudes. This can be seen if we split up the S-Matrix into a trivial and non trivial part $S=\mathbbm{1}+iT$ and then demand the unitarity condition 
\begin{gather}
SS^\dagger=(\mathbbm{1}+iT)(\mathbbm{1}-iT^\dagger) \equiv \mathbbm{1} \\
\rightarrow i(T-T^\dagger)=TT^\dagger.
\label{uni}
\end{gather}
The next step is to contract these matrices with a initial and final state and insert a complete set of intermediate states on the right-hand side
\begin{gather}
2 \text{Im}(\mathcal{M}_{i \rightarrow f}) = \sum_{k}{ \mathcal{M}^*_{f \rightarrow k} \mathcal{M}_{i \rightarrow k} } .
\label{setofstates}
\end{gather}
This is the well known optical theorem. If we perturbatively expand this equation in the coupling constant of the theory and compare terms order by order we see that we found a link between the imaginary part of a scattering amplitude and the product of two lower-loop amplitudes. This has been known since the 1960s in form of the Cutkosky rules. In recent years this technique has been extended to compute the coefficients of master integrals. This can be done by matching the unitarity cuts of an amplitude to the cuts of the master integrals. Since the coefficients of the master integrals are purely rational they are not affected by this procedure and one can simply read them off after all integrations are performed. \\
Given that we work at the integrand level it is sufficient to know that unitarity guarantees that an amplitudes factorizes into products of lower-loop amplitudes, which was first presented in \cite{Ellis:2007br}. With this information we can go back to the several cuts we have to perform to calculate a massless one-loop amplitude in four dimensions.  

\subsubsection{Generating a Quadruple-Cut Residue}
\begin{figure}[h]
	\centering
		\includegraphics[width=0.15\textwidth]{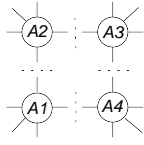}
	\caption{Factorization of an one-loop amplitude into a product of four tree amplitudes}
	\label{fig:uniboxcut}
\end{figure}

From the OPP reduction algorithm we know that the first residue in our decomposition (\ref{oppnum}) appears on top of four propagators $D_i,D_j,D_k$ and $D_l$. If we apply a quadruple-cut where we set all four propagators on-shell we will isolate this residue. On the side of the integrand we know from unitarity that setting the four propagators on-shell will lead to a factorization of the amplitude into a product of four tree amplitudes displayed in figure \ref{fig:uniboxcut}.
This gives us the following equation for the determination of the coefficients in the residue
\begin{gather}
 \text{Res}_{ijkl} \left\{  \sum_{\text{int. states}} { \left( \prod_{i=1}^{4} A_i \right) }   \right\} \equiv \Delta_{ijkl}(q) .
\label{quadrocut}
\end{gather}
Since there are two solutions for the loop momentum which set the four propagators $D_i,D_j,D_k$ and $D_l$ to zero we obtain two equations of the form of (\ref{quadrocut}). With these equations it is then possible to fit the two coefficients appearing in the parametrization of the quadruple-cut residue (\ref{4res}).

\subsubsection{Generating a Triple-Cut Residue}
\begin{figure}[h]
	\centering
		\includegraphics[width=0.15\textwidth]{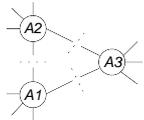}
	\caption{Factorization of an one-loop amplitude into a product of three tree amplitudes}
	\label{fig:unitricut}
\end{figure}

The next residue we want to determine sits on three propagators $D_i,D_j$ and $D_k$. If we set these propagators on-shell we will not only isolate the corresponding residue but also have a contribution of the quadruple-cut residue
\begin{gather}
N(q)= \sum_{l}{\Delta_{ijkl}(q)} + \Delta_{ijk}(q) D_l
\end{gather}
where the sum runs over a subset of the quadruple-cut residues which share the same propagators $D_i,D_j$ and $D_k$. Since the product of amplitudes includes all the propagators in the denominator we need to divide this equation by them to obtain the key equation to fit the residues
\begin{gather}
 \text{Res}_{ijk} \left\{  \sum_{\text{int. states}}{ \left( \prod_{i=1}^{3} A_i \right) }  - \sum_l{\frac{\Delta_{ijkl}(q)}{D_iD_jD_kD_l}} \right\} \equiv \Delta_{ijk}(q)
\label{tripplecut}
\end{gather}
where we used the numerator will factorize into a product of tree amplitudes is displayed in figure \ref{fig:unitricut}.
In this situation the cut equations $D_i=D_j=D_k=0$ are not able to freeze the four components of the loop momentum. Hence we have an infinite set of solutions on this cut. This means we can easily fit all coefficients appearing in the parametrization of the triple-cut residue (\ref{3res}).

\subsubsection{Generating a Double-Cut Residue}
\begin{figure}[h]
	\centering
		\includegraphics[width=0.15\textwidth]{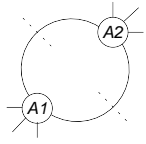}
	\caption{Factorization of an one-loop amplitude into a product of two tree amplitudes}
	\label{fig:unidicut}
\end{figure}

The double-cut residue sits on two propagators $D_i$ and $D_j$. Here the situation is similar to the triple-cut residue. Setting the two propagators on-shell not only isolates the residue but also the two higher residues. Subtracting these higher residues to obtain an equation for the determination of the pure double-cut residue we arrive at 
\begin{gather}
 \text{Res}_{ij} \left\{ \sum_{\text{int. states}} { \left( \prod_{i=1}^{2} A_i  \right) } - \sum_{l,k}{\frac{\Delta_{ijkl}(q)}{D_iD_jD_kD_l}} -\sum_k{\frac{\Delta_{ijk}(q)}{D_iD_jD_kD_l}} \right\} \equiv \Delta_{ij}(q).
\label{doublecut}
\end{gather}
Here unitarity ensures that our numerator will factorize into a product of tree amplitudes as displayed in figure \ref{fig:unidicut}.

In this situation the cut equations are only able to freeze two components of the loop momentum and therefore we are also able to fit all the coefficients in the parametrization of the double-cut residue (\ref{2res}).

\subsubsection{Generating a Single-Cut Residue}
\begin{figure}[h]
	\centering
		\includegraphics[width=0.15\textwidth]{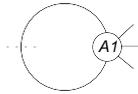}
	\caption{Factorization of an one-loop amplitude into a product of a tree amplitude}
	\label{fig:unimonocut}
\end{figure}

For the single-cut residue we only have on propagator $D_i$ which we can set on-shell. Thus we will get contributions from all higher residues, which share the propagator $D_i$
\begin{gather}
 \text{Res}_{i} \left\{ \sum_{\text{int. states} } { \left(  A_1  \right)  } - \sum_{j,l,k}{\frac{\Delta_{ijkl}(q)}{D_iD_jD_kD_l}} -\sum_{j,k}{\frac{\Delta_{ijk}(q)}{D_iD_jD_kD_l}}- \sum_{j}{\frac{\Delta_{ij}(q)}{D_iD_jD_kD_l}} \right\} \equiv \Delta_{i}(q) .
\label{singlecut}
\end{gather}
As in the other cases we replaced the numerator with a product of tree amplitudes as displayed in figure \ref{fig:unimonocut}.

With one cut equation we have three free components for the loop momentum. Therefore we are able to fit all coefficients appearing in the parametrization of the single-cut residue (\ref{1res}). \\[0.1cm]

As we have seen in the discussion above we always need the higher residues to obtain the lower ones. Therefore we can systematically obtain an one-loop amplitude by starting with all maximum cuts and then moving step by step down to the lower cuts. Where at each step we will have to subtract the higher residues and then obtain the corresponding residue by a polynomial fit of the coefficients in its parametrization. This algorithm has been implemented into two computer codes namely 'CutTools' \cite{Ossola:2007ax} and 'SAMURAI' \cite{Mastrolia:2010nb}. Both codes have been implemented into programs, 'MadLoop' \cite{madloop} and 'GoSam' \cite{GoSam} respectively, which provide efficient tools for automated NLO calculations at the LHC.   \\

\section{Calculating Loop Amplitudes with the Integrand-Reduction Algorithm}
If we want to extend the ideas presented in the OPP method to higher loop amplitudes we are confronted with one major problem. There is no systemized approach to the determination of the scalar products which can appear in the residues. Especially since at higher loops we have additional scalar products appearing in the residues which will not vanish upon integration. These scalar products will then lead to a non-scalar master integral. \\
Here we will discuss the integrand-reduction through multivariate polynomial division \cite{Mastrolia:2012an} which will solve this problem for any loop amplitude and at one-loop will give rise to the OPP parametrization of the integrand.  

\subsection{Multivariate Polynomial Division}
The aim of this integrand-reduction is to rewrite the integrand in its multi-pole expansion
\begin{gather}
\begin{split}
\frac{N}{D_1...D_n}=\frac{\Delta_{1..n}}{D_1...D_n}+\frac{\Delta_{2..n}}{D_2...D_n}+...+\frac{\Delta_{1..n-1}}{D_1...D_{n-1}} \\
\frac{\Delta_{3..n}}{D_3...D_n}+\frac{\Delta_{145..n}}{D_1D_4D_5...D_n}...+ \frac{\Delta_{12..n-2}}{D_1D_2..D_{n-2}} .
\end{split}
\label{mmop}
\end{gather}
The basic principle of that algorithm is to iteratively perform a multivariate polynomial division which allows us to split up the numerator
\begin{gather}
N_{12..n}= \Gamma_{12..n} + \Delta_{12..n}
\end{gather}
into a part which is a linear combination of the propagators
\begin{gather}
\Gamma_{12..n}= \sum_i{N_{1..i-1i+1..n} D_i}
\label{linearcombi}
\end{gather}
and a part which can not be written as a linear combination of the propagators the so called remainder $\Delta_{1..n}$. As the notation suggests all the prefactors of the propagators in eq. (\ref{linearcombi}) will be starting points for another recursion step. The recursion will automaticly stop when we only have one propagator left in the denominator $\frac{N_i}{D_i}$. Because we know from the renormalizibilty condition that the corresponding numerator is linear in the loop momentum, we will therefore not be able to express it in terms of one quadratic propagator $D_i$. Hence the whole numerator will be the remainder. \\
In order to understand how we can use a multivariate polynomial division to reduce our integrand we first need to discuss some basic ideas of algebraic geometry.

\subsubsection{A small Excursion into Algebraic Geometry}
\begin{itemize}
	\item Ideal
\end{itemize}
If we consider the space of all multivariate polynomials in z $P[z]$ where z is $z=(z_1,..,z_l)$ we can form a special subspace called ideal $\mathcal{I}$. The ideal is special in a way that if we take any polynomial from it $a \in \mathcal{I}$ and multiply it by an polynomial of the full space $b \in P[z]$ we end up in the ideal again $b \times a \in \mathcal{I}$. We can generate such an subspace by forming all linear combinations of its generators
\begin{gather}
\mathcal{I}= \langle a_1,..,a_n  \rangle = \left\{ \sum_{\kappa =1}^n{ b_\kappa(z) a_\kappa(z) \ : \ b_{\kappa} \in P[z]  } \right\}
\label{ideal}
\end{gather}
where the coefficients of the linear combination are itself polynomials in $z$.
\begin{itemize}
	\item Gr\"obner Basis
\end{itemize}
If we choose a monomial ordering namely decide if $z_1 z_2$ is bigger or smaller then $z_1^2$ then we can construct a special set of generators out of an ideal by the Buchberger algorithm. This special set is called Gr\"obner Basis
\begin{gather}
\mathcal{I} = \langle g_1,..,g_m \rangle
\end{gather}
where in general $m$ the number of generators in the Gr\"obner Basis is not equal to the number of generators in a different set e.g. $n$ in equation (\ref{ideal}).
Another important attribute is that if we consider a point in the ideal $x=(x_1,..x_l) \in \mathcal{I}$ where all generators vanish simultaneously $a_1(x)=a_2(x)=..=a_n(x)=0$ then this point has the same effect one the Gr\"obner Basis namely $g_1(x)=g_2(x)=..=g_m(x)=0$.
\begin{itemize}
	\item Multivariate Division with by a Gr\"obner Basis 
\end{itemize} 
Working in the Gr\"obner basis has the advantage that the multivariate polynomial division of an expression by an ideal is well defined. Meaning that if we take a polynomial $b \in P[z]$ and perform a division by a Gr\"obner Basis $\mathcal{G}$ we will obtain an unique remainder $\Delta \notin \mathcal{I}$ and an unique quotient $Q \in \mathcal{I}$
\begin{gather}
b / \mathcal{G}= Q + \Delta  .
\end{gather}
This division can be understood as splitting up the polynomial $b$ into two parts One which belongs to the ideal which is called the quotient $Q$ and one that does not which is called the remainder $\Delta$. \\
\begin{itemize}
	\item Hilbert's weak Nullstellensatz
\end{itemize}
Another important theorem is Hilbert's weak Nullstellensatz. It simply states that if we can not find a point where all generators vanish simultaneously then one is in the ideal $1 \in \mathcal{I}$, meaning that the ideal contains the whole space of polynomials.

\subsubsection{Applying Algebraic Geometry to the Integrand}
In order to apply these concepts to an integrand let us consider an $l$-loop numerator sitting on top of $n$ propagators
\begin{gather}
I_{i_1..i_n}(q_1,..,q_l) = \frac{N_{i_1..i_n}(q_1,..,q_l)}{D_{i_1}(q_1,..,q_l)..D_{i_n}(q_1,..,q_l)} .
\end{gather}
The first step is to express all our Lorentz vectors in their components. Therefore we can reexpress the numerator and the propagators as polynomials in the components of the loop momenta $z=(z_1,..,z_{4l})$
\begin{gather}
I_{i_1..i_n}(q_1,..,q_l) = \frac{N_{i_1..i_n}(q_1,..,q_l)}{D_{i_1}(q_1,..,q_l)..D_{i_n}(q_1,..,q_l)} \\
\rightarrow \frac{ N_{i_1..i_n}(z) } {D_{i_1}(z)..D_{i_n} (z)} .
\end{gather}
The next thing to note is that the propagators form an ideal 
\begin{gather}
\begin{split}
\mathcal{I}_{i_1..i_n}= \langle D_{i_1},..,D_{i_n} \rangle \\
=\left\{ \sum_{\kappa =1}^n{ b_\kappa(z) D_{i_\kappa}(z) \ : \ b_{\kappa} \in P[z]  } \right\} .
\end{split}
\end{gather}
From the ideal we can construct its Gr\"obner Basis by the Buchberger algorithm
\begin{gather}
\mathcal{I}_{i_1..i_n}= \langle g_1,..,g_m \rangle = \mathcal{G}_{i_1..i_n}
\end{gather}  
where we chose here and in the following chapters the lexicographic order.\\
With the help of the Gr\"obner Basis we can now perform a multivariate polynomial division which splits the integrand into a part which belongs to the ideal $\Gamma$ and one which is not in the ideal $\Delta$
\begin{gather}
N_{i_1..i_n}(z)= \Gamma_{i_1..i_n}(z) + \Delta_{i_1..i_n}(z) .
\end{gather}
But since gamma is in the ideal we can rewrite it as a linear combination of the propagators
\begin{gather}
\Gamma_{i_1,..,i_n}(z)=\sum^n_\kappa N_{i_1..i_{\kappa-1}i_{\kappa+1}..i_n}(z)D_{i_\kappa}(z) .
\end{gather}
If we combine this with the whole integrand we find the following formula 
\begin{gather}
I_{i_1..i_n}= \sum_{\kappa=1}^n I_{i_1..i_{\kappa-1}i_{\kappa+1}..i_n} + \frac{\Delta_{i_1..i_n}}{D_{i_1}..D_{i_n}} .
\label{mmoprec} 
\end{gather}
This equation provides us a recursion relation since we connected an integrand with n propagators with its multi-pole channel and a sum over integrands with $n-1$ propagators. The recursion relation will automaticly stop when we only have one propagator in the denominator left since there the whole numerator will be identified as the remainder.\\
We can find the starting point of the integrand-reduction by introducing the reducibility criterion \cite{Mastrolia:2012an}. We will call a numerator reducible if the remainder of the polynomial division vanishes, which implies that the whole numerator is part of the ideal formed by the propagators. Therefore we can rewrite the numerator as a linear combination of lower-point numerator times propagators. \\
\begin{prop}
 The integrand $I_{i_1..i_n}$ is reducible iff the remainder of the division modulo a Gr\"obner basis vanishes, i.e. iff $N_{i_1..i_n} \in \mathcal{I}_{i_1..i_n}$
 \label{prop1}
 \end{prop}
Proposition \ref{prop1} allows us to prove a second proposition
\begin{prop}
An integrand $I_{i_1..i_n}$ is reducible if the cut $(i_1,..,i_n)$ leads to a system of equations with no solution.
\label{propred}
\end{prop}
We can prove Proposition \ref{propred} with the help of the weak Nullstellensatz.\\
If a system of cut equations has no solutions i.e. 
\begin{gather}
D_{i_1}=..=D_{i_n}=0
\end{gather}  
does not have a solution, the weak Nullstellensatz tells us that $1$ is part of the ideal
\begin{gather}
1= \sum_{\kappa =1}^n{ b_\kappa(z) D_{i_\kappa}(z)} .
\label{oneinideal}
\end{gather}  
But if $1$ is part of the ideal the whole space of polynomials is part of the ideal. Therefore also the numerator is part of the ideal. Hence we can rewrite it as a combination of denominators
\begin{gather}
N_{i_1..i_n}= \sum_{\kappa =1}^n{ b'_\kappa(z) D_{i_\kappa}(z)}
\end{gather}
where the coefficients $b'_{\kappa(z)}$ will play the role of a lower-point integrand $N_{i_1..i_{\kappa-1}i_{\kappa+1}..i_n}$ thus the integrand is reducible.\\
Now that we have identified the starting and end point of our reduction we can see that equation (\ref{mmoprec}) will produce an integrand-decompositions of the form presented in (\ref{mmop}). \\
In order to show that the remainders of the polynomial division are really the multi-pole channels of the amplitude let us consider the first non vanishing remainder. From the reducible criterion we know that this will happen for an $l$-loop amplitude when we have $4l$ propagators in the denominator. At this point the corresponding cut equations 
\begin{gather}
D_{i_1}=D_{i_2}=..=D_{i_{4l}}=0
\end{gather}
will freeze the loop momenta completely. We will refer to this situation as a maximum cut. If we want to identify the corresponding remainder with the multi-pole channel where all these propagators are on-shell then the remainder should not get any contributions from lower-point remainders. \\
We can easily verify this by reminding ourselves that all the lower-point remainders will originate from the ideal constructed from the propagators we set on-shell. Therefore all lower-point remainders must vanish if we set all propagators to zero. Furthermore simply by the definition of the integrand-reduction algorithm all irreducible scalar products will appear in the remainders. Therefore we are really able to identify these remainders with the residue at the corresponding cut. \\ 

\subsubsection*{Maximum Cut Theorem}
The first important consequence of the reduction algorithm manifests itself in the form of the maximum cut theorem \cite{Mastrolia:2012an}. Under the assumption that the cut equations of the maximum cut provide us with a finite number of solutions $n_s$ with multiplicity one the maximum cut theorem states \\
\begin{theo}[Maximum cut]
The residue at the maximum-cut is a polynomial parametrized by $n_s$ coefficients, which admits a univariate representation of degree $(n_s-1)$.
\end{theo}
In other words the number of coefficients we have to fit to determine the remainder of a maximum cut will always coincide with the number of solutions of the cut equations. This ensures that at any cut we will always have enough values of loop momenta to fit all the coefficients in the residue. Because if we consider a cut where less then the maximum number of propagators are on-shell we will always find an infinite number of solutions. This theorem underlines the identification of the remainder with the multi-pole channel of the amplitude and ensures that every residue can be fixed with values of the loop momentum which correspond to this channel. \\
As an example of this theorem we can look back at the maximum cut of an one-loop amplitude. From the discussion of the quadruple-cut residue we know that we have two unknowns in our parametrization [see equation (\ref{4res})]. The cut equations of the maximum cut involve one quadratic equation and three linear ones, therefore this system of equations has two solutions. This means that as stated from the maximum cut theorem the number of cut solutions and coefficients in the residue coincidences.

\subsection{Application of the Reduction Algorithm}
The presented integrand-reduction algorithm can be used in two ways. The first approach is called 'fit-on-the-cut'. Here one basicly follows the same strategy we presented in the OPP reduction. First we use a generic numerator which includes all possible scalar products up to the renormalizibilty condition and then reduces 
 integrand with the reduction algorithm. From this we will obtain the parametric form of the residues which can appear. The parametric form only depends on the form of the corresponding propagators and therefore we can reuse it for any other amplitude which shares this residue. \\
After we obtained the parametric form of all residues the calculation of the loop amplitude is reduced to the task of a polynomial fit. The maximum cut theorem ensures that we are able to fit all the coefficients of a residue on its corresponding cut. Furthermore also enables us to use the unitarity method before integration for an even easier generation of these coefficients. \\

The second approach is called 'divide-and-conquer'. The first step is to generate the full integrand of the amplitude, which can be done through any generator of loop amplitudes. The second step is to use integrand-reduction directly on the integrand. The reduction algorithm will directly produce the scalar products including their coefficients. Besides this nice feature there is another advantage. At no step of the reduction the explicit solutions to the cut equations is needed. Therefore it becomes possible to reduce amplitudes which have higher powers of propagators in the denominator \cite{mp}. 
\\
In chapter six and seven we will present two examples where we used the 'fit-on-the-cut' approach for loop amplitudes. These two examples are the five-point one- and two-loop amplitudes in $\mathcal{N}$=4 sYM.  In order to use unitarity method before integration to fit the coefficients in the residues we need to discuss the tree amplitudes of this theory first, which will be the subject of the next chapter.
\chapter{Scattering Amplitudes in $\mathcal{N}$=4 sYM}
\begin{center}
$\mathcal{N}$=4 sYM is the harmonic oscillator of four dimensional gauge theories...
\end{center}
\begin{flushright}
David Gross
\end{flushright}
The harmonic oscillator is of extreme importance for the understanding of modern physics and especially quantum mechanics. Furthermore it is one of the few quantum mechanical systems where an exact analytic solution is known. It therefore led to important insights into quantum mechanics, such as the second quantization. There is reasonable hope that $\mathcal{N}$=4 super Yang-Mills can play the same role for gauge theories. \\
The recent years have shown that $\mathcal{N}$=4 sYM has a remarkable simple structure in its results. This is always a sign that there are many hidden symmetries at work in this theory and indeed in the planar limit an infinitesimal Yangian symmetry \cite{yangiana,yangianb,yangianc } was found. The Yangian symmetry is a combination of both the conformal and the dual conformal invariance of the theory. Furthermore it was possible to reformulate the theory in the planar limit in an on-shell formulation \cite{ArkaniHamed:2012nw} which made these symmetries manifest. The generalization of BCFW recursion relations made it possible to write down the all loop integrand in the planar limit \cite{allloopint}. \\
Even beyond the integrand there has been remarkable progress. The technology of symbols of transcendental functions is able to strongly constrain the polylogarithms which appear in the final result \cite{symbola,symbolb}. \\
After a short introduction to $\mathcal{N}$=4 sYM we will describe on-shell methods to construct any MHV or $\overline{MHV}$ tree amplitude in $\mathcal{N}$=4 sYM \cite{drummond}. Later in this chapter we will discuss how this description also helps us with the sum over all internal states in the unitarity construction of our integrands  \cite{n=4cuts}.

\section{An Introduction to $\mathcal{N}$=4 sYM}
Similar to Yang-Mills $\mathcal{N}$=4 super Yang-Mills (sYM) is a gauge theory with a non-abelian gauge group $SU(N)$ living in four dimensions. The difference to Yang-Mills comes from an additional symmetry between the bosons and fermions called supersymmetry. In fact here we have four supersymmtry generators as denoted by $\mathcal{N}$=4. \\
In order to be invariant under this fourfolded supersymmetry we need to introduce a scalar field s and a spin-$\frac{1}{2}$ fermion field f besides the normal gluon field g.
\begin{center}
$g^+, \ \ \ \ f_a, \ \ \ \ s_{ab}, \ \ \ \ \bar{f}_a, \ \ \ \ g^-$
\end{center} 
For $\mathcal{N}$=4 sYM we have one gluon field $g$ with helicity $\pm1$, four fermion fields $f$ with helicity $\pm\frac{1}{2}$ and six real scalar fields $s_{ab}$ with helicity $0$. 
If we consider the generators of the fourfolded supersymmetry the supercharges $Q_a$ we can see that they transform a bosonic field into a fermionic one and vice versa
\begin{gather}
[Q_a(q,\theta),g^+(k)]= \theta \spb{k}.q f_a \\
[Q_a(q,\theta),f_b(k)]= \theta \delta_{ab} \spa{k}.q g^+ + \theta \spb{k}.q s_{ab} \\
[Q_a(q,\theta),s_{bc}(k)]= \theta \delta_{ab} \spa{k}.q f_c - \theta \delta_{ac} \spa{k}.q f_b + \theta \spb{q}.k \epsilon_{abcd} \bar{f}_d  \\
[Q_a(q,\theta),\bar{f}_b(k)]= \theta \delta_{ab} \spb{q}.k g^- + \frac{1}{2} \theta \spa{q}.k \epsilon_{abcd} s_{cd} \\
[Q_a(q,\theta),g^-(k)]= \theta \spa{q}.k \bar{f}_a .
\end{gather}
A second thing to note is that the supersymmetry connects all three fields we introduced. That was the reason why we had to introduce the additional fields to our gluon field. \\
In order to make this fourfolded supersymmetry manifest we combine the three fields into one superfield $\Phi$ 
\begin{gather}
 \begin{split}
 \Phi(p,\eta)=g^+ +\eta^a f_a +\frac{1}{2}\eta^a \eta^b s_{ab}+ \frac{1}{3!}\eta^a\eta^b\eta^c \epsilon_{abcd} \bar{f}^d\\
 +\frac{1}{4!}\eta^a\eta^b\eta^c\eta^d \epsilon_{abcd}g^- 
 \end{split}
\end{gather}     
and consider scattering amplitudes of these superfields \cite{drummond}. \\
In the literature $\mathcal{N}$=4 sYM is often referred to as a maximal supersymmetric theory because we have the maximum number of supersymmetries without having to introduce a spin-two particle in the superfield. This maximum amount of supersymmtry has an important consequence. It leads to a vanishing $\beta$-function. Therefore the coupling constant does not run in the theory and this is a sufficient condition for the UV finiteness of the theory.\\
Despite these differences Yang-Mills theory and $\mathcal{N}$=4 sYM share several properties. Since the latter is a much simpler theory it serves as an ideal testing ground for new calculations methods for Yang-Mills theory and QCD. Furthermore at tree-level the all gluon part of $\mathcal{N}$=4 sYM is identical to the Yang-Mills amplitude. At one-loop the situation becomes a little bit more complicated but we can still split up the one-loop Yang-Mills amplitude into a part in $\mathcal{N}$=4 sYM, one in $\mathcal{N}$=1 sYM and one where a scalar runs in the loop
\begin{gather}
\mathcal{A}^{YM}_g= \mathcal{A}^{\mathcal{N}=4}_g + 4\mathcal{A}^{\mathcal{N}=4}_f+3\mathcal{A}^{\mathcal{N}=4}_s-4(\mathcal{A}^{\mathcal{N}=1}_f+\mathcal{A}^{\mathcal{N}=1}_s)+\mathcal{A}_s
\end{gather}
which is usually simpler than the original amplitude. \\
Another reason to study $\mathcal{N}$=4 sYM comes form the connection to $\mathcal{N}$=8 SUGRA through the BCJ equations \cite{Bern:2008qj} we discussed earlier. This connection enabled calculation of higher loop amplitudes in the latter which would otherwise be considered impossible. In addition these loop amplitudes turned out to be UV finite in four dimensions. This is in contradiction with arguments which were based solely on power counting and raised the hope that $\mathcal{N}$=8 SUGRA might be UV finite too.\\

\section{Three-Point Generating Functions}
The introduction to on-shell methods as input for loop level calculations mainly follows a lecture given at the Arnold Sommerfeld School \cite{amplitudeschool} and \cite{drummond}. \\
The first scattering amplitudes which can appear are the three-point ones, and these amplitudes are rather special. At three-points momentum conservation is prohibiting any kinematical invariants for real massless momenta
\begin{gather}
p_1+p_2+p_3=0 \Rightarrow 0=p_k^2=(p_i+p_j)^2= s_{ij}
\end{gather}
where $i\neq j \neq k$. The solution for that problem is to go to complex momenta and treat the momentum and its complex conjugate separately. Later this will play an important role in our analysis. \\
With the big particle content we described before we get a huge variety of color-ordered three-point amplitudes
\begin{gather}
A_3(1^-_{g},2^-_{g},3^+_g)= \frac{\spa1.2 ^4}{\spa1.2 \spa2.3 \spa3.1 } \\
A_3(1^-_{g},2^-_{\bar{f}},3^+_{f})= \frac{\spa3.1 ^3 \spa1.2 }{\spa1.2 \spa2.3 \spa3.1 } \\
A_3(1^-,2_s,3_s)= \frac{ \spa1.2 ^2 \spa3.1 ^2}{\spa1.2 \spa2.3 \spa3.1 }\\
A_3(1^+_{g},2^+_{g},3^-_g)=\frac{\spb1.2 ^4}{ \spb1.2 \spb2.3 \spb3.1} .
\end{gather}
In order to treat this abundance efficiently we should introduce a generating function, which encodes the whole particle and helicity content. From the example amplitudes we can already see that MHV three-point amplitudes are functions only of the angle brackets and $\overline{MHV}$ amplitudes are functions only of the square brackets. Thus we should distinguish between generating functions of MHV and $\overline{MHV}$ amplitudes. Let's focus on the MHV generating functions for now. \\
Looking back at the example amplitudes we see that all MHV three-point amplitudes sit on the same denominator $\frac{1}{\spa1.2 \spa2.3 \spa3.1}$ and only have different spinor products as a numerator. From dimensional analysis we know that the total power of angle products in the numerator must be four. Since this power is finite it is convenient to introduce Grassmann variables, because they also support only finite powers. In detail that means we introduce four Grassmann numbers $\bar{\eta}_j^a$ for each particle, where $j$ is the particle label and $a$ is the SU(4) group index which runs form $1$ to $4$ according to the maximum power in the polynomial. So putting everything together we get the following generating function 
\begin{gather}
A_3^{MHV}(1,2,3)=\frac{\prod_{a=1}^{4}{\left( \spa1.2 \bar{\eta}_3^a+ \spa2.3 \bar{\eta}_1^a+ \spa3.1 \bar{\eta}_2^a  \right)}}{\spa1.2 \spa2.3 \spa3.1 } .
\end{gather}
For Grassmann numbers the delta function and polynomials are the same object so we can replace the product with an delta function raised to the fourth power
\begin{gather}
A_3^{MHV}(1,2,3)=\frac{\delta^{4}{\left( \spa1.2 \bar{\eta}_3+ \spa2.3 \bar{\eta}_1+ \spa3.1 \bar{\eta}_2  \right)}}{\spa1.2 \spa2.3 \spa3.1 }
\end{gather}
where I dropped the $SU(4)$ label. If we now want to compute a specific amplitude we need to integrate over all Grassmann numbers with the appropriate measure. In order to understand this better let's obtain the all gluon amplitude
\begin{gather}
A_3(1^-_{g},2^-_{g},3^+_g)=\int{d^4\bar{\eta}_1 \: (\bar{\eta}_1)^4 d^4\bar{\eta}_2 \: (\bar{\eta}_2)^4 d^4\bar{\eta}_3 A^{MHV}(1,2,3)}\\
= \int{d^4\bar{\eta}_1 \: (\bar{\eta}_1)^4 d^4\bar{\eta}_2 \: (\bar{\eta}_2)^4 d^4\bar{\eta}_3 \frac{\delta^{4}{ \left( \spa1.2 \bar{\eta}_3+ \spa2.3 \bar{\eta}_1+ \spa3.1 \bar{\eta}_2  \right)}}{ \spa1.2 \spa2.3 \spa3.1 }}\\
= i \frac{\spa1.2 ^4}{\spa1.2 \spa2.3 \spa3.1 }
\end{gather}
where we have used the notation $(\bar{\eta}_i)^4=\bar{\eta}^1_i\bar{\eta}^2_i\bar{\eta}^3_i\bar{\eta}^4_i$. Furthermore the last lines follows because it is the only expression with no $\bar{\eta}_1$'s or $\bar{\eta}_2$'s. \\
In general to obtain an amplitude of specific particles we need to use the following measure for the integral.
\begin{center}
\begin{tabular}{c|c}
helicity & measure \\
\hline
$+1$ & 1 \\
$+\frac{1}{2}$ & $\bar{\eta}^a$\\
$0$ & $\bar{\eta}^a \bar{\eta}^b$ \\
$-\frac{1}{2}$ & $\bar{\eta}^a \bar{\eta}^b \bar{\eta}^c$ \\
$-1$ & $(\bar{\eta})^4$
\end{tabular}
\end{center}
As we have stated above this generating function only encodes MHV amplitudes. So what happens if we try to obtain an $\overline{MHV}$ amplitude $A(1_g^+,2_g^+,3_g^-)$ from our generating function
\begin{gather}
A_3(1_g^+,2_g^+,3_g^-)= \int{d^4\bar{\eta}_1 \:  d^4\bar{\eta}_2 \:  d^4\bar{\eta}_3 (\bar{\eta}_3)^4 \frac{\delta^{4}{\left( \spa1.2  \bar{\eta}_3^a+ \spa2.3  \bar{\eta}_1^a+ \spa3.1 \bar{\eta}_2^a  \right)}}{\spa1.2 \spa2.3 \spa3.1 }}\\
=0
\end{gather}
where the last line follows because we only have eight powers of $\bar{\eta}$ and integrate with 12 powers. This is what we should expect since the introduced generating function only encodes MHV amplitudes. Following a similar strategy as in the MHV case we find a $\overline{MHV}$ generating function
\begin{gather}
A_3^{\overline{MHV}}(1,2,3)= \frac{\delta^4\left( \spb1.2 \eta_3+\spb2.3 \eta_1+\spb3.1 \eta_2\right)}{\spb1.2 \spb2.3 \spb3.1 }
\label{mhvbar3}
\end{gather}
which is the complex conjugated of the MHV generating functions. Note that the generating functions now depends on different Grassmann numbers $\eta$. In order to obtain a specific amplitude with the $\eta$'s we need to integrate with a different measure.
\begin{center}
\begin{tabular}{c|c}
helicity & measure \\
\hline
$+1$ & $(\eta)^4$ \\
$+\frac{1}{2}$ & $\eta^a \eta^b \eta^c$\\
$0$ & $\eta^a \eta^b$ \\
$-\frac{1}{2}$ & $\eta^a$ \\
$-1$ & $1$
\end{tabular}
\end{center}
The measures can be motivated if we remind ourselves of the superfield $\Phi$
\begin{gather}
 \begin{split}
 \Phi(p,\eta)=g^+(p)+\eta^a f_a +\frac{1}{2}\eta^a \eta^b s_{ab}+ \frac{1}{3!}\eta^a\eta^b\eta^c \epsilon_{abcd} \bar{f}^d\\
 +\frac{1}{4!}\eta^a\eta^b\eta^c\eta^d \epsilon_{abcd}g^-.
 \end{split}
\end{gather}
Instead of considering a scattering amplitude of certain particles we will now consider a scattering amplitude of super fields
\begin{gather}
A(1,2,..n) \rightarrow A(\Phi_1,\Phi_2,..,\Phi_n) .
\end{gather}
If we plug in the expression for the superfields we get a sum over scattering amplitudes of the different component fields
\begin{gather}
\begin{split}
A(\Phi_1,\Phi_2,..,\Phi_n)=A(1_g^+,2_g^+,..,n_g^+)+ \eta_1^a A(1_f^+,2_g^+,3_g^+,..,n_g^+)\\
+ \frac{1}{2} \eta^a_1\eta^b_1A(1_s,2_g^+,3_g^+,..,n_g^+) + \frac{1}{3!} \eta_1^a \eta_1^b \eta_1^c A(1_f^-,2_g^+,3_g^+,..,n_g^+) \\
+ \left(\eta_1 \right)^4 A(1_g^-,2_g^+,..n_g^+)+ .. \\
+ \left(\eta_1 .. \eta_n \right)^4 A(1_g^-,..,n_g^-) .
\end{split}
\end{gather}
One way to extract a certain component field is to integrate over all Grassmann variables and introduce a measure which is exactly opposite to the powers of Grassmann variables in the superfield. This exactly corresponds to the rules we introduced above. To obtain the integration measure for the $\bar{\eta}$s one has to take the complex conjugate of the superfield
\begin{gather}
 \begin{split}
 \bar{\Phi}(p,\bar{\eta})=g^-(p)+\bar{\eta}^a \bar{f}_a +\frac{1}{2}\bar{\eta}^a \bar{\eta}^b \bar{s}_{ab}+ \frac{1}{3!}\bar{\eta}^a\bar{\eta}^b\bar{\eta}^c \epsilon_{abcd} f^d\\
 +\frac{1}{4!}\bar{\eta}^a\bar{\eta}^b\bar{\eta}^c\bar{\eta}^d \epsilon_{abcd}g^+.
 \end{split}
\end{gather}
Comparing this expression to the one before we see that the measure for the $\bar{\eta}$s is exactly the opposite of the $\eta$s. The only additional information we put in the generating functions is that we immediately used that some amplitudes vanish. \\
 If we want to combine these amplitudes through let's say a BCFW recursion relation they need to live in the same superspace, which can be done by Fourier transforming the generating function from the $\bar{\eta}$ superspace to the $\eta$ superspace or vice versa.  Here we decide to work in the $\eta$ superspace so we need to transform the MHV generating function
\begin{gather}
\int{d^4\bar{\eta_1}d^4\bar{\eta_2}d^4\bar{\eta_3} e^{\eta_1 \bar{\eta}_1}e^{\eta_2 \bar{\eta}_2}e^{\eta_3 \bar{\eta}_3}A_3^{MHV}(1,2,3)}\\
= \int{d^4\bar{\eta_1}d^4\bar{\eta_2}d^4\bar{\eta_3} e^{\eta_1 \bar{\eta}_1}e^{\eta_2 \bar{\eta}_2}e^{\eta_3 \bar{\eta}_3} \frac{\delta^{4}{ \left( \spa1.2 \bar{\eta}_3+ \spa2.3 \bar{\eta}_1+ \spa3.1 \bar{\eta}_2  \right)}}{\spa1.2 \spa2.3 \spa3.1 }} \\
= \frac{\delta^4\left( \spa1.2  \eta_1 \eta_2+  \spa2.3 \eta_2 \eta_3 + \spa3.1 \eta_3 \eta_1 \right)}{ \spa1.2 \spa 2.3 \spa3.1 }\\
=\frac{\delta^{(8)}(\lambda_1 \eta_1+\lambda_2 \eta_2+\lambda_3 \eta_3)}{\spa1.2 \spa 2.3 \spa3.1 } .
\label{mhv3}
\end{gather}
At the moment we only talked about three-point generating functions, but as we have already seen in the paragraph about superfields this concept can be generalized to higher-point superamplitudes. One way to obtain a higher point generating function is through a BCFW recursion. In the next part we will use that to construct the four-point generating function and then see how this generalizes.

\section{$n$-Point Generating Function}
The first thing we need to discuss before using the BCFW recursion relations is the merging of the generating functions. To merge two generating functions we have to set all parameters of a leg equal $(\lambda,\tilde{\lambda},\eta)$ and integrate over the shared Grassmann variable, which takes care of the internal helicities and particles that can propagate. To see this mechanism at work lets use the BCFW recursion to obtain a four-point MHV generating function.
In order to obtain an overall MHV generating function we need to combine a MHV three-point generating function and an $\overline{MHV}$ one, which can be seen by looking at the all gluon part of the amplitude. There a MHV three-point amplitude has two gluons with negative helicity and a $\overline{MHV}$ amplitude has one gluon with negative helicity. A propagator connects a gluon with negative helicity with one of positive helicity. Thus we can write down the following formula for the overall number of external negative gluons
\begin{gather}
\kappa=2M+\bar{M}-i
\end{gather} 
where $M$ is the number of MHV amplitudes, $\bar{M}$ the number of $\overline{MHV}$ amplitudes and $i$ is the number of internal legs. So in our case we want to have an overall MHV generating function so we have $\kappa$=2 and one internal leg $i=1$, hence we need $M=\bar{M}=1$. \\
In order to use the BCFW recursion formula we will introduce a complex shift on the external  momenta $p_1$ and $p_4$\footnote{When we combine three-point functions we can always choose the shift to be in the square brackets of the MHV generating function and in the angle brackets of the $\overline{MHV}$ generating function. But since the MHV generator only depends on the angle brackets and the $\overline{MHV}$ generator on the square brackets the dependence on the shift of the external particles drops out. The only remainder is the shift of the internal momenta $l$.}
\begin{gather}
\hat{p}_1=p_1 + z \epsilon_{14} \\
\hat{p}_4=p_4 - z \epsilon_{14}.
\end{gather} 
This shift gives us the following three-point generating functions
\begin{gather}
A^{MHV}_3(\hat{1},2,\hat{l})=\frac{\delta^8(\lambda_1 \eta_1+\lambda_2 \eta_2+\hat{\lambda}_l \eta_l )}{ \spa1.2 \spa2.{\hat{l}} \spa{\hat{l}}.1 }\\
A^{\overline{MHV}}_3( 3,\hat{4},-\hat{l})=\frac{ \delta^4 \left( -\spb3.4 \eta_l+\spb3.{\hat{l}} \eta_4+ \spb{\hat{l}}.4  \eta_3 \right) } {\spb3.4 \spb4.{\hat{l}} \spb{\hat{l}}.3 } 
\end{gather}
where we used that the shift only effects angle brackets  of $p_1$ and only square brackets of $p_4$.
We can now merge these two generating functions by the BCFW recursion formula and by integrating over the shared Grassmann variable 
\begin{gather}
A_4^{MHV}=\int{d^4 \eta_l A^{MHV}_3(\hat{1},2,\hat{l}) \frac{1}{s_{34}} A^{\overline{MHV}}_3(3,\hat{4},-\hat{l}) } \\
= \int d^4 \eta_l \frac{\delta^8 \left( \lambda_1 \eta_1+\lambda_2 \eta_2+\hat{\lambda}_l \eta_l \right)  } {\spa1.2 \spa2.{\hat{l}} \spa{\hat{l}}.1 } \frac{1}{s_{34}} \frac{\delta^4 \left( - \spb3.4 \eta_l+ \spb3.{\hat{l}} \eta_4+ \spb{\hat{l}}.4 \eta_3 \right) } {\spb3.4 \spb4.{\hat{l}} \spb{\hat{l}}.3 }\\
=\frac{\spb3.4 ^4 \delta^8 \left( \lambda_1 \eta_1+\lambda_2 \eta_2+\hat{\lambda}_l \left(\frac{ \spb3.{\hat{l}} }{ \spb3.4 } \eta_4+ \frac{\spb{\hat{l}}.4}{ \spb3.4 } \eta_3   \right) \right) } { \spa1.2  \spab2.{\hat{\slashed{l}}}.4  \spab1.{\hat{\slashed{l}}}.3 \spb3.4 } \frac{1}{s_{34}} .
\end{gather}
The shift in the momenta $\hat{l}$ will drop out because of the Fierz identity and if we use momentum conservation in the delta function in the form of
\begin{gather}
\tilde{\lambda}_i \spa{i}.k +\tilde{\lambda}_j   \spa{j}.k =0
\end{gather}
for MHV generators and 
\begin{gather}
\lambda_i \spb{i}.k  + \lambda_j \spb{j}.k =0
\end{gather}
for $\overline{MHV}$ generators we arrive at
\begin{gather}
A_4^{MHV}(1,2,3,4)=\frac{\delta^8(\lambda_1 \eta_1+\lambda_2 \eta_2+\lambda_3 \eta_3+\lambda_4 \eta_4)}{\spa1.2 \spa2.3 \spa3.4 \spa4.1 } 
\label{mhv4} 
\end{gather}
which is the four-point generating function for an MHV amplitude. We can now use more BCFW shifts to build up higher generating functions, but I will just state the general formula which by now should be well motivated
\begin{gather}
A_n^{MHV}(\eta_1,...\eta_n)=\frac{\delta^{(8)}(Q)}{\spa1.2 \spa2.3 ..\spa{n}.1 }
\end{gather}
where $Q= \sum_{i=1}^n \lambda_{i \alpha} \eta_i^a$. We also should note that q is one of the generators of the $\mathcal{N}=4$ Poincaré supersymmetry algebra. We can check our result by obtaining the n-point gluon function
\begin{gather}
\begin{split}
A_n^{MHV}(1^+_g,...i^-_g,...,j^-_g,...n^+_g)=\\
\int d^4\eta_1 (\eta_1)^4...d^4\eta_id^4\eta_{i+1}(\eta_{i+1})^4...d^4\eta_jd^4\eta_{j+1}(\eta_{j+1})^4...d^4\eta_n(\eta_n)^4\\ \frac{ \delta^{(8)}\left(\sum_i^n\lambda^\alpha_i\eta^a_i \right)}{ \spa1.2 \spa2.3 .. \spa{n}.1 } 
\end{split}\\
\begin{split}
=\int d^4\eta_1 (\eta_1)^4...d^4\eta_id^4\eta_{i+1}(\eta_{i+1})^4...d^4\eta_jd^4\eta_{j+1}(\eta_{j+1})^4...d^4\eta_n(\eta_n)^4 \\
\frac{ \delta^4\left( {\sum_{i<j}^n{ \spa{i}.j  \eta_i \eta_j }} \right) }  { \spa1.2 \spa2.3 .. \spa{n}.1 }
\end{split}\\
= \frac { \spa{i}.j ^4}{ \spa1.2 \spa2.3 .. \spa{n}.1 }
\end{gather}
which is indeed the result found by Parke and Taylor \cite{park taylor}.\\
In order to also obtain the general $\overline{MHV}$ amplitude we can complex conjugate our result
\begin{gather}
A_n^{\overline{MHV}}=\frac{\delta^{(8)}(\sum_{i=1}^n{\tilde{\lambda}_{i\dot{\alpha}} \bar{\eta}_{ia}})}{ \spb1.2 \spb2.3 .. \spb{n}.1 } .
\end{gather}
But this generating function is now in the superspace generated by $\bar{\eta}$. Therefore we need to Fourier-transform the result to go back to the superspace generated by $\eta$
\begin{gather}
A_n^{\overline{MHV}}(p,\eta)=\int{\prod_{i=1}^n\left({d^4\bar{\eta}e^{-\eta^a_i\bar{\eta}^a_i}}\right)A_n^{\overline{MHV}}(p,\bar{\eta})} \\
=\frac{1}{\left[12\right]\left[23\right]...\left[n1\right]}\int{\prod_{i=1}^n\left({d^4\bar{\eta}e^{-\eta^a_i\bar{\eta}^a_i}}\right)\delta^{(8)}\left(\sum_i \tilde{\lambda}_i^{\dot{\alpha}} \bar{\eta}_{a \: i} \right)} \\
=\frac{1}{ \spb1.2 \spb2.3 .. \spb{n}.1 }\int{\prod_{i=1}^n\left({d^4\bar{\eta}e^{-\eta^a_i\bar{\eta}^a_i}}\right)\int{d^8w e^{w_{\dot{\alpha}}^a\sum_i \tilde{\lambda}_i^{\dot{\alpha}} \bar{\eta}_{ai} }}} \\
=\frac{1}{\spb1.2 \spb2.3 .. \spb{n}.1 } \int{\prod_{i=1}^n\left(d^4\bar{\eta}_i\right)d^8w \: e^{-\sum_i \bar{\eta}_i(\eta^a_i-w_{\dot{\alpha}}^a \tilde{\lambda}_i^{\dot{\alpha}}  )}} \\
=\frac{1}{ \spb1.2 \spb2.3 .. \spb{n}.1 }\int{d^8w \: \prod_{i=1}^n{ \delta^4\left(\eta^a_i-w_{\dot{\alpha}}^a \tilde{\lambda}_i^{\dot{\alpha}}\right)}} .
\end{gather}

\section{Generating Functions as Input for Unitarity Cuts}
The amplitudes we discussed here are sufficient to reconstruct all  MHV and $\overline{MHV}$ loop amplitudes up to six-points in $\mathcal{N}$=4 sYM. This is the case because the numerator in this theory is only linear in the loop momentum. Thus the numerator can be completely reconstructed from maximum and near-maximum cuts. The amplitudes in these cuts involve at most five-particles and are therefore either MHV or $\overline{MHV}$ amplitudes.
For the treatment of non-MHV amplitudes I refer to \cite{n=4cuts}. The key fact is that these amplitudes can be written as products of MHV amplitudes and therefore the here discussed methods can be extended to the non MHV sector. \\
The calculation of an unitarity cut is nothing else then the merging of all appearing tree-level amplitudes. The advantage of using the Grassmann formulations lies in the fact that we can take care of the complicated sum over internal particles and helicities by simply integrating over the internal Grassmann variable. Therefore we are now ready to discuss two examples for the integrand-reduction through multivariate polynomial division where we will directly obtain the residues through unitarity cuts.

\chapter{The One-Loop Five-Point Amplitude in $\mathcal{N}$=4 sYM}
The one- and two-loop five-point amplitude in $\mathcal{N}$=4 sYM have been the subject of previous analyses. The planar part of the two-loop amplitude was already calculated by Bern, Czakon, Kosower, Roiban and Smirnov \cite{Bern:2006vw}. Later Carrasco and Johansson \cite{Carrasco:2011mn} used the conjectured BCJ equations to obtain the full five-point one- and two-loop amplitudes. Furthermore Mastrolia, Mirabella, Ossola and Peraro \cite{twoloopisp} used the known integrands by Carrasco and Johansson of the two-loop amplitudes to perform an integrand-reduction through multivariate polynomial division. \\
The calculation of the one- and two-loop amplitude in \cite{Carrasco:2011mn} is done in four steps. First they use the BCJ equations between different graphs to reduce the number of graphs down to a set of master graphs. Secondly they write down an ansatz for each master graph in terms of all possible scalar products. In the third step this ansatz is reduced through symmetries of graphs and BCJ equations which connect only master graphs. In the last step they fix the remaining parameters with unitarity cuts. \\
In this thesis we will follow a slightly different approach. We will also use the fact that the BCJ reduces the number of graphs down to a set of master graphs, but then use the integrand-reduction to determine the graph as far as possible through its unitarity cuts. In the last we will fix the remaining freedom through BCJ equations which only involve the master graph. Furthermore we will also present a way of obtaining these amplitudes which does not rely on the BCJ equations. \\
 
\begin{figure}[h]
	\centering
		\includegraphics[width=0.45\textwidth]{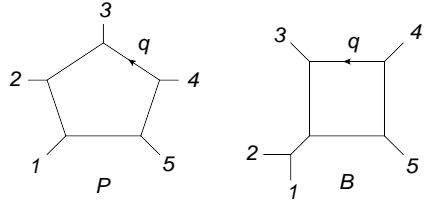}
	\caption{The two topologies for a one-loop five-point amplitude}
	\label{fig:boxandpenta}
\end{figure}

Our goal is to reconstruct the one-loop MHV five-point amplitude with the integrand-reduction method. Instead of sampling a known numerator on the residues we will use the fact that the integrand factorizes into a product of tree amplitudes in the corresponding channel. \\ 
If we absorb the four-point interactions into the three-point ones by multiplying and dividing with the corresponding propagator and if we do not consider any graphs with triangle subgraphs since they will sum up to zero for $\mathcal{N}$=4 sYM \cite{BjerrumBohr:2006yw,Bern:2007xj} we arrive at two graphs for the one-loop amplitude: a box and a pentagon. Both of these graphs can be seen in figure \ref{fig:boxandpenta}.\\
In $\mathcal{N}$=4 the numerator of an integrand can be at most linear in the loop momentum. This means that our integrand-reduction will naturally stop after the first step. Therefore the pentagon integrand has the following expansion in its residues
\begin{gather}
\begin{split}
I^P(1,2,3,4,5,q)=\frac{\Delta^P_{12345}(1,2,3,4,5,q)}{D_1D_2D_3D_4D_5}+ \frac{\Delta^P_{1234}(51,2,3,4)}{D_1D_2D_3D_4}+ \frac{\Delta^P_{1235}(45,1,2,3,)}{D_1D_2D_3D_5} \\
+ \frac{\Delta^P_{1245}(34,5,1,2)}{D_1D_2D_4D_5}
 + \frac{\Delta^P_{1345}(23,4,5,1)}{D_1D_3D_4D_5} + \frac{\Delta^P_{2345}(12,3,4,5)}{D_2D_3D_4D_5}
\end{split}
\label{pentint}
\end{gather}
with the following list of propagators
\begin{gather}
D_1=(q-p_2-p_3)^2 \\
D_2=(q-p_3)^2 \\
D_3=q^2\\
D_4=(q+p_4)^2\\
D_5=(q+p_4+p_5)^2 .
\end{gather} 
Before we proceed with the integrand-reduction we should notice a contradiction. The integrand-reduction had a reducibility criterion which told us that an one-loop integrand with five propagators should be rewritable as a sum of integrands with four propagators. Therefore the introduction of the pentagon topology is somewhat artificial. This will obscure some features of integrand-reduction but we will try to follow the strategy we discussed earlier as close as possible. \\ 
The box integrand in $\mathcal{N}$=4 sYM has the following expansion in its propagator multipoles
\begin{gather}
I^B(12,3,4,5,q)= \frac{\Delta^B_{2345}(12,3,4,5)}{s_{12}D_2D_3D_4D_5} .
\label{boxint}
\end{gather}
We will now use maximum cuts of the one-loop amplitude to fit the coefficients in the residues of the pentagon and the box numerator.

\section{Residues from Unitarity Cuts}
In four dimensions the maximum number of propagators we can set on-shell is four. As we can see from the expressions for the box (\ref{boxint}) and pentagon (\ref{pentint}) integrand we will have contributions from both graphs in this case. Therefore we are only able to sample the mixed residues of the pentagon and the box numerator. Generally the quadruple-cut has two solutions, $q_1$ and $q_2$ so we will solve the following two by two system 
\begin{gather}
 I(q_1)=\frac{\Delta^P_{abcde}(a,b,c,d,e,q_1)}{D_a(q_1)}+c_{bcde}(ab,c,d,e)   \\
 I(q_2)=\frac{\Delta^P_{abcde}(a,b,c,d,e,q_2)}{D_a(q_2)}+c_{bcde}(ab,c,d,e).
 \label{cut}
\end{gather}
As stated above the $c_{bcde}(ab,c,d,e)$ has contributions from the pentagon and the box numerator
\begin{gather}
 I^P(a,b,c,d,e,q_i)+I^B(ab,c,d,e,q_i)\\
 =\frac{\Delta^P_{abcde}(a,b,c,d,e,q_i)}{D_a} + \Delta^P_{bcde}(ab,c,d,e)+ \frac{\Delta^B_{bcde}(ab,c,d,e)}{s_{ab}} \\
 = \frac{\Delta^P_{abcde}(a,b,c,d,e,q_i)}{D_a} + c_{bcde}(ab,c,d,e).
\end{gather}
As we have mentioned earlier the introduction of the pentagon obscures some features of the integrand-reduction. This fact becomes visible here. Normally we would use the quadruple-cut to fit the constant and the linear term of quadruple-cut residue. But here the pentagon numerator replaces the linear term and a mix of the pentagon and the box will build up the constant.\\
Since we do not expect that the maximum cut theorem is violated the residue of the pentagon $\Delta^P_{abcde}(a,b,c,d,e)$ and the mixed residue of the pentagon and the box $\Delta^B_{bcde}(ab,c,d,e)$ are both constants 
\begin{gather}
 \Delta^P_{12345}(1,2,3,4,5,q)=a^P_{12345}(1,2,3,4,5)\\ 
 \Delta^P_{2345}(12,3,4,5)=a^P_{2345}(12,3,4,5) \\
 \Delta^P_{1345}(23,4,5,1)=a^P_{1345}(23,4,5,1) \\
 \Delta^P_{1245}(34,5,1,2)=a^P_{1245}(34,5,1,2) \\
 \Delta^P_{1235}(45,1,2,3)=a^P_{1235}(45,1,2,3) \\
 \Delta^P_{1234}(51,2,3,4)=a^P_{1234}(51,2,3,4) \\
 \Delta^B_{2345}(12,3,4,5)=b^B_{2345}(12,3,4,5).
\end{gather}
We will verify this assumption on the corresponding unitarity cuts.

\subsection{Quadruple-Cut of $D_2,D_3,D_4$ and $D_5$}
\begin{figure}[h]
	\centering
		\includegraphics[width=0.20\textwidth]{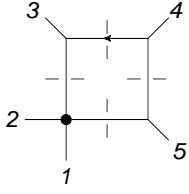}
	\caption{Quadruple-cut of an one-loop amplitude}
	\label{fig:box}
\end{figure}

First we will consider a quadruple-cut where we have a four-point amplitude involving the external legs one and two as displayed in figure \ref{fig:box}. 
This means our product of generating functions involves one four-point generator and three three-point ones. As we discussed while merging generating functions we can use the following formula to determine how many MHV and $\overline{MHV}$ vertices we need
\begin{gather}
\kappa=2M+\bar{M}-i
\end{gather}
where the four-point generating function has also two negative gluons so it can be counted as a three-point MHV generating function. Applying this formula for a MHV amplitude with $\kappa=2$ and with four internal legs $i=4$ we find that we have two MHV and two $\overline{MHV}$ amplitudes to distribute. To find all ways to distribute the three-point generating functions we should also note that we cannot connect two external legs with two generating functions of the same type. This follows from the fact that the $\lambda$s or the $\tilde{\lambda}$s of the two neighboring vertices would be proportional to each other which means that the two external momenta would not be independent of each other. \\
With this restriction and the fact that the four-point amplitude is already fixed we only have one way of distributing the generating functions
\begin{gather}
\begin{split}
I(q)=A_4^{MHV}(1,2,-l_2,l_5)A_3^{\overline{MHV}}(l_2,3,-l_3)A_3^{MHV}(l_3,4,-l_4)\\A_3^{\overline{MHV}}(l_4,5,-l_5) .
\end{split}
\end{gather}
Plugging in our generating functions (\ref{mhv4}, \ref{mhv3}, \ref{mhvbar3}) and integrating over the shared Grassmann variables we find\footnote{The complete calculation can be found in the appendix \ref{derivationintegrand}.}
\begin{gather}
I(q) =  - \delta^{(8)}\left(\sum_{i=1}^{5}{\lambda_i \eta_i} \right)\frac{ \spab3.\slashed{l_5}.5 \spb3.4} {\spa1.2 \spa2.3 \spa3.4 \spa5.1} .
\label{boxcuteq}
\end{gather} 

The $l_i$ corresponds to the cut propagator $D_i$. The next step is to find the solutions of the cut equations on which we will evaluate this integrand. This can be done with the following decomposition for the loop momentum
\begin{gather}
q^\mu=x_1p_3^\mu+x_2p_4^\mu+ x_3 \spab3.\gamma_\mu.4 +x_4 \spab4.\gamma_\mu.3 .
\end{gather}
Solving the first three equations we find that
\begin{gather}
D_2=(q-p_3)^2=0  \rightarrow x_1=0 \\
D_4=(q+p_4)^2=0 \rightarrow x_2=0 \\
D_3=(q)^2=0 \rightarrow x_1x_2=x_3x_4.
\end{gather}
This leads us to the two possible solutions that either $x_3$ or $x_4$ has to be zero. With the help of the last cut equation we can then determine $x_4$ or $x_3$ respectively. Doing this we arrive at the two solutions of the quadruple-cut
\begin{gather}
q_1^\mu=-\frac{ \spa4.5 }{2 \spa3.5 } \spab3.\gamma^\mu.4  \\
q_2^\mu=-\frac{\spb4.5 }{2 \spb3.5 } \spab4.\gamma^\mu.3 .
\label{4cutsol}
\end{gather}
Evaluating the product of tree amplitudes (\ref{boxcuteq}) on the two cut solutions we find the following two expressions
\begin{align}
I(q_1)= s_{34}s_{45} A^{tree}(1,2,3,4,5)\\
I(q_2)=0 .
\end{align}
These two expressions span a system of equations with which we can obtain the two coefficients of the maximum cut as we mentioned in (\ref{cut}) 
\begin{align}
s_{34}s_{45} A^{tree}(1,2,3,4,5)=\frac{a^P_{12345}(1,2,3,4,5)}{D_1(q_1)}+c_{2345}(12,3,4,5)   \\
0=\frac{a^P_{12345}(1,2,3,4,5)}{D_1(q_2)}+c_{2345}(12,3,4,5) .
\end{align}
Solving this system we find the two constants $a^P_{12345}$ and $c_{2345}$
\begin{gather}
a^P_{12345}(1,2,3,4,5)= \beta_{12345}= \delta^{(8)}(Q)\frac{ \spb1.2 \spb2.3 \spb3.4 \spb4.5 \spb5.1 }{\spab1.{\slashed{2} \slashed{3} \slashed{5}}.1  - \spba1.{\slashed{2} \slashed{3} \slashed{5}}.1 }
\label{penta} \\
c_{2345}(12,3,4,5)= \frac{\gamma_{12345}}{s_{12}}= \delta^{(8)}(Q)\frac{ \spb1.2 \spb3.4 \spb4.5 \spb3.5  }{ \spa2.1 \left(\spab1.{\slashed{2} \slashed{3} \slashed{5}}.1  - \spba1.{\slashed{2} \slashed{3} \slashed{5}}.1   \right)} .
\end{gather}
We should note that the denominator of beta and gamma can be written as
\begin{gather}
\spab1.{\slashed{2} \slashed{3} \slashed{5}}.1 - \spba1.{\slashed{2} \slashed{3} \slashed{5}}.1 = 4 \epsilon(1,2,3,5)
\end{gather} 
where epsilon is defined through the equation
\begin{gather}
\spab{i}.{\slashed{j} \slashed{k} \slashed{l}}.i = Tr(\frac{1}{2}(1-\gamma_5)\slashed{p}_i\slashed{p}_j\slashed{p}_k\slashed{p}_l)\\
=\frac{1}{2} \left[s_{ij}s_{lm} - s_{il}s_{jm}+s_{im}s_{jl}- 4\epsilon(i,j,k,l)  \right] .
\end{gather}
In the special case of a five-point kinematic we find that $\epsilon(1,2,3,4)= \epsilon(2,3,4,5)$. \\
The other unitarity cuts can be obtained by relabeling the results $c_{2345}(12,3,4,5)$ and $a^P_{12345}(1,2,3,4,5)$ according to the labels in the parenthesis. The pentagon coefficient $a^P_{12345}(1,2,3,4,5)$ is invariant under these relabellings. Therefore we can conclude that the maximum cut theorem really holds even for this obscured case. The introduction of the pentagon can be understood as an absorption of the linear term of the box into the introduced pentagon. 

\section{The Duality Equations}
\begin{figure}[h]
	\centering
		\includegraphics[width=0.60\textwidth]{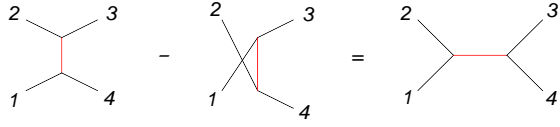}
	\caption{The Jacobi Identity for a product of two structure constants}
	\label{fig:jacobicolor}
\end{figure}

\begin{figure}[h]
	\centering
		\includegraphics[width=0.65\textwidth]{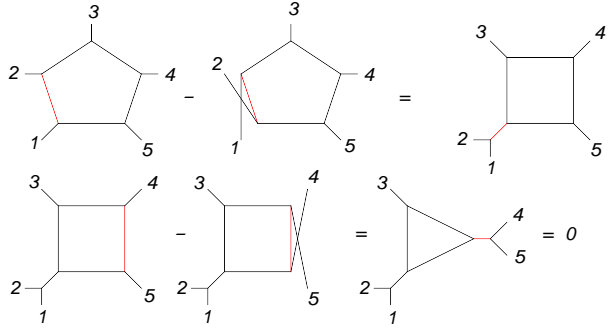}
	\caption{BCJ equations for the box and pentagon numerator}
	\label{fig:bcjbox}
\end{figure}

We obtained a mix of the pentagon numerator and the box numerator. This means we have some freedom where we can move terms between these two graphs as long as the unitarity cuts are satisfied. In order to disentangle the two graphs in a BCJ conform manner we will use the following BCJ equations
\begin{gather}
N^B(12,3,4,5,q)=N^P(1,2,3,4,5,q)-N^P(2,1,3,4,5,q)\\
N^B(12,3,4,5,q)-N^B(12,3,5,4,q)=0\\
N^B(12,3,4,5,q)-N^B(12,4,3,5,q)=0 . 
\label{bcj1loop}
\end{gather}
The easiest way to understand these equations is to remind ourselves of the Jacobi Identity displayed in figure \ref{fig:jacobicolor}
and then consider the corresponding graphs of the equations displayed in \ref{fig:bcjbox}

\subsection{Residues and BCJ equations}
By plugging in the expansion of the numerators into the BCJ equations we can translate these equations into constraints on our residues.  While the equations on the boxes do this in a straightforward manner the first equation needs a more careful analysis. We should start by noting that
\begin{gather}
a^P_{12345}(1,2,3,4,5)-a^P_{1'2345}(2,1,3,4,5)\\
=c_{2345}(12,3,4,5)s_{12}= a^P_{2345}(12,3,4,5)s_{12}+b^B_{2345}(12,3,4,5) .
\label{1steq}
\end{gather}
With that we can go into the first equation to see that
\begin{gather}
N^B(12,3,4,5,q)=N^P(1,2,3,4,5,q)-N^P(2,1,3,4,5,q)  \\
\begin{split}
 =  b^B_{2345}(12,3,4,5) + \left( a^P_{2345}(12,3,4,5)+a^P_{2345}(21,3,4,5)\right)s_{12} \\
 + \left(a^P_{2345}(12,3,4,5)+a^P_{2345}(21,3,4,5) \right)D_1\\
  \left(a^P_{1345}(23,4,5,1) -a^P_{1'345}(13,4,5,2) \right)D_2 \\
+\left(a^P_{1245}(34,5,1,2)-a^P_{1'245}(34,5,2,1)-a^P_{2345}(21,3,4,5) \right)D_3\\
 + \left(a^P_{1235}(45,1,2,3)-a^P_{1'235}(45,2,1,3) \right)D_4 \\ +\left(a^P_{1234}(51,2,3,4)-a^P_{1'234}(52,1,3,4)-a^P_{2345}(21,3,4,5) \right)D_5
\end{split}
\end{gather}
where we already collected all the terms belonging to the propagators with the equation $D'_1=D_5-D_1+D_3-s_{12}$. As we have stated earlier in the discussion of the one-loop amplitude we expect the box to be a constant. Therefore all the prefactors of the denominators must vanish giving us the following five equations
\begin{gather}
\begin{split}
a^P_{2345}(12,3,4,5)+a^P_{2345}(2,1,3,4,5)=0 \\
a^P_{1345}(23,4,5,1,2) -a_{1'345}(13,4,5,2)=0 \\
a^P_{1245}(34,5,1,2)-a^P_{1'245}(34,5,2,1)-a^P_{2345}(21,3,4,5)=0 \\
a^P_{1235}(45,1,2,3)-a^P_{1'235}(45,1,2,3)=0  \\
a^P_{1234}(51,2,3,4)-a^P_{1'234}(52,1,3,4)-a^P_{2345}(21,3,4,5) =0  .
\end{split}
\label{5cons1bcj}
\end{gather}
Together with the two constraints from the BCJ equations on the box numerators and the conditions that the unitarity cut has to be satisfied we arrive at a set of 13 constraints on the residues
\begin{gather}
\begin{split}
a^P_{2345}(12,3,4,5)+a^P_{2345}(2,1,3,4,5)=0 \\
a^P_{1345}(23,4,5,1,2) -a_{1'345}(13,4,5,2)=0 \\
a^P_{1245}(34,5,1,2)-a^P_{1'245}(34,5,2,1)-a^P_{2345}(21,3,4,5)=0 \\
a^P_{1235}(45,1,2,3)-a^P_{1'235}(45,1,2,3)=0  \\
a^P_{1234}(51,2,3,4)-a^P_{1'234}(52,1,3,4)-a^P_{2345}(21,3,4,5) =0 \\
b^B_{2345}(12,3,4,5)-b^B_{234'5}(12,3,5,4)=0\\
b^B_{2345}(12,3,4,5)-b^B_{23'45}(12,4,3,5)=0 \\
c_{2345}(12,3,4,5)=a^P_{2345}(1,2,3,4,5)+\frac{b^B_{2345}(12,3,4,5)}{s_{12}} .
\end{split}
\label{constraints}
\end{gather}

\subsection{$\gamma$-Decomposition}

In order to solve these functional equations we should formulate an ansatz for the appearing functions. This can be done by realizing that all the dependence on the external momenta can be captured by the gammas 
\begin{gather}
\gamma_{12345}= \delta^{(8)}(Q)\frac{ \spb1.2 ^2 \spb3.4 \spb4.5 \spb3.5  }{ \spab1.{\slashed{2} \slashed{3} \slashed{5}}.1  - \spba1.{\slashed{2} \slashed{3}  \slashed{5}}.1 } .
\end{gather}
The gammas have two properties which can be easily proven. First they are completely symmetric in the last three indices. Secondly they are antisymmetric in the first two indices. This makes it sufficient to label the gammas by the first two indices. Therefore there are only $\frac{5*4}{2}=10$ independent gammas, but there are five more identities of the form
\begin{gather}
\gamma_{1a}+\gamma_{2a}+\gamma_{3a}+\gamma_{4a}+\gamma_{5a}=0
\end{gather} 
which reduce the number of independent gammas down to six, since one of the identities is not independent. \\
Choosing six gammas we can formulate ans\"atze for $a^P_{bcde}(ab,c,d,e)$ and $b^B_{bcde}(ab,c,d,e)$
\begin{gather}
a^P_{bcde}(ab,c,d,e)=\alpha_1\gamma_{ab}+\alpha_2\gamma_{bc}+\alpha_3\gamma_{cd}+\alpha_4\gamma_{de}+\alpha_5\gamma_{ea}+\alpha_6\gamma_{ac}\\
b^B_{bcde}(ab,c,d,e)=\beta_1\gamma_{ab}+\beta_2\gamma_{bc}+\beta_3\gamma_{cd}+\beta_4\gamma_{de}+\beta_5\gamma_{ea}+\beta_6\gamma_{ac} 
\end{gather}
with 12 rational coefficients $\alpha_i$ and $\beta_i$. Expanding the six equations (\ref{constraints}) in terms of the six gammas and comparing the coefficients of the gammas we find $6*6=36$ equations. But 24 of the 36 equations are not linearly independent. Therefore we can still find a solution which can be verified with any algebraic program.

\subsection{Solution}
Imposing the last two duality equations on the boxes
\begin{gather}
b^B_{bcde}(ab,c,d,e)-b^B_{bcd'e}(ab,c,e,d)=0\\
b^B_{bcde}(ab,c,d,e)-b^B_{bc'de}(ab,d,c,e)=0
\end{gather}
tells us that the box constant $b^B(ab,c,d,e)$ has to be completely symmetric in the last three arguments $c,d$ and $e$. 
\begin{gather}
b^B_{bcde}(ab,c,d,e) = \kappa_1 \gamma_{ab}
\end{gather}
Therefore the unitarity cut equation
\begin{gather}
\frac{\gamma_{ab}}{s_{ab}}=a^P_{bcde}(ab,c,d,e)+\frac{b^B_{bcde}(ab,c,d,e)}{s_{ab}}
\end{gather}
directly implies that 
\begin{gather}
a^P_{bcde}(ab,c,d,e) = (1-\beta_1)\frac{\gamma_{ab}}{s_{ab}}.
\end{gather}
Now looking at the five constraints from the first BCJ equation (\ref{5cons1bcj})
\begin{gather}
a^P_{2345}(12,3,4,5)+a^P_{2345}(2,1,3,4,5)=0 \\
a^P_{1345}(23,4,5,1) -a_{1'345}(13,4,5,2)=0 \\
a^P_{1245}(34,5,1,2)-a^P_{1'245}(34,5,2,1)-a^P_{2345}(21,3,4,5)=0 \\
a^P_{1235}(45,1,2,3)-a^P_{1'235}(45,1,2,3)=0  \\
a^P_{1234}(51,2,3,4)-a^P_{1'234}(52,1,3,4)-a^P_{2345}(21,3,4,5) =0 
\end{gather}
we find that the only solutions is 
\begin{gather}
a^P_{bcde}(ab,c,d,e)=0
\end{gather} 
which immediately gives us
\begin{gather}
b^B_{bcde}(ab,c,d,e)=\gamma_{ab}.
\end{gather}
Plugging these solutions into the numerator expansions of the pentagon and the box we arrive at the following result
\begin{gather}
I^P(1,2,3,4,5,q)=\frac{\beta_{12345}}{D_1D_2D_3D_4D_5} \\
I^B(12,3,4,5,q)= \frac{\gamma_{12}}{s_{12}D_2D_3D_4D_5}
\end{gather}
which is in agreement with the results from \cite{Carrasco:2011mn}.

\section{The Full One-Loop Amplitude}
We can obtain the full color dressed amplitude if we sum over all permutations of legs, dress up each graph with the corresponding product of structure constants and include a factor for each graph to compensate for over counting
\begin{gather}
\mathcal{A}^{\text{1-loop}}=ig^5 \sum_{\text{all perm}} \frac{1}{10} \beta_{12345} C^P \text{Int}^P + \frac{1}{4} \frac{\gamma_{12}}{s_{12}}C^B \text{Int}^B
\label{oneloopamp}
\end{gather}
where the color factors are given by
\begin{gather}
C^P=f^{g1b}f^{b2c}f^{c3d}f^{d4e}f^{e5g}\\
C^B=f^{12b}f^{bcg}f^{c3d}f^{d4e}f^{e5g}
\end{gather}
and the corresponding integrals are 
\begin{gather}
\text{Int}^P=\int \frac{d^4q} {(2\pi)^4} \frac{1}{D_1D_2D_3D_4D_5} \\
\text{Int}^B=\int \frac{d^4q}{(2\pi)^4} \frac{1}{D_2D_3D_4D_5} .
\label{int1loop}
\end{gather}
In the next section we are going to discuss the two-loop five-point amplitude. At two-loops we will not have to introduce a reducible graph to apply the BCJ equations as we have done here. This has the advantage that the integrand-reduction can be used in a more normal fashion.

\newpage
\chapter {The Two-Loop Five-Point Amplitude in $\mathcal{N}$=4 sYM}
\begin{figure}[h]
	\centering
		\includegraphics[width=0.90\textwidth]{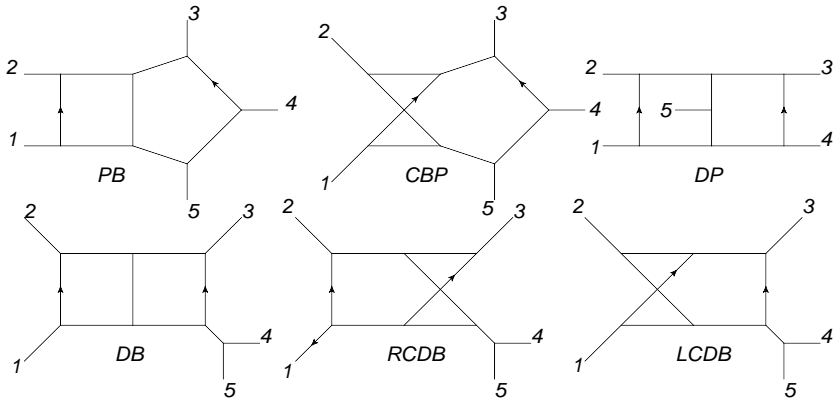}
	\caption{All graphs of the two-loop five-point amplitude}
	\label{Figurefulltwoloop}
\end{figure}

In order to obtain all graphs for the two-loop five-point amplitude we are going to follow the same strategy as in the one-loop case. This means drawing all possible two-loop graphs with only three-point vertices and no loops with less then four legs attached. Doing so we can find the six graphs displayed in figure \ref{Figurefulltwoloop}. \\  
The remarkable thing is that all these graphs can be connected by BCJ equations
\begin{gather}
N^{PB}(1,2,3,4,5,q,k)=N^{CPB}(1,2,3,4,5,q,k) \\
\begin{split}
N^{DP}(1,2,3,4,5,q,k)=N^{PB}(1,2,5,3,4,-q-3,k)\\
-N^{PB}(3,4,1,2,5,-k,q)
\end{split} \\
N^{DB}(1,2,3,4,5,q,k)=N^{PB}(1,2,3,4,5,q,k)-N^{PB}(1,2,3,5,4,q,k) \\
N^{DB}(1,2,3,4,5,q,k)=N^{LCDB}(1,2,3,4,5,q,k) \\
N^{DB}(1,2,3,4,5,q,k)=N^{RCDB}(1,2,3,4,5,q,k) .
\end{gather}
This drasticly reduces the work that has to be done in order to determine this amplitude. Here we will choose to promote the pentabox to the master graph and then obtain the rest through the BCJ equations.

\section{Two-Loop Pentabox Diagram}
As in the one-loop case we are going to use the integrand reduction technique and the unitarity method to constraint the pentabox diagram as much as possible. After that we will end up with some leftover freedom how to disentangle the seven-pole cuts which can be used to apply BCJ equations.\\ 
The integrand of the pentabox has the following expansion in its residues
\begin{gather}
\begin{split}
I^{PB}(1,2,3,4,5,q,k)=\frac{\Delta^{PB}_{12345678}(1,2,3,4,5,q,k)}{D_1D_2D_3D_4D_5D_6D_7D_8}+\frac{\Delta^{PB}_{1235678}(1,2,3,4,5)}{D_1D_2D_3D_5D_6D_7D_8}\\
+\frac{\Delta^{PB}_{1234678}(1,2,34,5)}{D_1D_2D_3D_4D_6D_7D_8}+\frac{\Delta^{PB}_{1234578}(1,2,3,45)}{D_1D_2D_3D_4D_5D_7D_8}+\frac{\Delta^{PB}_{1234568}(1,2,3,4,5)}{D_1D_2D_3D_4D_5D_6D_8}
\end{split}
\label{cutsystem}
\end{gather}
with the following list of propagators
\begin{gather}
D_1=(k+p_1)^2 \\
D_2=k^2 \\
D_3=(k-p_2)^2 \\
D_4=(q-p_3)^2\\
D_5=q^2 \\
D_6=(q+p_4)^2 \\
D_7=(q+p_4+p_5)^2\\
D_8=(q+k-p_2-p_3)^2 .
\end{gather}
In order to determine the residues we will first evaluate the integrand at the solutions of the eightfold-cut to isolate the eightfold-cut residue $\Delta^{PB}_{12345678}(1,2,3,4,5,q,k)$ and to fit all the coefficients in the eightfold-cut residue. In the next step we will proceed to the sevenfold-cuts to determine the rest of the residues.

\subsection{The Eightfold-Cut}
\begin{figure}[h]
	\centering
		\includegraphics[width=0.30\textwidth]{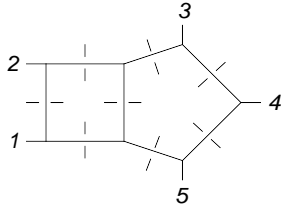}
	\caption{Eightfold-cut of the pentabox}
	\label{fig:eightfoldcut}
\end{figure}

The eightfold-cut is displayed in figure \ref{fig:eightfoldcut}

The maximum cut theorem \cite{Mastrolia:2012an} tells us that the number of cut solutions is equal to the number of free parameters in the residue. Indeed in \cite{twoloopisp} we can find that the polynomial of the eightfold-cut residue of the pentabox has four coefficients
\begin{gather}
\begin{split}
\Delta^{PB}_{12345678}(1,2,3,4,5,q,k)=c^{PB}_{0;12345678}(1,2,3,4,5)+c^{PB}_{1;12345678}(1,2,3,4,5) q_1\cdot p_1 \\
+c^{PB}_{2;12345678}(1,2,3,4,5) k_1\cdot p_4+c^{PB}_{3;12345678}(1,2,3,4,5) k_1\cdot p_3
\end{split}
\end{gather}
and by solving the cut equations we find four solutions. So in total we will solve the following system
\begin{gather}
\begin{split}
I^{PB}(1,2,3,4,5,q_1,k_1)=c^{PB}_{0;12345678}(1,2,3,4,5)+c^{PB}_{1;12345678}(1,2,3,4,5) q_1\cdot p_1\\
+c^{PB}_{2;12345678}(1,2,3,4,5) k_1\cdot p_4+c^{PB}_{3;12345678}(1,2,3,4,5) k_1\cdot p_3 \\
I^{PB}(1,2,3,4,5,q_2,k_2)=c^{PB}_{0;12345678}(1,2,3,4,5)+c^{PB}_{1;12345678}(1,2,3,4,5) q_2\cdot p_1\\
+c^{PB}_{2;12345678}(1,2,3,4,5) k_2\cdot p_4+c^{PB}_{3;12345678}(1,2,3,4,5) k_2\cdot p_3 \\
I^{PB}(1,2,3,4,5,q_3,k_3)=c^{PB}_{0;12345678}(1,2,3,4,5)+c^{PB}_{1;12345678}(1,2,3,4,5) q_3\cdot p_1\\
+c^{PB}_{2;12345678}(1,2,3,4,5) k_3\cdot p_4+c^{PB}_{3;12345678}(1,2,3,4,5) k_3\cdot p_3 \\
I^{PB}(1,2,3,4,5,q_4,k_4)=c^{PB}_{0;12345678}(1,2,3,4,5)+c^{PB}_{1;12345678}(1,2,3,4,5) q_4\cdot p_1\\
+c^{PB}_{2;12345678}(1,2,3,4,5) k_4\cdot p_4+c^{PB}_{3;12345678}(1,2,3,4,5) k_4\cdot p_3
\end{split}
\label{8polecut}
\end{gather}
where $q_i$ and $k_i$ denote the $i$th solution of the cut equations. In order to solve this system of equations we will use factorization of the integrand through unitarity. This means that instead of sampling the cut solutions on the integrand we will sample them on the product of tree generating functions and sum over all choices of internal helicities. In our example we can find four ways of distributing the generating functions
\begin{gather}
\begin{split}
I^{PB}_1(1,2,3,4,5,q,k)=A_3^{\overline{MHV}}(1,l_2,-l_1)A_3^{MHV}(-l_2,2,l_3)A_3^{\overline{MHV}}(-l_3,-l_4,l_8)\\A_3^{\overline{MHV}}(l_4,3,-l_5)
A_3^{MHV}(l_5,4,-l_6)A_3^{\overline{MHV}}(l_6,5,-l_7)A_3^{MHV}(l_7,-l_8,l_1)
\end{split}\\
\begin{split}
I^{PB}_2(1,2,3,4,5,q,k)=A_3^{MHV}(1,l_2,-l_1)A_3^{\overline{MHV}}(-l_2,2,l_3)A_3^{MHV}(-l_3,-l_4,l_8)\\A_3^{\overline{MHV}}(l_4,3,-l_5) 
A_3^{MHV}(l_5,4,-l_6)A_3^{\overline{MHV}}(l_6,5,-l_7)A_3^{\overline{MHV}}(l_7,-l_8,l_1)
\end{split}\\
\begin{split}
I^{PB}_3(1,2,3,4,5,q,k)=A_3^{MHV}(1,l_2,-l_1)A_3^{\overline{MHV}}(-l_2,2,l_3)A_3^{\overline{MHV}}(-l_3,-l_4,l_8)\\A_3^{MHV}(l_4,3,-l_5) 
A_3^{\overline{MHV}}(l_5,4,-l_6)A_3^{MHV}(l_6,5,-l_7)A_3^{\overline{MHV}}(l_7,-l_8,l_1)
\end{split}\\
\begin{split}
I^{PB}_4(1,2,3,4,5,q,k)=A_3^{\overline{MHV}}(1,l_2,-l_1)A_3^{MHV}(-l_2,2,l_3)A_3^{\overline{MHV}}(-l_3,-l_4,l_8)\\A_3^{MHV}(l_4,3,-l_5)
A_3^{\overline{MHV}}(l_5,4,-l_6)A_3^{MHV}(l_6,5,-l_7)A_3^{\overline{MHV}}(l_7,-l_8,l_1).
\end{split}
\end{gather}
The next step is to solve the cut equations so we can later plug their solutions into the product of trees. Using the same decomposition for the loop momentum as in the one-loop case
\begin{gather}
q^\mu=x_1p_3^\mu+x_2p_4^\mu+ x_3 \spab3.\gamma_\mu.4 +x_4 \spab4.\gamma_\mu.3 \\
k^\mu=y_1p_1^\mu+y_2p_2^\mu+ y_3 \spab3.\gamma_\mu.4 +x_4 \spab4.\gamma_\mu.3  
\end{gather}
we find the following four solutions of the cut equations.
\begin{center}
\begin{tabular}{c|c|c}
solution & $k^\mu$ & $q^\mu$ \\
\hline
1st & $\frac{ \spa5.1 } {2 \spa2.5  } \spab2.\gamma^\mu.1 $ & $\frac{ \spa5.4 }{2 \spa3.5  } \spab3.\gamma^\mu.4  $ \\ 
2nd & $\frac{ \spb2.3  }{2 \spb1.3 } \spab2.\gamma^\mu.1$ & $\frac{ \spb5.4 }{2 \spb3.5 }\spab4.\gamma^\mu.3 $\\
3rd & $\frac{ \spa2.3 }{2 \spa1.3 }\spab1.\gamma^\mu.2 $ & $ \frac{ \spa5.4 }{2 \spa3.5 } \spab3.\gamma^\mu.4 $\\
4th & $\frac{ \spb5.1 }{2 \spb2.5 } \spab1.\gamma^\mu.2 $ & $\frac{ \spb5.4 }{2 \spb3.5} \spab4.\gamma^\mu.3 $
\end{tabular}
\end{center}
These solutions can now be plugged into the product of tree amplitudes 
\begin{center}
\begin{tabular}{c|c}
solution & Integrand \\
\hline
$I^{PB}(1,2,3,4,5,q_1,k_1)$ & $ \delta^{(8)}(Q)\frac { \spa1.2 ^2 \spb3.4 \spb5.4 } { \spa3.5 }$  \\ 
$I^{PB}(1,2,3,4,5,q_2,k_2)$ & $ 0 $\\
$I^{PB}(1,2,3,4,5,q_3,k_3)$ & $\delta^{(8)}(Q) \frac { \spa1.2 ^2 \spb3.4 \spb5.4 } { \spa3.5 } $\\
$I^{PB}(1,2,3,4,5,q_4,k_4)$ & $0$
\end{tabular}
\end{center}
to span the left-hand side of the system of equations in (\ref{8polecut})
\begin{align}
\begin{split}
 \delta^{(8)}\left(q\right) \frac { \spa1.2 ^2 \spb3.4 \spb5.4 } { \spa3.5 }= c^{PB}_{0;12345678}(1,2,3,4,5)+c^{PB}_{1;12345678}(1,2,3,4,5) q_1\cdot p_1 \\
 +c^{PB}_{2;12345678}(1,2,3,4,5) k_1\cdot p_4+c^{PB}_{3;12345678}(1,2,3,4,5) k_1\cdot p_3 \\
 0=c^{PB}_{0;12345678}(1,2,3,4,5) +c^{PB}_{1;12345678}(1,2,3,4,5) q_2\cdot p_1 \\
 +c^{PB}_{2;12345678}(1,2,3,4,5) k_2\cdot p_4+c^{PB}_{3;12345678}(1,2,3,4,5) k_2\cdot p_3 \\
 \delta^{(8)}\left(q\right) \frac { \spa1.2 ^2 \spb3.4 \spb5.4 } { \spa3.5 }= c^{PB}_{0;12345678}(1,2,3,4,5)+c^{PB}_{1;12345678}(1,2,3,4,5) q_3\cdot p_1 \\
 +c^{PB}_{2;12345678}(1,2,3,4,5) k_3\cdot p_4+c^{PB}_{3;12345678}(1,2,3,4,5) k_3\cdot p_3 \\
 0=c^{PB}_{0;12345678}(1,2,3,4,5) +c^{PB}_{1;12345678}(1,2,3,4,5) q_4\cdot p_1 \\
 +c^{PB}_{2;12345678}(1,2,3,4,5) k_4\cdot p_4+c^{PB}_{3;12345678}(1,2,3,4,5) k_4\cdot p_3 .
 \end{split}
\end{align}
Solving this system we find the coefficients of the maximum-cut residue
\begin{gather}
c^{PB}_{0;12345678}(1,2,3,4,5)=  \delta^{(8)}\left(q\right) \frac{\spb1.2 ^2 \spb4.5 ^2 \spbb3.\slashed{4}.\slashed{1}.3 }{\spab1.{ \slashed{2} \slashed{3} \slashed{5}}.1  - \spba1.{ \slashed{2} \slashed{3} \slashed{5}}.1} \equiv \lambda_{12345} \\
c^{PB}_{1;12345678}(1,2,3,4,5)=2 \delta^{(8)}\left(q\right) \frac{ \spb1.2 ^2 \spb3.4 \spb4.5 \spb3.5 }{\spab1.{ \slashed{2} \slashed{3} \slashed{5}}.1  - \spba1.{ \slashed{2} \slashed{3} \slashed{5}}.1 } \equiv 2\gamma_{12}\\
c^{PB}_{2;12345678}(1,2,3,4,5)=0 \\
c^{PB}_{3;12345678}(1,2,3,4,5)=0 .
\end{gather}
This means the residue of the eightfold-cut is completely determined by the maximum cut and has the following form
\begin{gather}
\Delta^{PB}(1,2,3,4,5,q,k)=\lambda_{12345}+2\gamma_{12} q \cdot p_1 .
\end{gather}
In the next step we will calculate all sevenfold-cuts of the pentabox in order to determine the corresponding residues.  

\subsection{The Sevenfold-Cut without $D_6$}
\begin{figure}[h]
	\centering
		\includegraphics[width=0.85\textwidth]{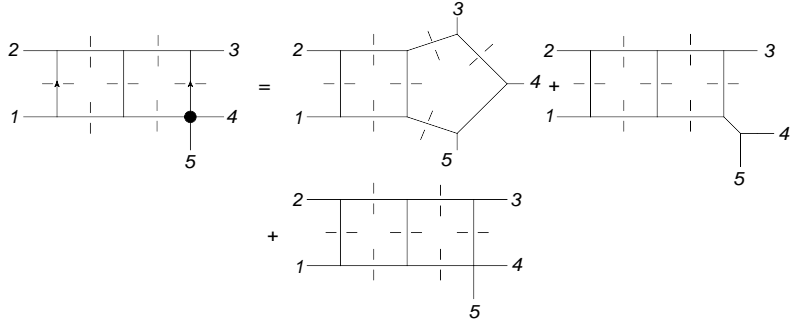}
	\caption{Sevenfold-cut leaving $D_6$ uncut}
	\label{Figure7polemassive}
\end{figure}

The considered sevenfold-cut is displayed in figure \ref{Figure7polemassive} where we simply expanded the four-point amplitude in its channels.
From the three different graphs in the figure we can see the different residues which will contribute to our cut. The integrand-reduction algorithm tells us that we need to subtract all higherfold-cut residues from the integrand. Since the doublebox has only seven propagators we only have one eightfold-cut residue to subtract we find
\begin{gather}
I^{PB}(1,2,3,4,5,q,k)-\frac{\Delta^{PB}_{12345678}(1,2,3,4,5,q,k)}{D_6} \\
=\Delta^{PB}_{1234578}(1,2,3,45)+s_{45} N^{DB}(1,2,3,45,q,k) =d^{PB}_{1234578}(1,2,3,45) .
\end{gather}
In general the sevenfold-cut residue $d^{PB}_{1234578}(1,2,3,45)$ would be a polynomial of the loop momentum with 32 coefficients\footnote{The polynomial is given in the appendix in table \ref{Tab:Pentabox}} \cite{twoloopisp}. But in $\mathcal{N}$=4 we only have numerators which are linear in the loop momentum therefore the sevenfold-cut residues are only constants. We will prove this by fitting the constant from several solutions of the sevenfold cuts.\\  
On the solution of the sevenfold-cut equations the integrand factorizes into product tree amplitudes. In order to take all helicity choices into account we have to sum over the three ways of distributing the MHV and $\overline{MHV}$ generating functions
\begin{gather}
\begin{split}
I^{PB}_1(1,2,3,4,5,q,k)= A_3^{\overline{MHV}}(1,l_2,-l_1)A_3^{MHV}(-l_2,2,l_3)A_3^{MHV}(-l_3,-l_4,l_8)\\A_3^{\overline{MHV}}(l_4,3,-l_5) 
A_4^{MHV}(l_5,4,5,-l_7)A_3^{\overline{MHV}}(l_7,-l_8,l_1)
\end{split} \\
\begin{split}
I^{PB}_2(1,2,3,4,5,q,k)= 
A_3^{\overline{MHV}}(1,l_2,-l_1)A_3^{MHV}(-l_2, 2,l_3)A_3^{\overline{MHV}}(-l_3,-l_4,l_8)\\A_3^{\overline{MHV}}(l_4,3,-l_5)
A_4^{MHV}(l_5,4,5,-l_7)A_3^{MHV}(l_7,-l_8,l_1)
\end{split} \\
\begin{split}
I^{PB}_3(1,2,3,4,5,q,k)= 
A_3^{MHV}(1,l_2,-l_1)A_3^{\overline{MHV}}(-l_2, 2,l_3)A_3^{MHV}(-l_3,-l_4,l_8)\\A_3^{\overline{MHV}}(l_4,3,-l_5) 
A_4^{MHV}(l_5,4,5,-l_7)A_3^{\overline{MHV}}(l_7,-l_8,l_1) .
\end{split}
\end{gather}
Now we can solve the cut equations numericly and evaluating these products of generating functions at the solutions and obtaining the mentioned constant we find
\begin{gather}
d^{PB}_{1234578}(1,2,3,45)=\frac{s_{12}}{s_{45}}\gamma_{45}.
\end{gather}

\subsection{The Sevenfold-Cut without $D_5$}
\begin{figure}[h]
	\centering
		\includegraphics[width=0.30\textwidth]{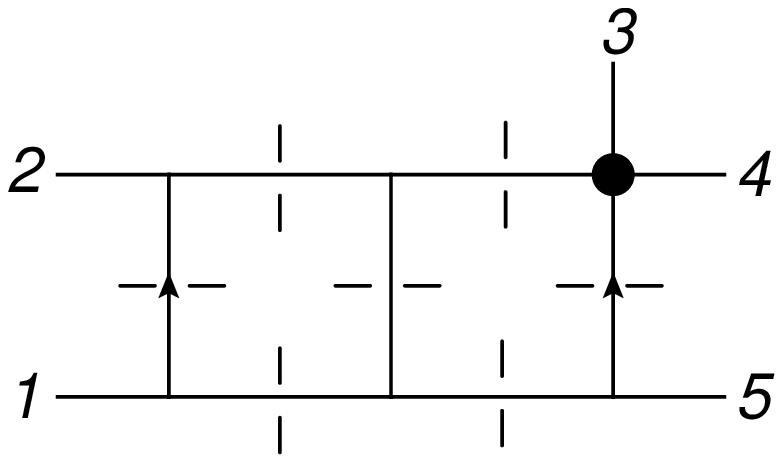}
	\caption{Sevenfold-cut leaving $D_5$ uncut}
	\label{Figure7polemassive2}
\end{figure}

As we can see from figure \ref{Figure7polemassive2}, which is very similar to the one before.
This time we want to solve the following equation
\begin{gather}
I^{PB}(1,2,3,4,5,q,k)-\frac{\Delta^{PB}_{12345678}(1,2,3,4,5,q,k)}{D_5} \\
=\Delta^{PB}_{1234678}(1,2,34,5)+s_{34} N^{DB}(1,2,34,5,q,k) =d^{PB}_{1234678}(1,2,34,5) .
\end{gather}
Since there is no second eightfold-cut residue to subtract we can immediately focus on the integrand.
There are three possibilities to distribute the MHV and $\overline{MHV}$ generating functions
\begin{gather}
\begin{split}
I^{PB}_1(1,2,3,4,5,q,k)=
A_3^{\overline{MHV}}(1,l_2,-l_1)A_3^{MHV}(-l_2,2,l_3)A_3^{\overline{MHV}}(-l_3,-l_4,l_8)\\A_4^{MHV}(l_4,3,4,-l_6) 
A_3^{\overline{MHV}}(l_6,5,-l_7)A_3^{MHV}(l_7,-l_8,l_1)
\end{split} \\
\begin{split}
I^{PB}_2(1,2,3,4,5,q,k)= 
A_3^{MHV}(1,l_2,-l_1)A_3^{\overline{MHV}}(-l_2, 2,l_3)A_3^{\overline{MHV}}(-l_3,-l_4,l_8)\\A_4^{MHV}(l_4,3,4,-l_6)
A_3^{\overline{MHV}}(l_6,5,-l_7)A_3^{MHV}(l_7,-l_8,l_1)
\end{split} \\
\begin{split}
I^{PB}_3(1,2,3,4,5,q,k)= 
A_3^{MHV}(1,l_2,-l_1)A_3^{\overline{MHV}}(-l_2, 2,l_3)A_3^{\overline{MHV}}(-l_3,-l_4,l_8)\\A_4^{MHV}(l_4,3,4,-l_6)
A_3^{\overline{MHV}}(l_6,5,-l_7)A_3^{MHV}(l_7,-l_8,l_1) .
\end{split}
\end{gather}
Using the super amplitudes techniques as before and then evaluating the product of generating functions at the cuts we find the following result
\begin{gather}
d^{PB}_{1234678}(1,2,34,5)=\frac{s_{12}}{s_{34}}\gamma_{34} .
\end{gather}

\subsection{The Sevenfold-Cut without $D_7$}
\begin{figure}[h]
	\centering
		\includegraphics[width=0.90\textwidth]{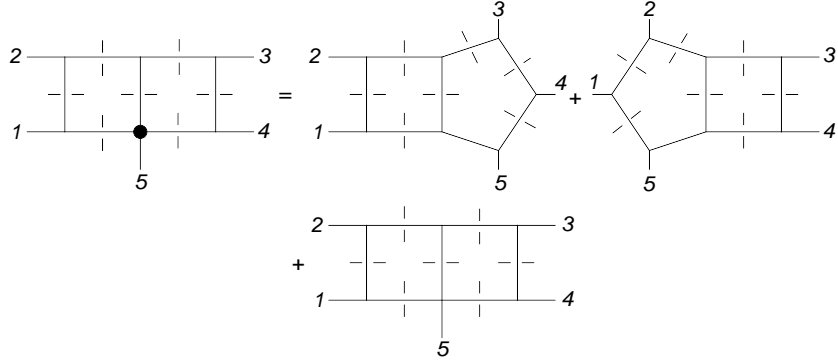}
	\caption{Sevenfold-cut of the pentabox where all propagators but $D_7$ are cut}
	\label{Figurepbd7exp}
\end{figure}

The corresponding unitarity cut is displayed in figure \ref{Figurepbd7exp}. 
In this example we have two graphs with eightfold-cuts contributing. Thus we have to subtract a second maximum cut coming from the pentabox with a different ordering of external legs
\begin{gather}
\label{77cut}
I^{PB}(1,2,3,4,5,q,k)-\frac{\Delta^{PB}_{12345678}(1,2,3,4,5,q,k)}{D_7}-\frac{\Delta^{PB}_{1234567'8}(3,4,5,1,2,k+p_1,q)}{D_7'} \\
=\Delta^{PB}_{1234568}(1,2,3,4,5)+\Delta^{PB}_{1235678}(3,4,5,1,2) =d^{PB}_{1234568}(1,2,3,4,5).
\end{gather}
Here we used that the two pentaboxes share all propagators except $D_7'=(k+p_1+p_5)^2$.\\
The second eightfold-cut residue appearing in (\ref{77cut}) is just a relabeling of the one we obtained earlier
\begin{gather}
\Delta^{PB}_{1234567'8}(3,4,5,1,2,k+p_1,q)=\lambda_{34512}+2\gamma_{34} \left( k + p_1 \right) \cdot p_3.
\end{gather}
With these informations we can now focus on the integrand. As before we will use the product of tree generating functions to sample the constant. In this cut there are three ways of distributing them
\begin{gather}
\begin{split}
I^{PB}_1(1,2,3,4,5,q,k)= 
A_3^{\overline{MHV}}(1,l_2,-l_1)A_3^{MHV}(-l_2,2,l_3)A_3^{\overline{MHV}}(-l_3,-l_4,l_8)\\A_3^{\overline{MHV}}(l_4,3,-l_5) 
A_3^{MHV}(l_5,4,-l_6)A_4^{MHV}(l_6,5,-l_8,l_1)
\end{split} \\
\begin{split}
I^{PB}_2(1,2,3,4,5,q,k)= 
A_3^{\overline{MHV}}(1,l_2,-l_1)A_3^{MHV}(-l_2,2,l_3)A_3^{\overline{MHV}}(-l_3,-l_4,l_8)\\A_3^{MHV}(l_4,3,-l_5) 
A_3^{\overline{MHV}}(l_5,4,-l_6)A_4^{MHV}(l_6,5,-l_8,l_1)
\end{split} \\
\begin{split}
I^{PB}_3(1,2,3,4,5,q,k)= 
A_3^{MHV}(1,l_2,-l_1)A_3^{\overline{MHV}}(-l_2,2,l_3)A_3^{\overline{MHV}}(-l_3,-l_4,l_8)\\A_3^{MHV}(l_4,3,-l_5)
A_3^{\overline{MHV}}(l_5,4,-l_6)A_4^{MHV}(l_6,5,-l_8,l_1) .
\end{split}
\end{gather}
Solving the cut equations and plug the solutions into product of generating functions to obtain the constant
\begin{gather}
d^{PB}_{1234568}(1,2,3,4,5)=-\frac{1}{2}\left(\gamma_{52}+\gamma_{51}-\gamma_{12}+\gamma_{34}  \right) .
\end{gather}

\subsection{The Sevenfold-Cut without $D_4$}
Leaving propagator $D_4$ uncut leaves us with a very similar situation to the sevenfold-cut we solved earlier.
\begin{figure}[h]
	\centering
		\includegraphics[width=0.30\textwidth]{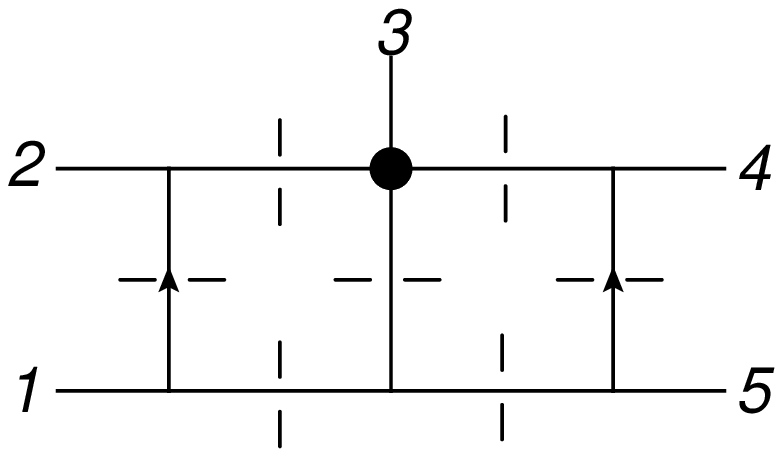}
	\caption{Sevenfold-cut leaving $D_4$ uncut}
	\label{Figuredoubleboxnrom2}
\end{figure}
Again we have two graphs with eightfold-cut residues contributing which gives the following equation
\begin{gather}
I^{PB}(1,2,3,4,5,q,k)-\frac{\Delta^{PB}_{12345678}(1,2,3,4,5,q,k)}{D_4}-\frac{\Delta^{PB}_{1234'5678}(4,5,1,2,3,k,q)}{D_4'} \\
=\Delta^{PB}_{1235678}(1,2,3,4,5)+\Delta^{PB}_{1234568}(4,5,1,2,3) =d^{PB}_{1235678}(1,2,3,4,5)
\end{gather}
with the propagator $D_4'=(k-p_2-p_3)^2$ and the second eightfold-cut residue
\begin{gather}
\Delta^{PB}_{1234'5678}(4,5,1,2,3,k,q)=\lambda_{45123}-2\gamma_{45} k \cdot p_4 .
\end{gather}
As an input for the integrand we will use the product of generating functions, which has three ways of distributing the different three-point generators
\begin{gather}
\begin{split}
I^{PB}_1(1,2,3,4,5,q,k)= 
A_3^{\overline{MHV}}(1,l_2,-l_1)A_3^{MHV}(-l_2,2,l_3)A_4^{MHV}(-l_3,3,-l_5,l_8)\\A_3^{\overline{MHV}}(l_5,4,-l_6) 
A_3^{MHV}(l_6,5,-l_7)A_3^{\overline{MHV}}(l_7,-l_8,l_1)
\end{split} \\
\begin{split}
I^{PB}_2(1,2,3,4,5,q,k)= 
A_3^{MHV}(1,l_2,-l_1)A_3^{\overline{MHV}}(-l_2,2,l_3)A_4^{MHV}(-l_3,3,-l_5,l_8)\\A_3^{\overline{MHV}}(l_5,4,-l_6) 
A_3^{MHV}(l_6,5,-l_7)A_3^{\overline{MHV}}(l_7,-l_8,l_1)
\end{split} \\
\begin{split}
I^{PB}_3(1,2,3,4,5,q,k)= 
A_3^{MHV}(1,l_2,-l_1)A_3^{\overline{MHV}}(-l_2,2,l_3)A_4^{MHV}(-l_3,3,-l_5,l_8)\\A_3^{MHV}(l_5,4,-l_6)
A_3^{\overline{MHV}}(l_6,5,-l_7)A_3^{\overline{MHV}}(l_7,-l_8,l_1) .
\end{split}
\end{gather}
Now we can use the techniques from the super amplitudes to merge the generating functions, plug in the solutions of the cut equations and obtain the constant mentioned in the beginning of the section. Doing so we find the following expression
\begin{gather}
d^{PB}_{1235678}(1,2,3,4,5)=-\frac{1}{2}\left(\gamma_{34}+\gamma_{35}+\gamma_{12}-\gamma_{45} \right).
\end{gather}

\subsection{The Duality Equations}
\begin{figure}[h]
	\centering
		\includegraphics[width=0.80\textwidth]{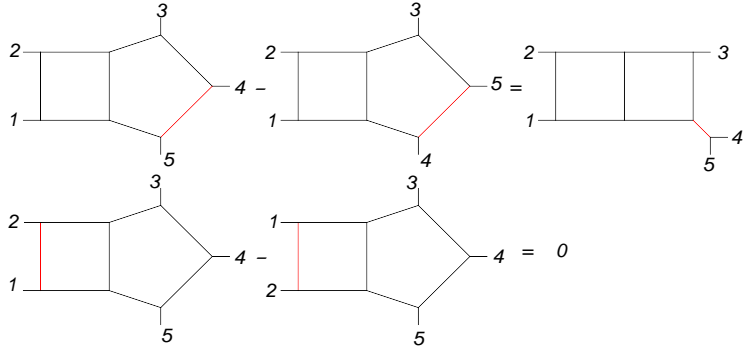}
	\caption{BCJ equations for the pentabox}
	\label{fig:bcj2loop}
\end{figure}

As in the one-loop case we arrived at a point where we have freedom to distribute the seven-pole cuts into different graphs. Once more we are able to use this freedom to apply the BCJ equations.\\
Even though we have mentioned the doublebox numerator we haven't looked at its expansion in residues yet, which is only a constant
\begin{gather}
N^{DB}(1,2,3,45,q,k)=\frac{\Delta^{DB}_{1234578}(1,2,3,45)}{D_1D_2D_3D_4D_5D_7D_8s_{45}} .
\end{gather}
With the help of the following duality equations we can disentangle the seven-pole cuts
\begin{gather}
N^{PB}(1,2,3,4,5,q,k)-N^{PB}(1,2,3,5,4,q,k)=N^{DB}(1,2,3,45,q,k)\\
N^{PB}(1,2,3,4,5,q,k)-N^{PB}(1,2,4,3,5,q,k)=N^{DB}(1,2,34,5,q,k)\\
N^{PB}(1,2,3,4,5,q,k)-N^{PB}(2,1,3,4,5,q,k)=0 .
\end{gather}
These equations are also displayed in figure \ref{fig:bcj2loop}

Plugging in the expansion in residues and demanding that the doubleboxes are a constants we arrive at the following 12 equations
\begin{gather}
\begin{split}
\Delta^{PB}_{1234578}(1,2,3,45)+\Delta^{PB}_{1234578}(1,2,3,54)=0\\
\Delta^{PB}_{1235678}(1,2,3,4,5)-\Delta^{PB}_{12356'78}(1,2,3,5,4)=0\\
\Delta^{PB}_{1234678}(1,2,34,5)-\Delta^{PB}_{12346'78}(1,2,35,4)-\Delta^{PB}_{1234578}(1,2,3,54)=0\\
\Delta^{PB}_{1234568}(1,2,3,4,5)-\Delta^{PB}_{123456'8}(1,2,3,5,4)-\Delta^{PB}_{1234578}(1,2,3,54)=0\\
\Delta^{PB}_{1234678}(1,2,34,5)-\Delta^{PB}_{1234678}(1,2,43,5)=0\\
\Delta^{PB}_{1234568}(1,2,3,4,5)+\Delta^{PB}_{12345'68}(1,2,4,3,5)=0\\
\Delta^{PB}_{1235678}(1,2,3,4,5)-\Delta^{PB}_{1235'678}(1,2,4,3,5)-\Delta^{PB}_{1234678}(1,2,43,5)=0\\
\Delta^{PB}_{1234578}(1,2,3,45)-\Delta^{PB}_{12345'78}(1,2,4,35)-\Delta^{PB}_{1234678}(1,2,43,5)=0\\
\Delta^{PB}_{1234568}(1,2,3,4,5)-\Delta^{PB}_{12'34568}(2,1,3,4,5)=\gamma_{12}\\
\Delta^{PB}_{1235678}(1,2,3,4,5)-\Delta^{PB}_{12'35678}(2,1,3,4,5)=-\gamma_{12}\\
\Delta^{PB}_{1234678}(1,2,34,5)-\Delta^{PB}_{12'34678}(2,1,34,5)=0\\
\Delta^{PB}_{1234578}(1,2,3,45)-\Delta^{PB}_{12'34578}(2,1,3,45)=0 .
\end{split}
\label{2cons}
\end{gather}
Now we can form ans\"atze for the all the sevenfold-cuts in terms of the six independent gammas
\begin{gather}
\Delta^{PB}_{1234568}(1,2,3,4,5)=a_1\gamma_{12}+a_2\gamma_{23}+a_3\gamma_{34}+a_4\gamma_{45}+a_5\gamma_{51}+a_6\gamma_{13} \\
\Delta^{PB}_{1235678}(1,2,3,4,5)=b_1\gamma_{12}+b_2\gamma_{23}+b_3\gamma_{34}+b_4\gamma_{45}+b_5\gamma_{51}+b_6\gamma_{13} \\
\Delta^{PB}_{1234678}(1,2,34,5)=c_1\gamma_{12}+c_2\gamma_{23}+c_3\gamma_{34}+c_4\gamma_{45}+c_5\gamma_{51}+c_6\gamma_{13} \\
\Delta^{PB}_{1234578}(1,2,3,45)=d_1\gamma_{12}+d_2\gamma_{23}+d_3\gamma_{34}+d_4\gamma_{45}+d_5\gamma_{51}+d_6\gamma_{13} .
\end{gather}
Plugging the ans\"atze into our $12$ constraints (\ref{2cons}) and comparing the coefficients we arrive at $6*12=72$ equations. But since 48 equations are not linearly independent we can find one solution to the system of constraints
\begin{gather}
\Delta^{PB}_{1234568}(1,2,3,4,5)=\frac{1}{4}\left(\gamma_{15}+\gamma_{25}+2\gamma_{12} \right)\\
\Delta^{PB}_{1234578}(1,2,3,45)=\frac{1}{4}\left(-\gamma_{34}+\gamma_{35}+2\gamma_{45} \right)\\
\Delta^{PB}_{1234678}(1,2,34,5)=\frac{1}{4}\left(-\gamma_{45}+\gamma_{35}+2\gamma_{34} \right)\\
\Delta^{PB}_{1235678}(1,2,3,4,5)=\frac{1}{4}\left(\gamma_{31}+\gamma_{32}-2\gamma_{12} \right).
\end{gather}

\section{The Full Two-Loop Amplitude}

Putting all our pentabox results together we find the following numerator
\begin{gather}
\begin{split}
N^{PB}(1,2,3,4,5,q,k)=\Delta^{PB}_{12345678}(1,2,3,4,5,q,k) + \Delta^{PB}_{1235678}(1,2,3,4,5)D_4 \\
+\Delta^{PB}_{1234678}(1,2,34,5)D_5+\Delta^{PB}_{1234578}(1,2,3,45)D_6+\Delta^{PB}_{1234568}(1,2,3,4,5)D_7 
\end{split} \\
\begin{split}
= \beta_{12345}+2\gamma_{12} q\cdot p_1+ \frac{1}{4}\left(\gamma_{31}+\gamma_{32}-2\gamma_{12} \right) D_4\\
+\frac{1}{4}\left(-\gamma_{45}+\gamma_{35}+2\gamma_{34} \right) D_5+\frac{1}{4}\left(-\gamma_{34}+\gamma_{35}+2\gamma_{45} \right)D_6\\
+\frac{1}{4}\left(\gamma_{15}+\gamma_{25}+2\gamma_{12} \right)D_7  
\end{split}
\label{pentabox}
\end{gather}
and in order to fit the unitarity cuts we need the following expression for the doublebox
\begin{gather}
N^{DB}(1,2,3,4,5,q,k)=\frac{s_{12}}{s_{45}} \gamma_{45}-\frac{1}{4}\left(2\gamma_{45}+\gamma_{35}+\gamma_{34}\right).
\end{gather}
With the complete result of the pentabox we can now obtain all other numerators through the BCJ equations
\begin{gather}
N^{PB}(1,2,3,4,5,q,k)=N^{CPB}(1,2,3,4,5,q,k) \\
\begin{split}
N^{DP}(1,2,3,4,5,q,k)=N^{PB}(1,2,5,3,4,-q-p_3,k)\\
-N^{PB}(3,4,1,2,5,-k,q)
\end{split} \\
N^{DB}(1,2,3,4,5,q,k)=N^{LCDB}(1,2,3,4,5,q,k) \\
N^{DB}(1,2,3,4,5,q,k)=N^{RCDB}(1,2,3,4,5,q,k) .
\end{gather}
The only non trivial equation is the second one where we obtain the numerator of the double penta $N^{DP}(1,2,3,4,5,q,k)$. Plugging in the expressions for the numerators and then reconstructing the propagators of the double penta out of the scalar products we can find the following numerator
\begin{gather}
\begin{split}
N^{DP}(1,2,3,4,5,q,k)= \frac{1}{4} \left[ \gamma_{34}(2s_{13}-2s_{23}+2s_{35}+s_{12}+s_{34}) \right. \\
\gamma_{12}(2s_{13}-2s_{23}+2s_{15}+s_{12}+s_{34}) \\
\left. (\gamma_{53}+\gamma_{54})s_{35}-(\gamma_{52}+\gamma_{51})(s_{23}-s_{14}-s_{15}) \right] \\
+ 2 \gamma_{34} k \cdot p_3 + \left( \gamma_{34}+\gamma_{35}+\gamma_{45}-\gamma_{12} \right) q \cdot p_2 \\
+ \left(\gamma_{34}+\gamma_{35}+\gamma_{45}+\gamma_{12} \right) q \cdot p_1 +\frac{1}{4} \left(2\gamma_{34}+\gamma_{12}+\gamma_{15} \right)D_1 \\
\frac{1}{4} \left(-2\gamma_{12}+\gamma_{25}-\gamma_{15} \right) D_2 + \frac{1}{4} \left(-2\gamma_{34}+\gamma_{12}+\gamma_{52} \right)D_3 \\
+ \frac{1}{4} \left(-3 \gamma_{34}-3\gamma_{35}-2\gamma_{45} \right)D_4
 +  \frac{1}{4} \left(2 \gamma_{34} + \gamma_{35} -\gamma_{45} \right) D_5\\
  + \frac{1}{4} \left( \gamma_{34}+2\gamma_{35} + 3 \gamma_{45} \right) D_6 +\frac{1}{4} \left(2\gamma_{34}+\gamma_{35}+\gamma_{45} \right)D_7 \\
  +  \frac{1}{4} \left(-2\gamma_{34}-\gamma_{35}-\gamma_{45} \right)D_8 .
\end{split}
\end{gather}
Since we obtained all expressions for the numerators, there are only two steps left: dressing up each diagram with the corresponding color factor, and then summing over all permutations of external legs. The full two-loop amplitude reads
\begin{gather}
\begin{split}
\mathcal{A}^{2-loop}=-g^7 \sum_{\text{all perm}} \left(\frac{1}{2}C^{PB}\text{Int}^{PB}+\frac{1}{4}C^{CPB}\text{Int}^{CPB}+\frac{1}{4}C^{DP}\text{Int}^{DP} \right. \\ 
\left. +\frac{1}{2}C^{DB}\text{Int}^{DB} +\frac{1}{4}C^{LCDB}\text{Int}^{LCDB}+\frac{1}{4}C^{RCDB}\text{Int}^{RCDB} \right)
\end{split}
\end{gather}
where the sum runs over all permutation of external legs, the factors in front of the integrals account for over counting in the sum and the integrals are given by
\begin{gather}
\text{Int}^{PB}= \int \frac{d^D k}{(2\pi)^D} \frac{d^D q}{(2\pi)^D} \frac{N^{PB}}{D_1D_2D_3D_4D_5D_6D_7D_8} \\
\text{Int}^{CPB} = \int \frac{d^D k}{(2\pi)^D} \frac{d^D q}{(2\pi)^D} \frac{N^{CPB}}{D_1D_2D_3D_4D_5D_6D_7D_8} \\
\text{Int}^{DP}= \int \frac{d^D k}{(2\pi)^D} \frac{d^D q}{(2\pi)^D} \frac{N^{DP}}{D_1D_2D_3D_4D_5D_6D_7D_8} \\
\text{Int}^{DB} = \int \frac{d^D k}{(2\pi)^D} \frac{d^D q}{(2\pi)^D} \frac{N^{DB}}{s_{45}D_1D_2D_3D_4D_5D_7D_8} \\
\text{Int}^{LCDB} =  \int \frac{d^D k}{(2\pi)^D} \frac{d^D q}{(2\pi)^D} \frac{N^{LCDB}}{s_{45}D_1D_2D_3D_4D_5D_7D_8} \\
\text{Int}^{RCDB} = \int \frac{d^D k}{(2\pi)^D} \frac{d^D q}{(2\pi)^D} \frac{N^{RCDB}}{s_{45}D_1D_2D_3D_4D_5D_7D_8}.
\end{gather}
The propagators appearing in the integrals can be read off the graphs in figure \ref{Figurefulltwoloop} and the color factors in the full two-loop amplitude are given by
\begin{gather}
C^{PB}= f^{a1b}f^{b2c}f^{cdh}f^{d3e}f^{e4f}f^{f5g}f^{gah} \\
C^{CPB} = f^{a1b}f^{h2c}f^{cdb}f^{d3e}f^{e4f}f^{f5g}f^{gah} \\
C^{DP}= f^{a1b}f^{b2c}f^{cdh}f^{d3e}f^{e4f}f^{fag}f^{g5h} \\ 
C^{DB} = f^{a1b}f^{b2c}f^{cdh}f^{d3e}f^{efg}f^{f45}f^{gah} \\
C^{LCDB} =f^{a1b}f^{b2c}f^{cdh}f^{d3e}f^{hfg}f^{f45}f^{gah} \\
C^{RCDB} = f^{a1b}f^{h2c}f^{cdb}f^{d3e}f^{efg}f^{f45}f^{gah} .
\end{gather}
In the next section we will present a way of obtaining the one- and two-loop numerators without the use of the BCJ equations.

\chapter{Alternative Approach to Color-Kinematic conform Graphs}
In this section we will analyze which assumptions in the derivation of the BCJ conform numerators lead to which consequences for the graphs. In order to do this we will start from the unitarity cuts of the amplitude with a minimal set of graphs. From there we will start imposing the assumptions one by one.

\section{The One-Loop Five-Point Amplitude}
To arrive at the BCJ pentagon and box numerator for the one-loop amplitude we assumed that we have a linear dependence in the loop momentum, that no triangles appear in the graphs and we only used graphs with cubic vertices. \\
Our starting point is the pentagon numerator and all its unitarity cuts. As a remainder we have the following numerator decomposition for the pentagon
\begin{gather}
\begin{split}
N^P(1,2,3,4,5,q)=a_{12345}(1,2,3,4,5)+ a_{1234}(51,2,3,4)D_5+ a_{1235}(45,1,2,3)D_4 \\
+ a_{1245}(34,5,1,2)D_3
 + a_{1345}(23,4,5,1)D_2 + a_{2345}(12,3,4,5)D_1
\end{split}
\label{intpenta}
\end{gather}
and the five quadruple-cuts which correspond fix the quadruple-cut residues in the integrand-decomposition
\begin{gather}
a_{12345}(1,2,3,4,5) = \beta_{12345}  \\
c_{2345}(12,3,4,5)=\frac{\gamma_{12}} {s_{12}} = a_{2345}(12,3,4,5) \\
c_{1345}(23,4,5,1)=\frac{\gamma_{23}} {s_{23}} =a_{1345}(23,4,5,1)\\
c_{1245}(34,5,1,2)=\frac{\gamma_{34}} {s_{34}} = a_{1245}(34,5,1,2)\\
c_{1235}(45,1,2,3)=\frac{\gamma_{45}} {s_{45}}= a_{1235}(45,1,2,3)\\
c_{1234}(51,2,3,4)=\frac{\gamma_{51}} {s_{51}} = a_{1234}(51,2,3,4)
\end{gather}
where all the factors just been obtained by the unitarity cuts and we have no leftover freedom. Now we will start to impose the assumptions from the derivation of the BCJ conform answer and look at their consequences.

\subsection{Linear Dependence on the Loop-Momenta}
The first assumption is that the pentagon numerator is linear in the loop momentum, which comes from the fact that we are working in $\mathcal{N}$=4 sYM. The pentagon has five vertices which can give us at most five powers of loop-momentum. But now we expect the four-fold supersymmetry to translate four of these powers into the over all supermomentum delta function $\delta^8(Q)$.\\
The integrand-decomposition of the pentagon
\begin{gather}
\begin{split}
N^P(1,2,3,4,5,q)=a^P_{12345}(1,2,3,4,5)+ a^P_{1234}(51,2,3,4) D_5 + a^P_{1235}(45,1,2,3)D_4 \\
+ a^P_{1245}(34,5,1,2)D_3
 + a^P_{1345}(23,4,5,1)D_2 + a^P_{2345}(12,3,4,5)D_1
\end{split}
\end{gather}
has in general a quadratic term. This can be seen if we expand all denominators in their scalar products. The coefficient of this quadratic term is then 
\begin{gather}
\begin{split}
a^P_{1234}(51,2,3,4) + a^P_{1235}(45,1,2,3)+ a^P_{1245}(34,5,1,2)
 + a^P_{1345}(23,4,5,1)\\
  + a^P_{2345}(12,3,4,5)
 \end{split}
\end{gather} 
since it has contributions from all quadruple-cut residues. To get rid of the quadratic dependence we need to demand that its coefficient vanishes
\begin{gather}
\begin{split}
a^P_{1234}(51,2,3,4)+a^P_{1235}(45,1,2,3,)+a^P_{1245}(34,5,1,2)+a^P_{1345}(23,4,5,1)\\
+a^P_{2345}(12,3,4,5)=0 
\end{split}
\label{linpentagon}
\end{gather}
which is not satisfied by the expression (\ref{intpenta}). In order to get a linear numerator we need to introduce a set of new graphs namely the box numerators with a massive leg. Since the quadruple-cuts are a combination of the introduced box and the pentagon we are able to absorb the unitarity cuts into the boxes and guarantee a linear numerator. This introduces the following integrand-decomposition for the box numerators
\begin{gather}
N^B(ij,k,l,m,q)=\gamma_{ij}-s_{ij} a^P_{jklm}(ij,k,l,m)
\end{gather}
where we used another assumption namely that we do not expect to have triangles present in $\mathcal{N}$=4 sYM and indeed a triangle cut leads to a vanishing unitarity cut.\\
Together with the pentagon the choice of the box numerator gives us the right unitarity cuts
\begin{gather}
\frac{N^B(ij,k,l,m,q)}{s_{ij}}+a^P_{jklm}(ij,k,l,m)= \frac{\gamma_{ij}}{s_{ij}} .
\end{gather}
Now we have some leftover freedom in the form of the $a$s. Since they are constants we can parametrize them in terms of the six independent gammas
\begin{gather}
a^P_{jklm}(ij,k,l,m)=\alpha_1\gamma_{ij}+\alpha_2\gamma_{jk}+\alpha_3\gamma_{kl}+\alpha_4\gamma_{lm}+\alpha_5\gamma_{mi}+\alpha_6\gamma_{ik}.
\end{gather}
At this point we went through all assumptions of the BCJ equations. We arrived at an ansatz with some leftover freedom which we could parameterize in terms of the gammas. This freedom can now be used to apply the BCJ equations. Instead of doing that we will go a different way and use graph automorphism to fix this freedom.

\subsection{Graph Automorphism}
A graph automorphism is a symmetry of an graph where we relabel the external particles in a way that all internal propagators stay the same. Secondly, the color factors of the two graphs must also be the same up to a minus sign, which provides us with an equation between two different relabellings of the same graph. \\ 
For the box numerator we find such an automorphism if we exchange the two external particles on the massive leg
\begin{gather}
N^B(12,3,4,5,q)=-N^B(21,3,4,5,q).
\end{gather}
Plugging in the integrand-decomposition this translates into the following constraints on our residues
\begin{gather}
a^P_{jklm}(ij,k,l,m)=-a^P_{jklm}(ji,k,l,m) .
\end{gather}
Solving this equation and the linearity condition (\ref{linpentagon}) with our ansatz we find that the only solution is 
\begin{gather}
a^P_{jklm}(ij,k,l,m)=0\\
\end{gather}
and therefore we arrived at the same result as the BCJ equations gave us. Looking back at our list of assumptions we saw that introducing no graphs with a four-point vertices was not a strong restriction. This can be understood in the way that every four-point vertex has a part with a color structure of an s-,t- and u-channel and by simply multiplying and dividing the corresponding kinematical part by the propagator of the channel we can absorb these contribution to corresponding three-point ones.\\
The assumption that we don't have triangles appearing in our theory and that we have a linear numerator are connected to each other, which comes from the fact that the unitarity cuts of triangles vanish if we subtract the contributions from the box and the pentagon. Therefore the only non trivial way of introducing graphs with triangles is to allow higher order terms in the integrand-reduction. But having higher order terms in the numerators conflicts with our expectation that four powers of loop-momentum are converted by the supersymmetry into the overall delta function $\delta^{8}(Q)$. On the other hand introducing triangles would correspond to even more freedom in our numerators. But as we have seen in this section introducing the box numerator already provided enough freedom to apply the BCJ equations. Even more demanding a linear integrand-reduction of the pentagon and using graph automorphisms was enough to arrive at the BCJ numerators and therefore to fulfill all the BCJ equations.

\section{The Two-Loop Five-Point Amplitude}
\begin{figure}[h]
	\centering
		\includegraphics[width=0.80\textwidth]{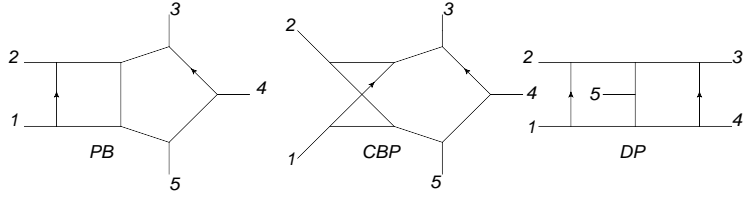}
	\caption{Graphs contributing to the two-loop five-point amplitude}
	\label{fig:twoloopnobcj}
\end{figure}

In the two-loop case we will follow the same strategy as in the one-loop case. We will start by the minimal set of graphs which are the pentabox, the crossed pentabox and the double penta displayed in fig \ref{fig:twoloopnobcj} and then step by step impose the assumptions.
In the section about the BCJ equations at two-loops it was sufficient to discuss only the integrand-decomposition of the pentabox and then obtain the rest of the graphs with the help of the BCJ equations. But here we also need the unitarity cuts of the other diagrams.

\subsection{The Pentabox}
As a short reminder we will first list the sevenfold-cuts of the pentabox
\begin{gather}
 \Delta^{PB}_{1235678}(1,2,3,4,5)+\Delta^{PB}_{1234568}(4,5,1,2,3)=d^{PB}_{1235678}(1,2,3,4,5)\\
=-\frac{1}{2}\left(\gamma_{34}+\gamma_{35}+\gamma_{12}-\gamma_{45} \right)\\[0.25cm]
\Delta^{PB}_{1234678}(1,2,34,5)=d_{1234678}(1,2,34,5)\\
=\frac{s_{12}}{s_{34}}\gamma_{34} \\[0.25cm]
\Delta^{PB}_{1234578}(1,2,3,45)=d_{1234578}(1,2,3,45)\\
=\frac{s_{12}}{s_{45}}\gamma_{45} \\[0.25cm]
\Delta^{PB}_{1234568}(1,2,3,4,5)+\Delta^{PB}_{1235678}(1,2,3,4,5)=d_{1234568}(1,2,3,4,5)\\
=-\frac{1}{2}\left(\gamma_{52}+\gamma_{51}-\gamma_{12}+\gamma_{34} \right) .
\end{gather}

\subsection{The Crossed Pentabox}
The crossed pentabox can be obtained in the same way as the pentabox in the two-loop section. Starting from the integrand-decomposition we will apply a maximum cut to determine the eightfold-cut delta and then move on to the sevenfold-cut ones. \\
For the crossed pentabox we find the following expansion in its residues
\begin{gather}
\begin{split}
N^{CPB}(1,2,3,4,5,q,k)=\Delta^{CPB}_{12345678}(1,2,3,4,5,q,k)+\Delta_{1235678}^{CPB}(1,2,3,4,5)D_4\\
+\Delta^{CPB}_{1234678}(1,2,34,5)D_5+ \Delta^{CPB}_{1234578}(1,2,3,45)D_6 + \Delta^{CPB}_{1234568}(1,2,3,4,5)D_7
\end{split}
\end{gather}
with the denominators
\begin{gather}
D_1=(k+p_1)^2 \\
D_2=k^2 \\
D_3=(q+k-p_3)^2 \\
D_4=(q-p_3)^2\\
D_5=q^2 \\
D_6=(q+p_4)^2 \\
D_7=(q+p_4+p_5)^2 \\
D_8=(q+k-p_2-p_3)^2 .
\end{gather}
Solving the cut equations and plugging them into the products of tree generating functions we can find the following solution for the eightfold-cut delta 
\begin{gather}
\Delta^{CPB}_{12345678}(1,2,3,4,5,q,k)=\beta_{12345}+ 2 \gamma_{12} q \cdot p_1
\end{gather}
which is the same as the eight-point delta of the pentabox. Now we can move on to the sevenfold-cut which again can have contributions from two graphs
\begin{gather}
\Delta^{CPB}_{1235678}(1,2,3,4,5)+\Delta^{DP}_{1235678}(1,2,3,4,5)=d^{CPB}_{1235678}(1,2,3,4,5)\\
=-\frac{1}{2}\left(\gamma_{34}+\gamma_{35}+\gamma_{12}-\gamma_{45} \right) \\[0.25cm]
\Delta^{CPB}_{1234678}(1,2,34,5)=d^{CPB}_{1234678}(1,2,34,5)\\
=\frac{s_{12}}{s_{34}}\gamma_{34} \\[0.25cm]
\Delta^{CPB}_{1234578}(1,2,3,45)=d^{CPB}_{1234578}(1,2,3,45) \\[0.25cm]
\Delta^{CPB}_{1234568}(1,2,3,4,5)+\Delta^{DP}_1(5,1,3,4,2)=d^{CPB}_{1234568}(1,2,3,4,5)\\
=-\frac{1}{2}\left(\gamma_{52}+\gamma_{51}-\gamma_{12}+\gamma_{34} \right) .
\end{gather}
One should note again that all unitarity cuts are the same as in the pentabox case. The only point where the two diagrams are differing are the contributing diagrams in the sevenfold-cuts. Nevertheless it becomes feasible that the two graphs could be equal as demanded by the BCJ equations.

\subsection{The Double Penta}
The numerator of the double penta has the following expansion in its residues
\begin{gather}
\begin{split}
N^{DP}(1,2,3,4,5,q,k)=\Delta^{DP}_{12345678}(1,2,3,4,5,q,k)+\Delta^{DP}_{2345678}(12,3,4,5)D_1\\
+\Delta^{DP}_{1345678}(1,2,3,4,5)D_2+\Delta^{DP}_{1245678}(1,2,3,4,5)D_3 +
\Delta^{DP}_{1235678}(1,2,3,4,5)D_4 \\
+\Delta^{DP}_{1234678}(1,2,34,5)D_5+\Delta^{DP}_{1234578}(1,2,3,4,5)D_6+\Delta^{DP}_{1234568}(1,2,3,4,5)D_7 \\
+\Delta^{DP}_{1234567}(1,2,3,4,5)D_8
\end{split}
\end{gather}
with these propagators
\begin{gather}
D_1=(k+p_1)^2 \\
D_2=k^2 \\
D_3=(k-p_2)^2 \\
D_4=(q-p_3)^2\\
D_5=q^2 \\
D_6=(q+p_4)^2 \\
D_7=(q+k+p_4+p_1)^2\\
D_8=(q+k-p_2-p_3)^2 .
\end{gather}
From solving the system of equations of the maximum cut we find the following eightfold-cut residue
\begin{gather}
\begin{split}
\Delta^{DP}_{12345678}(1,2,3,4,5,q,k)=c^{DP}_{0;12345678}(1,2,3,4,5)+c^{DP}_{1;12345678}(1,2,3,4,5) k \cdot p_3\\
 +c^{DP}_{2;12345678}(1,2,3,4,5) q \cdot p_2 + c^{DP}_{3;12345678}(1,2,3,4,5) q \cdot p_1
\end{split}\\
\begin{split}
c^{DP}_{0;12345678}(1,2,3,4,5)=\frac{1}{4} \left(\gamma_{34}(2s_{13}-2s_{23}+2s_{35}+s_{12}+s_{34}) \right. \\
\gamma_{12}(2s_{13}-2s_{23}+2s_{15}+s_{12}+s_{34}) \\
\left. (\gamma_{53}+\gamma_{54})s_{35}-(\gamma_{52}+\gamma_{51})(s_{23}-s_{14}-s_{15}) \right) 
\end{split}\\
c^{DP}_{1;12345678}(1,2,3,4,5)=2 \gamma_{34} \\ 
c^{DP}_{2;12345678}(1,2,3,4,5)= \gamma_{34}+\gamma_{35}+\gamma_{45}-\gamma_{12} \\
c^{DP}_{3;12345678}(1,2,3,4,5)=\gamma_{34}+\gamma_{35}+\gamma_{45}+\gamma_{12} .
\end{gather}
Proceeding to the sevenfold-cuts we find the following two graphs contributing to the following unitarity cuts
\begin{gather}
\Delta^{DP}_{2345678}(1,2,3,4,5)+\Delta^{CPB}_{1234568}(5,2,3,4,1)=d^{DP}_{2345678}(1,2,3,4,5)\\
=\frac{1}{2}\left(\gamma_{34}+\gamma_{25}+\gamma_{12}+\gamma_{15} \right)\\[0.25cm]
\Delta^{DP}_{1345678}(12,3,4,5)=d^{DP}_{1345678}(12,3,4,5)\\
=\frac{s_{34}}{s_{12}}\gamma_{12} \\[0.25cm]
\Delta^{DP}_{1245678}(1,2,3,4,5)+\Delta^{CPB}_{1235678}(5,1,2,3,4)=d^{DP}_{1245678}(1,2,3,4,5)\\
=\frac{1}{2}\left(\gamma_{15}+\gamma_{23}+\gamma_{24}-\gamma_{34} \right) \\[0.25cm]
\Delta^{DP}_{1235678}(1,2,3,4,5)+\Delta^{CPB}_{1234568}(4,5,1,2,3)=d^{DP}_{1235678}(1,2,3,4,5)\\
=-\gamma_{34}-\gamma_{35} \\[0.25cm]
\Delta^{DP}_{1234678}(1,2,34,5)=d^{DP}_{1234678}(1,2,34,5) \\
\frac{s_{12}}{s_{34}}\gamma_{34} \\[0.25cm]
\Delta^{DP}_{1234578}(1,2,3,4,5) +\Delta^{CPB}_{1235678}(3,5,4,1,2)=d^{DP}_{1234578}(1,2,3,4,5) \\
= \gamma_{45} \\[0.25cm]
\Delta^{DP}_{1234568}(1,2,3,4,5)+\Delta^{PB}_{1234568}(1,2,3,4,5)=d^{DP}_{1234568}(1,2,3,4,5) \\
= -\frac{1}{2} \left(\gamma_{34}-\gamma_{12}+\gamma_{35}+\gamma_{45} \right) \\[0.25cm]
\Delta^{DP}_{1234567}(1,2,3,4,5)+\Delta^{PB}_{1235678}(1,2,5,3,4)=d^{DP}_{1234567}(1,2,3,4,5)\\
=\frac{1}{2} \left(\gamma_{34}-\gamma_{12}+\gamma_{35}+\gamma_{45} \right).
\end{gather}
We are now in a similar position as in the one-loop case. From these unitarity cuts our amplitude is well defined. One only has to be careful when summing over all diagrams with different ordering to not over count certain terms. First we will try to separate these graphs from each other and see that there is leftover freedom. After demanding that the numerators of these graphs are linear in the loop momentum we will increase this freedom even further. But after imposing some graph automorphism this freedom will be completely fixed.

\subsection{$\gamma$-Decomposition}
As we have seen above several unitarity cuts have two different graphs which contribute. In order to solve these functional equations we need to write down ans\"atze for all deltas in terms of our six gammas
\begin{gather}
\Delta^{PB}_{1234568}(1,2,3,4,5)=a_1\gamma_{12}+a_2\gamma_{23}+a_3\gamma_{34}+a_4\gamma_{45}+a_5\gamma_{51}+a_6\gamma_{13} \\
\Delta^{PB}_{1235678}(1,2,3,4,5)=b_1\gamma_{12}+b_2\gamma_{23}+b_3\gamma_{34}+b_4\gamma_{45}+b_5\gamma_{51}+b_6\gamma_{13} \\
\Delta^{CPB}_{1235678}(1,2,3,4,5)=f_1\gamma_{12}+f_2\gamma_{23}+f_3\gamma_{34}+f_4\gamma_{45}+f_5\gamma_{51}+f_6\gamma_{13} \\
\Delta^{CPB}_{1234568}(1,2,3,4,5)=i_1\gamma_{12}+i_2\gamma_{23}+i_3\gamma_{34}+i_4\gamma_{45}+i_5\gamma_{51}+i_6\gamma_{13}\\
\Delta^{DP}_{2345678}(1,2,3,4,5)=j_1\gamma_{12}+j_2\gamma_{23}+j_3\gamma_{34}+j_4\gamma_{45}+j_5\gamma_{51}+j_6\gamma_{13}\\
\Delta^{DP}_{1345678}(1,2,3,4,5)=l_1\gamma_{12}+l_2\gamma_{23}+l_3\gamma_{34}+l_4\gamma_{45}+l_5\gamma_{51}+l_6\gamma_{13}\\
\Delta^{DP}_{1235678}(1,2,3,4,5)=m_1\gamma_{12}+m_2\gamma_{23}+m_3\gamma_{34}+m_4\gamma_{45}+m_5\gamma_{51}+m_6\gamma_{13}\\
\Delta^{DP}_{1234578}(1,2,3,4,5)=o_1\gamma_{12}+o_2\gamma_{23}+o_3\gamma_{34}+o_4\gamma_{45}+o_5\gamma_{51}+o_6\gamma_{13} \\
\Delta^{DP}_{1234568}(1,2,3,4,5)=r_1\gamma_{12}+r_2\gamma_{23}+r_3\gamma_{34}+r_4\gamma_{45}+r_5\gamma_{51}+r_6\gamma_{13} \\
\Delta^{DP}_{1234567}(1,2,3,4,5)=w_1\gamma_{12}+w_2\gamma_{23}+w_3\gamma_{34}+w_4\gamma_{45}+w_5\gamma_{51}+w_6\gamma_{13} .
\end{gather}
This means we introduced $6*10=60$ free parameters. The first obvious constraint is to reproduce the unitarity cuts
\begin{gather}
\Delta^{PB}_{1234568}(1,2,3,4,5)+\Delta^{PB}_{1235678}(3,4,5,1,2)=-\frac{1}{2}\left(\gamma_{52}+\gamma_{51}-\gamma_{12}+\gamma_{34} \right) \\
\Delta^{PB}_{1235678}(1,2,3,4,5)+\Delta^{PB}_{1234568}(4,5,1,2,3)=-\frac{1}{2}\left(\gamma_{34}+\gamma_{35}+\gamma_{12}-\gamma_{45} \right)\\
\Delta^{CPB}_{1235678}(1,2,3,4,5)-\Delta^{DP}_{1245678}(2,3,4,5,1)=\frac{1}{2}\left(\gamma_{31}+\gamma_{32}-\gamma_{12}+\gamma_{45} \right)\\
\Delta^{CPB}_{1234568}(1,2,3,4,5)-\Delta^{DP}_{2345678}(5,1,3,4,2)=\frac{1}{2}\left(\gamma_{15}+\gamma_{25}+\gamma_{12}-\gamma_{34} \right)\\
\Delta^{DP}_{2345678}(1,2,3,4,5)+\Delta^{CPB}_{1234568}(5,2,3,4,1)=\frac{1}{2}\left(\gamma_{34}+\gamma_{25}+\gamma_{12}+\gamma_{15} \right)\\
\Delta^{DP}_{1245678}(1,2,3,4,5)+\Delta^{CPB}_{1235678}(1,5,2,3,4)=\frac{1}{2}\left(-\gamma_{34}-\gamma_{25}+\gamma_{12}+\gamma_{15} \right)\\
\Delta^{DP}_{1235678}(1,2,3,4,5)+\Delta^{CPB}_{1235678}(4,5,1,2,3)=-\gamma_{34}-\gamma_{53}\\
\Delta^{DP}_{1234578}(1,2,3,4,5)+\Delta^{CPB}_{1234568}(3,5,4,1,2)=\gamma_{45} \\
\Delta^{DP}_{1234568}(1,2,3,4,5)+\Delta^{PB}_{1234568}(1,2,3,4,5)+=\frac{1}{2}\left(-\gamma_{34}+\gamma_{12}+\gamma_{15}+\gamma_{25}\right) \\
\Delta^{DP}_{1234567}(1,2,3,4,5)+\Delta^{PB}_{1235678}(1,2,5,3,4)=-\frac{1}{2}\left(-\gamma_{34}+\gamma_{12}+\gamma_{15}+\gamma_{25}\right) .
\end{gather}
By plugging in the expansion in gammas and comparing their coefficients we obtain $10*6=60$ equations from the unitarity constraints reducing our 60 parameters down to 11, since not all of the equations are linear independent. To restrict these parameters even further we will demand that the numerators of the graphs are linear in the loop momenta.

\subsection{Linear Dependence on the Loop-Momenta}
As in the one-loop case we can obtain a linear integrand-decomposition by imposing that the coefficient of the quadratic term vanishes in the integrand-reduction. This gives five more equations on the sevenfold-cut residues
\begin{gather}
\begin{split}
\Delta^{PB}_{1235678}(1,2,3,4,5)+\Delta^{PB}_{1234678}(1,2,3,4,5)+\Delta^{PB}_{1234578}(1,2,3,4,5)\\
+\Delta^{PB}_{1234568}(1,2,3,4,5)=0
\end{split}\\
\begin{split}
\Delta^{CPB}_{1235678}(1,2,3,4,5)+\Delta^{CPB}_{1234678}(1,2,3,4,5)+\Delta^{CPB}_{1234578}(1,2,3,4,5)\\
+\Delta^{CPB}_{1234568}(1,2,3,4,5)=0
\end{split}\\
\Delta^{DP}_{2345678}(1,2,3,4,5)+\Delta^{DP}_{1345678}(1,2,3,4,5)+\Delta^{DP}_{1245678}(1,2,3,4,5)=0\\
\Delta^{DP}_{1235678}(1,2,3,4,5)+\Delta^{DP}_{1234678}(1,2,3,4,5)+\Delta^{DP}_{1234578}(1,2,3,4,5)=0\\
\Delta^{DP}_{1234568}(1,2,3,4,5)+\Delta^{DP}_{1234567}(1,2,3,4,5)=0 .
\end{gather}
It turns out that with our current ansatz these equations do not have a solution, so we need to introduce new graphs which absorb the $s_{kl}\frac{\gamma_{ij}}{s_{ij}}$ terms. These graphs are analog to the one-loop case the doubleboxes with the following expansion in their residues
\begin{gather}
N^{DB}(1,2,34,5,q,k)=\gamma_{34}s_{12}-s_{34}\Delta^{PB}_{1234678}(1,2,3,4,5)\\
N^{DB}(1,2,3,45,q,k)=\gamma_{45}s_{12}-s_{45}\Delta^{PB}_{1234578}(1,2,3,4,5) \\
N^{LCDB}(1,2,34,5,q,k)=\gamma_{34}s_{12}-s_{34}\Delta^{CDP}_{1234678}(1,2,3,4,5)\\
N^{LCDB}(1,2,3,45,q,k)=\gamma_{45}s_{12}-s_{45}\Delta^{CDP}_{1234578}(1,2,3,4,5)\\
N^{RCDB}(3,4,5,12,q,k)=\gamma_{12}s_{34}-s_{12}\Delta^{DP}_{1345678}(1,2,3,4,5) \\
N^{RCDB}(1,2,34,5,q,k)=\gamma_{34}s_{12}-s_{34}\Delta^{DP}_{1234678}(1,2,3,4,5)
\end{gather}
where we took into account the cancellation that can occur between this numerator and the sevenfold-cut residues at the unitarity cuts. But this means we have to introduce $6*6=36$ new parameters for the six new deltas
\begin{gather}
\Delta^{PB}_{1234678}(1,2,3,4,5)=c_1\gamma_{12}+c_2\gamma_{23}+c_3\gamma_{34}+c_4\gamma_{45}+c_5\gamma_{51}+c_6\gamma_{13} \\
\Delta^{PB}_{1234578}(1,2,3,4,5)=d_1\gamma_{12}+d_2\gamma_{23}+d_3\gamma_{34}+d_4\gamma_{45}+d_5\gamma_{51}+d_6\gamma_{13} \\
\Delta^{CPB}_{1234678}(1,2,3,4,5)=g_1\gamma_{12}+g_2\gamma_{23}+g_3\gamma_{34}+g_4\gamma_{45}+g_5\gamma_{51}+g_6\gamma_{13} \\
\Delta^{CPB}_{1234578}(1,2,3,4,5)=h_1\gamma_{12}+h_2\gamma_{23}+h_3\gamma_{34}+h_4\gamma_{45}+h_5\gamma_{51}+h_6\gamma_{13} \\
\Delta^{DP}_{12345678}(1,2,3,4,5)=k_1\gamma_{12}+k_2\gamma_{23}+k_3\gamma_{34}+k_4\gamma_{45}+k_5\gamma_{51}+k_6\gamma_{13} \\
\Delta^{DP}_{1234678}(1,2,3,4,5)=n_1\gamma_{12}+n_2\gamma_{23}+n_3\gamma_{34}+n_4\gamma_{45}+n_5\gamma_{51}+n_6\gamma_{13} . 
\end{gather}
Even though the linearity condition gives us $6*6=36$ new equations the number of free parameters goes up from 11 to 19. 

\subsection{Graph Automorphism}
Completely analog to the one-loop case we will use graph automorphism to fix the rest of our free parameters. First the doubleboxes must be antisymmetric under the exchange of the particles in the massive leg
\begin{gather}
\Delta^{PB}_{1234678}(1,2,34,5)=-\Delta^{PB}_{1234678}(1,2,43,5)\\
\Delta^{PB}_{1234578}(1,2,3,45)=-\Delta^{PB}_{1234578}(1,2,3,54)\\
\Delta^{CPB}_{1234678}(1,2,34,5)=-\Delta^{CPB}_{1234678}(1,2,43,5)\\
\Delta^{CPB}_{1234578}(1,2,3,45)=-\Delta^{CPB}_{1234578}(1,2,3,54)\\
\Delta^{DP}_{1345678}(12,3,4,5)=-\Delta^{DP}_{1345678}(21,3,4,5) \\
\Delta^{DP}_{1234678}(1,2,34,5)=-\Delta^{DP}_{1234678}(1,2,43,5) 
\end{gather}
and three other automorphisms for the doubleboxes and the crossed pentabox
\begin{gather}
N^{DB}(1,2,3,45,q,k)=N^{DB}(2,1,45,3,-q,-k) \\
N^{LCDB}(1,2,3,45,q,k)=N^{LCDB}(1,2,45,3,-q,k) \\
N^{CPB}(1,2,3,4,5,q,k)=-N^{CPB}(2,1,5,4,3,-q-p_4,k+q-p_2-p_3,) .
\end{gather}
Plugging in the integrand-decompositions and the ans\"atze we find $9*6=54$ new equations which fix all parameters
\begin{gather}
\Delta^{PB}_{1234568}(1,2,3,4,5)=\frac{1}{4}\left(\gamma_{15}+\gamma_{25}+2\gamma_{12} \right)\\
\Delta^{PB}_{1234578}(1,2,3,4,5)=\frac{1}{4}\left(-\gamma_{34}+\gamma_{35}+2\gamma_{45} \right)\\
\Delta^{PB}_{1234678}(1,2,3,4,5)=\frac{1}{4}\left(-\gamma_{45}+\gamma_{35}+2\gamma_{34} \right)\\
\Delta^{PB}_{1235678}(1,2,3,4,5)=\frac{1}{4}\left(\gamma_{31}+\gamma_{32}-2\gamma_{12} \right) \\
\Delta^{CPB}_{1234568}(1,2,3,4,5)=\frac{1}{4}\left(\gamma_{15}+\gamma_{25}+2\gamma_{12} \right)\\
\Delta^{CPB}_{1234578}(1,2,3,4,5)=\frac{1}{4}\left(-\gamma_{34}+\gamma_{35}+2\gamma_{45} \right)\\
\Delta^{CPB}_{1234678}(1,2,3,4,5)=\frac{1}{4}\left(-\gamma_{45}+\gamma_{35}+2\gamma_{34} \right)\\
\Delta^{CPB}_{1235678}(1,2,3,4,5)=\frac{1}{4}\left(\gamma_{31}+\gamma_{32}-2\gamma_{12} \right) \\
\Delta^{DP}_{2345678}(1,2,3,4,5)=\frac{1}{4} \left(2\gamma_{34}+\gamma_{12}+\gamma_{15} \right) \\
\Delta^{DP}_{1345678}(1,2,3,4,5)=\frac{1}{4} \left(-2\gamma_{12}+\gamma_{25}-\gamma_{15} \right)\\
\Delta^{DP}_{1245678}(1,2,3,4,5)= \frac{1}{4} \left(-2\gamma_{34}+\gamma_{12}+\gamma_{52} \right)\\
\Delta^{DP}_{1235678}(1,2,3,4,5)= \frac{1}{4} \left(-3 \gamma_{34}-3\gamma_{35}-2\gamma_{45} \right)\\
\Delta^{DP}_{1234678}(1,2,3,4,5)= \frac{1}{4} \left(2 \gamma_{34} + \gamma_{35} -\gamma_{45} \right) \\
\Delta^{DP}_{1234578}(1,2,3,4,5)= \frac{1}{4} \left( \gamma_{34}+2\gamma_{35} + 3 \gamma_{45} \right)\\
\Delta^{DP}_{1234568}(1,2,3,4,5)= \frac{1}{4} \left(2\gamma_{34}+\gamma_{35}+ \gamma_{45} \right)\\
\Delta^{DP}_{1234567}(1,2,3,4,5)= \frac{1}{4} \left(-2\gamma_{34}-\gamma_{35}- \gamma_{45} . \right) 
\end{gather}
The solutions are exactly the solutions of the BCJ conform numerators we found earlier. This means that in the one-loop and the two-loop case demanding a linear ansatz was the key step. It increased our minimal set of graphs to the one used by the BCJ conform presentation and distributed the terms from the unitarity cuts in a way that they could be disentangled by graph automorphisms. This was somehow enough to reorder to Feynman diagrams in a way that their corresponding graphs satisfy the BCJ equations. \\
But as we discussed already for the one-loop case this can not be the key ingredient for the BCJ equations, since we know that there also BCJ representation of amplitudes in theories with non-linear numerators e.g. \cite{bcjqcd}. Linearity merely simplifies the BCJ equations to make the identification of the corresponding amplitudes easier.

\chapter{Ultra-Violet Behavior}
As known since the 1980s $\mathcal{N}$=4 sYM is UV finite \cite{uvfinite}, however this is only true for four dimensions. Explicit calculations in higher dimensions \cite{n=4calca,n=4calcb}
established a finiteness bound in $D$ dimensions at $L$ loops as
\begin{gather}
D<4+\frac{6}{L} \ \ \ \ \ L \geq 2
\end{gather}
which conforms to the finiteness in four dimensions. One of the remaining questions is if this bound is saturated or not. The question is tightly bound to the issue if there are any hidden symmetries in $\mathcal{N}$=4 sYM which could further alter the finiteness bound. \\
Furthermore all amplitudes in $\mathcal{N}$=4 sYM are connected through the double copy procedure to amplitudes in $\mathcal{N}$=8 SUGRA. Here the question of the finiteness of the theory is not settled yet. Even though general arguments point out that $\mathcal{N}$=8 SUGRA should not be finite recent calculations \cite{4point6loop} show that the finiteness bound of $\mathcal{N}$=8 SUGRA seems to follow the one of $\mathcal{N}$=4 sYM and therefore point towards a finite theory. \\
The UV behavior of the one- and two-loop five-point $\mathcal{N}$=4 sYM were already studied in \cite{Carrasco:2011mn}. We can determine the critical dimension of any Feynman integral by a simple power counting argument. In the UV limit all external momenta become infinitesimal small compared to the loop momentum so we can neglect them. Doing so we are left with an integral of the following form 
\begin{gather}
\int \frac{d^D q}{(2\pi)^D} \frac{1}{(q^2+m^2)^x}
\end{gather}
where x can be obtained by setting all the external momenta to zero in e.g. equation (\ref{int1loop}).
Furthermore we introduced a regulator $m$ to prevent the integral from vanishing in dimensional regularization. Such an integral will diverge if $D \geq x$.

\section{The One-Loop Amplitude}
If we have a look at our one-loop amplitude we can see that the box numerator diverges in $D=8$ dimensions. \\
In order to extract the UV divergence we will use dimensional regularization, which means instead of calculating the loop integral in the critical $D=8$ dimension we will shift the dimensions by a $2\epsilon$ term and extract a pole in the shifting parameter epsilon. The epsilon pole then corresponds to the UV divergence in the critical dimensions. Here we will show three methods of calculating the UV divergence of the Box integral: Direct Integration with Feynman parametrization, Integration-by-parts Identities \cite{grozin} and Small Momentum Injection. For the first two methods we need to reintroduce a regulator to the integral, which is done by making the internal momenta massive
\begin{gather}
\text{Int}^B(12,3,4,5,q) \rightarrow \int \frac{d^D q}{(2\pi)^D}\frac {1}{(q^2+m^2)^4} .
\label{1loopuv}
\end{gather}
To obtain the right integral we should take the limit of the mass going to zero in the end of the calculation. \\
For the last method we will keep two small momenta and let them flow through the diagram. This small momentum then will be our regulator. As in the other two methods we will take the limit of the regulator going to zero in the end.

\subsection{Direct Integration with Feynman Parametrization}
As we can see from eq. (\ref{1loopuv}) the box integral is already in the Feynman parametrization. Therefore we can directly turn the integral over the Minkowski space into an Euclidean one by a Wick rotation
\begin{gather}
\int \frac{d^D q}{(2\pi)^D}\frac {1}{(q^2+m^2)^4}= i\int \frac{d^D q_E}{(2\pi)^D}\frac {1}{(q_E^2+m^2)^4} .
\end{gather}
The next step is to introduce spherical coordinates which gives us
\begin{gather}
i\int \frac{d^D q_E}{(2\pi)^D}\frac {1}{(q_E^2+m^2)^4}= i\int \frac{d q_E}{(2\pi)^D} \Omega_D \frac {q_E^{D-1}}{(q_E^2+m^2)^4}
\end{gather}
where $\Omega_D$ is the surface of the $D$ dimensional sphere and given by
\begin{gather}
\Omega_D=\frac{2\pi^{D/2}}{\Gamma(D/2)}.
\end{gather}
The fourth step is a parameter transformation of the form $x=q_E^2$
\begin{gather}
i\int \frac{d q_E}{(2\pi)^D} \Omega_D \frac {q_E^{D-1}}{(q_E^2+m^2)^4}=\frac{i \pi^{D/2}}{\Gamma(D/2)} \int \frac{d x}{(2\pi)^D}  \frac {x^{D/2-1}}{(x+m^2)^4} .
\end{gather} 
Now we notice that the remaining integral is actually a combination of Gamma functions
\begin{gather}
\frac{\Gamma(a)\Gamma(b)}{\Gamma(a+b)}=\int dx \frac{x^{a-1}}{(1+x)^{a+b}} 
\end{gather}
which gives us
\begin{gather}
\frac{i \pi^{D/2}}{\Gamma(D/2)} \frac{\Gamma(D/2)\Gamma(4-D/2)}{\Gamma(4)} (m^2)^{D/2-4}\\
= \frac{i \pi^{D/2} \Gamma(4-D/2) } {6(2\pi)^D } (m^2)^{D/2-4}.
\end{gather}
The regulator $m$ indicates that the critical dimension is $D=8$. As discussed before using dimensional regularization with $D=8-2\epsilon$ gives us
\begin{gather}
\frac{i \pi^{\epsilon} \Gamma(\epsilon) } {6(4\pi)^4 } m^{-2 \epsilon}.
\end{gather}
Now the pole in $\epsilon$ is located in the Gamma function, which can be seen by expanding the Gamma function in a small argument
\begin{gather}
\Gamma(\epsilon)=\frac{1}{\epsilon} -\gamma + \mathcal{O}(\epsilon).
\end{gather}
Together with the expansion of an exponential function
\begin{gather}
a^n=e^{n\text{ln}(a)}=1+n\text{ln}(a)+\mathcal{O}((n\text{ln}(a))^2)
\end{gather}
we can take the limit of $\epsilon$ going to zero
\begin{gather}
\frac{i \pi^{\epsilon} \Gamma(\epsilon) } {6(4\pi)^4 } m^{-2 \epsilon} \stackrel{\epsilon \rightarrow 0} {=} \frac{i } {6 \epsilon (4\pi)^4 }
\end{gather}
to obtain the leading UV pole of the one-loop amplitude.

\subsection{Integration-by-Parts Identities}

The Integration-by-Parts Identities \cite{IBP} (see \cite{grozin} for a nice review) give us the possibility to connect an integral of the form 
\begin{gather}
T(D,m^2,-\alpha)=\int \frac{d^D q}{(2\pi)^D} \frac{1}{(q^2+m^2)^\alpha} 
\end{gather} 
by a recursive relation to an integral with a lower power in the propagator
\begin{gather}
T(D,m^2,-\alpha)=\left( 1-\frac{D}{2(\alpha-1)} \right) \frac{1}{m^2}T(D,m^2-\alpha+1) .
\end{gather}
As a convention we choose to promote the integral with $\alpha=3$ to the master integral
\begin{gather}
T(D,m^2,-3)=i\frac{m^{D-6}\Omega(D)}{(2\pi)^D} \int \frac{dQ Q^{D-1} }{(Q^2+1)^3}\\
= -i\frac{2 m^{D-6}}{2^8(\pi)^{D/2} \Gamma(D/2)} \frac{1}{2} \frac{\Gamma(D/2)\Gamma(3-D/2)}{\Gamma(3)} \\
= -i\frac{m^{D-6} \Gamma(3-D/2) }{2^9(\pi)^{D/2}} .
\end{gather}
In our case, as we can see from eq. (\ref{1loopuv}), we have
\begin{gather}
\int \frac{d^D q}{(2\pi)^D}\frac {1}{(q^2+m)^4} = T(D,m^2,-4).
\end{gather}
Using the recursive formula once and then the known result for $T(D,m^2,-3)$ we find
\begin{gather}
T(D,m^2,-4)= -i\left( 1-\frac{D}{6} \right) \frac{1}{m^2} \frac{m^{D-6} \Gamma(3-D/2) }{2^9(\pi)^{D/2}}\\
=-i\left( 1-\frac{D}{6} \right) \frac{m^{D-8} \Gamma(3-D/2)} {2^9(\pi)^{D/2}} .
\end{gather}
From our regulator we can identify the critical dimensions $D=8-2\epsilon$. Plugging this into our expression and using an identity of the $\Gamma$s
\begin{gather}
\Gamma(\epsilon-n)=(-1)^{n-1} \frac{\Gamma(-\epsilon) \Gamma(1+\epsilon)}{\Gamma(n+1-\epsilon)} ,
\end{gather}
 we can expose the pole in $\epsilon$
\begin{gather}
-i\left( 1-\frac{8-2\epsilon}{6} \right) \frac{m^{-2 \epsilon} \pi^{\epsilon} \Gamma(-\epsilon) \Gamma(1+\epsilon)  } {2(4\pi)^{4}\Gamma(2-\epsilon)} .
\end{gather}
Taking the limit of $\epsilon$ going to zero and with the expansion for $\Gamma$ and the exponential function we obtain
\begin{gather}
-i\left( 1-\frac{8-2\epsilon}{6} \right) \frac{m^{-2 \epsilon} \pi^{\epsilon} \Gamma(-\epsilon) \Gamma(1+\epsilon) } {2(4\pi)^{4}\Gamma(2-\epsilon)} \stackrel{\epsilon \rightarrow 0} {=} \frac{i } {6 \epsilon (4\pi)^4 } .
\end{gather}

\subsection{Small-Momentum-Injection}
\begin{figure}[h]
	\centering
		\includegraphics[width=0.60\textwidth]{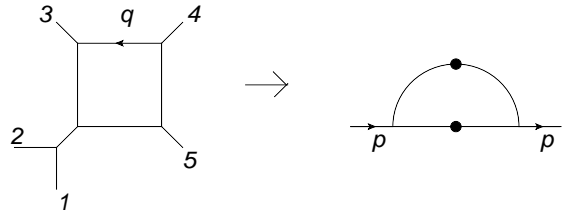}
	\caption{Small-momentum-injection of an one-loop box integral where the bubble integral has been obtained in the limit where $p_{3,5}\rightarrow 0$}
	\label{fig:oneloopvacua}
\end{figure}

In order to obtain the UV pole with the small-momentum-injection we won't directly send all external momenta to zero. Instead we will keep two external momenta, which reduces our box integral to a bubble one. The only information left from the legs where the external momenta are set to zero is the power the propagators are raised to, which is denoted by a dot in figure \ref{fig:oneloopvacua}. In our example we have the following integral
\begin{gather}
\int \frac{d^Dq}{(2\pi)^D} \frac{1}{(q-p_3)^2q^2(q+p_4)^2(q+p_4+p_5)^2}\stackrel{p_{3,5} \rightarrow 0}{=} \int \frac{d^Dq}{(2\pi)^D} \frac{1}{q^4(q+p_4)^4} \\
= \int \frac{d^D q} {(2\pi)^D} \frac{1}{D_{p1}^{2}D_{p2}^{2}}
\end{gather}
where $D_{p1}$ and $D_{p2}$ are
\begin{gather}
D_{p1}=(q+p)^2\\
D_{p2}=q^2. 
\end{gather}

The advantage of this method is that these integrals are known for each power of the propagators in form of the following function \cite{smallp}
\begin{gather}
\int \frac{d^D q} {(2\pi)^D} \frac{1}{D_{p1}^{n_1}D_{p2}^{n_2}}=i\frac{(p^2)^{D/2-n_1-n_2}} {(4\pi)^{D/2}}G(n_1,n_2)
\end{gather}
where $G(n1,n2)$ can be expressed in terms of Gamma functions 
\begin{gather}
G(n_1,n_2)=\frac{\Gamma(-D/2+n_1+n_2)\Gamma(D/2-n_1)\Gamma(D/2-n_2)}{\Gamma(n_1)\Gamma(n_2)\Gamma(D-n_1-n_2)}.
\end{gather}
Back to our case we find for our integral 
\begin{gather}
\int \frac{d^D q} {(2\pi)^D} \frac{1}{D_{p1}^{2}D_{p2}^{2}}=i\frac{(p^2)^{D/2-4}} {(4\pi)^{D/2}}G(2,2)
\end{gather}
where we can see from our regulator that the critical dimension is $D=8-2\epsilon$. Plugging this in we find
\begin{gather}
= i\frac{(p)^{-2\epsilon}(4\pi)^{\epsilon}}  {(4\pi)^{4}} \frac{\Gamma(\epsilon)\Gamma(2-\epsilon)\Gamma(2-\epsilon)}{\Gamma(4-2\epsilon)} .  
\end{gather}
In the last step we send epsilon to zero. Together with the expansion of the Gamma function we find 
\begin{gather}
i\frac{(p)^{-2\epsilon}(4\pi)^{\epsilon}}  {(4\pi)^{4}} \frac{\Gamma(\epsilon)\Gamma(2-\epsilon)\Gamma(2-\epsilon)}{\Gamma(4-2\epsilon)} \stackrel{\epsilon \rightarrow 0}{=} \frac{i} {6(4\pi)^{4} \epsilon}. 
\end{gather}
With this expression of the UV pole we can now go back to our full amplitude and expose the divergence of the full one-loop amplitude.

\subsection{UV poles of the Full One-Loop Amplitude}

In order to determine the UV Pole of the full amplitude we should first remind ourselves of the full one-loop amplitude
\begin{gather}
\mathcal{A}^{\text{1-loop}}=ig^5 \sum_{\text{all perm}} \frac{1}{10} \beta_{12345} C^P \text{Int}^P + \frac{1}{4} \frac{\gamma_{12}}{s_{12}}C^B \text{Int}^B .
\label{oneloopamp2}
\end{gather}
As we have argued earlier the leading UV pole comes only from the box integral, therefore we can neglect the pentagon integral $\text{Int}^P$. It is convenient to turn the structure constant of the box integral
\begin{gather}
C^B=f^{12b}f^{bcg}f^{c3d}f^{d4e}f^{e5g}.
\end{gather}
into traces over the generators like we did in the section about color-ordered amplitudes
\begin{gather}
\begin{split}
C^B=N(T^{12345}+T^{12543}-T^{21345}-T^{21543})+2T^{123}T^{45}+2T^{124}T^{35}\\
+2T^{125}T^{34}-2T^{213}T^{45}-2T^{214}T^{35}-2T^{215}T^{34}.
\end{split} 
\end{gather}
where we used the notation
\begin{gather}
T^{ijklm}=Tr(T^iT^jT^kT^lT^m).
\end{gather}
But since the sum in (\ref{oneloopamp2}) runs over all permutation of external legs we also have to take into account the different orderings of external legs can contribute to one trace. Collecting all the pieces from different orderings and taking into account the prefactor of $\frac{1}{4}$ of the box in (\ref{oneloopamp2}) we arrive at
\begin{gather}
\begin{split}
\left. \mathcal{A}^{1-loop} \right|_{\text{UV}}=-g^5 \frac{1}{6 (4\pi)^4 \epsilon} \left[N Tr(T^1T^2T^3T^4T^5)\left( \frac{\gamma_{12}}{s_{12}}+\frac{\gamma_{23}}{s_{23}}+\frac{\gamma_{34}}{s_{34}}+\frac{\gamma_{45}}{s_{45}}+\frac{\gamma_{51}}{s_{15}} \right) \right. \\
 \left. +6Tr(T^1T^2T^3)Tr(T^4T^5) \left( \frac{\gamma_{12}}{s_{12}}+\frac{\gamma_{23}}{s_{23}}+\frac{\gamma_{31}}{s_{31}} \right)+ \text{perms}  \right]
\end{split} 
\end{gather}
which is the leading UV divergence of the five-point one-loop amplitude at $D=8$ as previously presented in \cite{Carrasco:2011mn}.

\section{The Two-Loop Amplitude}
\begin{figure}[h]
	\centering
		\includegraphics[width=0.80\textwidth]{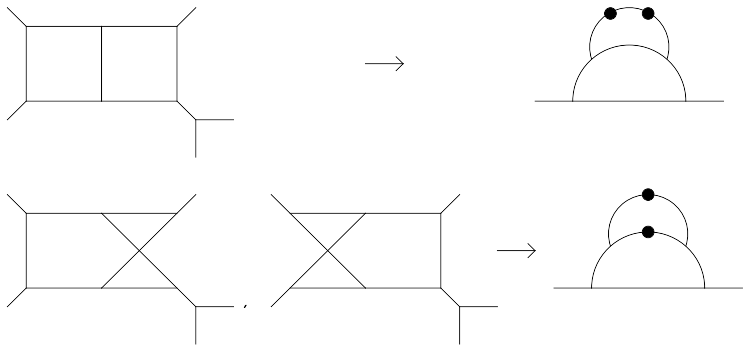}
	\caption{Two-loop planar and non-planar doubleboxes reduce to the corresponding bubble integrals in small-momentum-injection limit}
	\label{fig:2vacuaneu}
\end{figure}

The UV divergence at two-loop can be obtained in a similar way to the one-loop case. We will calculate the integral by a recursive use of the small-momentum-injection method and then obtain the divergence of the full amplitude by summing over all orderings of external legs.\\
By power counting we can conclude that the first UV divergence occurs at $D=7$. This divergence only occurs in the doubleboxes since the $q^2$ dependence in the propagators of the other graphs drops out. This comes from the fact that our numerators are at most linear in loop momentum and can be easily seen by expanding the denominators in terms of scalar products in eq. (\ref{pentabox}).\\
Other than in the one-loop case we have two two-point functions appearing in the UV limit. The planar doublebox reduces to a two-point function $V^P$ and the non-planar left and right crossed doublebox to a two-point function $V^{NP}$. Both integrals are displayed in figure \ref{fig:2vacuaneu}. \\

We can calculate these bubble integrals with the small-momentum-injection method. Therefore we will only keep two momenta and call the remaining momenta $p$
\begin{gather}
V^P=\int \frac{d^Dq}{(2\pi)^D}\frac{d^Dk}{(2\pi)^D} \frac{1}{(p-k)^2} \frac{1}{(k^2)^2} \frac{1}{q^2} \frac{1}{((k-q)^2)^3} \\
V^{NP}=\int \frac{d^Dq}{(2\pi)^D}\frac{d^Dk}{(2\pi)^D} \frac{1}{(p-k)^2} \frac{1}{(k^2)^2} \frac{1}{(q^2)^2} \frac{1}{((k-q)^2)^2}.
\end{gather}
In contrast to the one-loop case we now have two integrals over the loop momenta. But we can overcome this difficulty by recursively using the small-momentum injection. In detail this means we will first calculate the loop in the top of the diagram, absorb the momentum factor into the left over propagators and then calculate the second loop. For the planar case this takes the following form
\begin{gather}
V^P=i \int \frac{d^Dk}{(2\pi)^D} \frac{1}{(p-k)^2} \frac{1}{(k^2)^2} \frac{(k^2)^{D/2-4}}{(4\pi)^{D/2}} G(1,3)\\
=-\frac{1}{(4\pi)^D}(p^2)^{D-7}G(1,6-\frac{D}{2})G(1,3)\\
=-\frac{1}{(4\pi)^D}(p^2)^{D-7} \frac{\Gamma(7-D)(\Gamma(D/2-1))^2\Gamma(D-6)}{\Gamma(6-D/2)\Gamma(3/2D-7)} \frac{\Gamma(4-D/2)\Gamma(D/2-3)}{2\Gamma(D-4)}.
\end{gather}
From the regulator we can see that this integral diverges at $D=7$ as we expected from power counting. Using the shifting parameter epsilon we calculate the integral at $D=7-2\epsilon$ dimensions and taking the limit of epsilon going to zero we find for the planar integral
\begin{gather}
V^P \stackrel{\epsilon \rightarrow 0 } {=} - \frac{\pi}{20 (4\pi)^7 \epsilon }.
\end{gather}
The non-planar integral can be obtained in the same way as we have obtained the planar one
\begin{gather}
V^{NP}=i \int \frac{d^Dk}{(2\pi)^D} \frac{1}{(p-k)^2} \frac{1}{(k^2)^2} \frac{(k^2)^{D/2-4}}{(4\pi)^{D/2}} G(2,2)\\
= -\frac{1}{\pi^D} (k^2)^{D-7} G(1,6-\frac{D}{2}) G(2,2)\\
=-\frac{1}{\pi^D}(k^2)^{D-7} \frac{\Gamma(7-D)\Gamma(D/2-1)\Gamma(D-6)}{\Gamma(6-D/2)\Gamma(3/2D-7)} \frac{\Gamma(4-D/2)(\Gamma(D/2-2))^2}{\Gamma(D-4)}.
\end{gather}
As in the planar case this integral will diverge at $D=7$. Shifting the dimension by $D=7-2\epsilon$ we can extract the following pole in epsilon
\begin{gather}
V^{NP} \stackrel{\epsilon \rightarrow 0 } {=} - \frac{\pi}{30 (4\pi)^7 \epsilon }.
\end{gather}

\subsection{UV poles of the Full Two-Loop Amplitude}
In order to obtain the leading UV pole of the full amplitude we should first remind ourselves of the full two-loop amplitude.
\begin{gather}
\begin{split}
\mathcal{A}^{2-loop}=-g^7 \sum_{\text{all perm}} \left(\frac{1}{2}C^{PB}\text{Int}^{PB}+\frac{1}{4}C^{CPB}\text{Int}^{CPB}+\frac{1}{4}C^{DP}\text{Int}^{DP} \right. \\ 
\left. +\frac{1}{2}C^{DB}\text{Int}^{DB} +\frac{1}{4}C^{LCDB}\text{Int}^{LCDB}+\frac{1}{4}C^{RCDB}\text{Int}^{RCDB} \right)
\end{split}
\label{twoloopamp}
\end{gather}
Since the planar and non-planar doubleboxes will diverge first we can neglect the pentabox, crossed pentabox and the double penta for the leading UV divergence. Furthermore it is convenient to transform the product of structure into traces over generators
\begin{gather}
\begin{split}
C^{DB}= T^{12345}-2T^{13245}+T^{21345}-2T^{23145}+T^{31245}+T^{32145} \\
-T^{41235}+2T^{41325}-T^{42135}+2T^{42315}-T^{43125}-T^{43215}\\
+N\left(-3T^{12}T^{345} + 3T^{12}T^{435}\right) \\
 +N^2 \left(T^{12345}+ T^{32145}-T^{41235}-T^{43215} \right)
\end{split}\\
\begin{split}
C^{LCDB}= T^{12345}+2T^{13245}-T^{21345}+2T^{23145}-T^{31245}-T^{32145}\\
+T^{41235}-2T^{41325}
+T^{42135}-2T^{42135}+T^{43125}+T^{43215}\\
+N\left(T^{23}T^{145} + T^{13}T^{245}-2T^{12}T^{345} -T^{23}T^{415} \right.\\
\left. - T^{13}T^{425}+2T^{12}T^{435} \right)
\end{split}\\
\begin{split}
C^{RCDB}= T^{12345}+2T^{13245}-T^{21345}+2T^{23145}-T^{31245}-T^{32145} \\
+T^{41235}-2T^{41325}+T^{42135}-2T^{42135}+T^{43125}+T^{43215}\\
+N\left(T^{23}T^{145} + T^{13}T^{245}-2T^{12}T^{345} -T^{23}T^{415} \right.\\
\left. - T^{13}T^{425}+2T^{12}T^{435} \right) .
\end{split}
\end{gather}
The first thing to note is that the left and right crossed doublebox have the same color factor namely $C^{LCDB}=C^{RCDB}$. This means instead of keeping track of both we can just add them together. \\
As in the one-loop case the sum in (\ref{twoloopamp}) runs over all ordering of external legs, therefore we need to collect all the pieces that belong to a certain color trace. Doing so we can find the following UV divergence of the full amplitude
\begin{gather}
\begin{split}
\mathcal{A}^{2-loop}=-g^7\left[\left(N^2 V^P + 12(V^P+V^{NP}) \right) Tr(T^1T^2T^3T^4T^5) \right. \\
\left(5 \beta_{12345} + \frac{\gamma_{12}}{s_{12}}(s_{35}-2s_{12})+ \frac{\gamma_{12}}{s_{12}}(s_{35}-2s_{12})+ \frac{\gamma_{23}}{s_{23}}(s_{14}-2s_{23})\right. \\
\left. + \frac{\gamma_{34}}{s_{34}}(s_{25}-2s_{34})+ \frac{\gamma_{45}}{s_{45}}(s_{13}-2s_{45})+ \frac{\gamma_{51}}{s_{15}}(s_{24}-2s_{15})  \right) \\ -12N(V^P+V^{NP})Tr(T^1T^2T^3)Tr(T^4T^5)s_{45}\left(\frac{\gamma_{12}}{s_{12}}+\frac{\gamma_{23}}{s_{23}}+\frac{\gamma_{31}}{s_{31}} \right)\\
\left. + \text{perms}  \right] .
\end{split}
\end{gather}
The UV pole of the two-loop five-point amplitude is in full agreement with the results presented earlier in \cite{Carrasco:2011mn}. 
\newpage
\chapter{Conclusion}

This work focused on the development of new mathematical methods for computing multi-loop scattering amplitudes
in gauge theories, aiming at the combined use of {\it multiloop integrand-decomposition},
{\it unitarity-based methods}, {\it and color-kinematic duality}.

We have shown, for the first time, how the recently introduced multi-loop integrand-reduction technique 
can be combined with a unitarity-based construction of the integrand, yielding the decomposition of scattering amplitudes
in terms of independent integrals.

We discussed the generic features of this novel reduction algorithm, based on multivariate polynomial division and basic principles of algebraic geometry, and we applied it to the one- and two-loop five-point amplitudes in $\mathcal {N}=4$ sYM.
The integrands of the multiple-cuts were built as products of tree-level amplitudes within the super-amplitudes formalism, 
accounting for the whole particle content of the whole $\mathcal {N}=4$ super-multiplets which can run in the loops.
The reduction of the amplitudes in terms of independent integrals was achieved by fitting the multiple-cut residues analytically. 
The parametric form of the residue could be determined a priori by means of successive polynomial divisions 
using the Gr\"obner basis generated from the denominators that go simultaneously the on-shell. 

In this work, the integrand-reduction method has been exploited to investigate the color-kinematic duality for multi-loop $\mathcal{N}=4$ sYM scattering amplitudes. We aimed at understanding how the shape of the residues at the multiple-cuts can be made compatible with a cut-reconstruction of graph numerators which automatically satisfy the color-kinematic dualities.
Furthermore we provided the first step into a new research direction, namely how symmetries can restrict the number of monomials in a residue. Through the integrand-reduction we know that the each monomial in the integrand is corresponding to a potential master integral. Therefore we can use symmetries of the integr{\it and} to further reduce the number of appearing master integ{\it rals}.  
We finally extracted the leading ultra-violet divergences of the one- and two-loop five-point amplitudes in 
$\mathcal {N}=4$ sYM, which represent a paradigmatic example for studying the UV behavior of supersymmetric amplitudes.

\chapter*{Acknowledgements}
\label{sec:Acknowledgements}
\addcontentsline{toc}{chapter} {Acknowledgements}
I want to thank my Co-Supervisor Pierpaolo Mastolia for the huge amount of time he invested into my training and this thesis. Furthermore I want to thank him for all the useful discussion we had and the insight he provided me into this extremely interesting field. In addition I want to thank Tiziano Peraro and Johannes Schlenk for the fruitful discussions we had. Furthermore I want to thank my parents, who made it possible for me to study at the Technical University Munich (TUM) and for their endless support on the way. A special thanks goes to my girlfriend Kate for all the support and the motivation she gave me.
\newpage
\chapter{Appendix}
\section{Merging of the One-Loop Integrand}
\label{derivationintegrand}
The product of generating functions for the one-loop amplitude is given by
\begin{gather}
\begin{split}
I(q)=A_4^{MHV}(1,2,-l_2,l_5)A_3^{\overline{MHV}}(l_2,3,-l_3)A_3^{MHV}(l_3,4,-l_4)\\A_3^{\overline{MHV}}(l_4,5,-l_5) .
\end{split}
\end{gather}
Plugging in the generating functions for $\mathcal{N}$=4 sYM we find the following expression
\begin{gather}
\begin{split}
\int d^4\eta_{l_2}d^4\eta_{l_3}d^4\eta_{l_4}d^4\eta_{l_5} \: \delta^{(8)}\left(\lambda_1 \eta_1 + \lambda_2 \eta_2-\lambda_{l_2} \eta_{l_2}+\lambda_{l_5} \eta_{l_5} \right)\\
\delta^4(\spb3.{l_3} \eta_{l_2}+ \spb{l_3}.{l_2} \eta_{3}+ \spb{l_2}.3 \eta_{l_3}) \delta^{(8)}\left(\lambda_{l_3}\eta_{l_3}+\lambda_{4}\eta_{4}-\lambda_{l_4}\eta_{l_4} \right) \\
\delta^4\left(\spb5.{l_5} \eta_{l_4}+\spb{l_4}.{l_5} \eta_{5}+ \spb{l_4}.5 \eta_{l_5} \right) 
\end{split}\\
\begin{split}
=\left(\spb3.{l_3} \spb5.{l_5}\right)^4 \int d^4\eta_{l_2}d^4\eta_{l_3}d^4\eta_{l_4}d^4\eta_{l_5} \: \delta^{(8)}\left(\lambda_1 \eta_1 + \lambda_2 \eta_2- \lambda_{l_2} \eta_{l_2}+\lambda_{l_5} \eta_{l_5} \right)\\
\delta^4(\eta_{l_2}+\frac{\spb{l_3}.{l_2}}{\spb{3}.{l_3}}\eta_{3}+\frac{\spb{l_2}.3}{\spb{3}.{l_3}}\eta_{l_3}) \delta^{(8)}\left(\lambda_{l_3}\eta_{l_3}+\lambda_{4}\eta_{4}-\lambda_{l_4}\eta_{l_4} \right) \\
\delta^4\left(\eta_{l_4}+\frac{\spb{l_4}.{l_5}}{\spb{5}.{l_5}}\eta_{5}+\frac{\spb{l_4}.5}{\spb{5}.{l_5}}\eta_{l_5} \right)
\end{split}\\
\begin{split}
=\left(\spb3.{l_3} \spb5.{l_5}\right)^4 \int d^4\eta_{l_3}d^4\eta_{l_5} \\ \delta^{(8)}\left(\lambda_1 \eta_1 + \lambda_2 \eta_2+\lambda_{l_2} \left(\frac{\spb{l_3}.{l_2}}{\spb{3}.{l_3}}\eta_{3}+\frac{\spb{l_2}.3}{\spb{3}.{l_3}}\eta_{l_3}\right)+\lambda_{l_5} \eta_{l_5} \right)\\
\delta^{(8)}\left(\lambda_{l_3}\eta_{l_3}+\lambda_{4}\eta_{4}+\lambda_{l_4}\left(\frac{\spb{l_5}.{l_4}}{\spb{5}.{l_5}}\eta_{5}+\frac{\spb{l_4}.5}{\spb{5}.{l_5}}\eta_{l_5}\right) \right).
\end{split}
\end{gather}
From the momentum conservation at the all incoming $\overline{MHV}$ three point vertices we find the following identity
\begin{gather}
0=\slashed{p}_i+\slashed{p}_j+\slashed{p}_k \\ 
= \lambda_i^\alpha \spb{i}.k + \lambda_j^\alpha \spb{j}.k
\end{gather}
where we obtained the last line by contracting the expression with a $k]$.
With this identity we can now continue to calculate the quadruple-cut residue
\begin{gather}
\begin{split}
\left(\spb3.{l_3} \spb5.{l_5}\right)^4 \int d^4\eta_{l_3}d^4\eta_{l_5} \\ \delta^{(8)}\left(\lambda_1 \eta_1 + \lambda_2 \eta_2+\lambda_{l_2} \left(\frac{\spb{l_3}.{l_2}}{\spb{3}.{l_3}}\eta_{3}+\frac{\spb{l_2}.3}{\spb{3}.{l_3}}\eta_{l_3}\right)+\lambda_{l_5} \eta_{l_5} \right)\\
\delta^{(8)}\left(\lambda_{l_3}\eta_{l_3}+\lambda_{4}\eta_{4}+\lambda_{l_4}\left(\frac{\spb{l_5}.{l_4}}{\spb{5}.{l_5}}\eta_{5}+\frac{\spb{l_4}.5}{\spb{5}.{l_5}}\eta_{l_5}\right) \right)
\end{split}\\
\begin{split}
=\left(\spb3.{l_3} \spb5.{l_5}\right)^4 \int d^4\eta_{l_3}d^4\eta_{l_5} \\ \delta^{(8)}\left(\lambda_1 \eta_1 + \lambda_2 \eta_2+\lambda_3 \eta_{3}-\lambda_{l_3} \eta_{l_3}+\lambda_{l_5} \eta_{l_5} \right)\\
\delta^{(8)}\left(\lambda_{l_3}\eta_{l_3}+\lambda_{4}\eta_{4}+\lambda_{5}\eta_{5}-\lambda_{l_5}\eta_{l_5} \right)
\end{split}\\
\begin{split}
=\left(\spb3.{l_3} \spb5.{l_5}\right)^4 \int d^4\eta_{l_3}d^4\eta_{l_5} \\ \delta^{(8)}\left(\lambda_1 \eta_1 + \lambda_2 \eta_2+\lambda_3 \eta_{3}+\lambda_{4}\eta_{4}+\lambda_{5}\eta_{5} \right)\\
\delta^{(8)}\left(\lambda_{l_3}\eta_{l_3}+\lambda_{4}\eta_{4}+\lambda_{5}\eta_{5}-\lambda_{l_5}\eta_{l_5} \right)
\end{split}
\end{gather}
where we have obtained the last line by the identity $\delta(A)\delta(B)=\delta(A+B)\delta(B)$.
Continuing our calculation we find
\begin{gather}
\begin{split}
\left(\spb3.{l_3} \spb5.{l_5}\right)^4 \int d^4\eta_{l_3}d^4\eta_{l_5} \delta^{(8)}\left(\sum_{i=1}^{5}{\lambda_i \eta_i} \right)\\
\delta^{(8)}\left(\lambda_{l_3}\eta_{l_3}+\lambda_{4}\eta_{4}+\lambda_{5}\eta_{5}-\lambda_{l_5}\eta_{l_5} \right)
\end{split} \\
\begin{split}
=\left(\spb3.{l_3} \spb5.{l_5}\right)^4 \delta^{(8)}\left(\sum_{i=1}^{5}{\lambda_i \eta_i} \right) \int d^4\eta_{l_3}d^4\eta_{l_5}  \\ 
\prod_{a=1}^4 \left(\spa{l_3}.4 \eta^a_{l_3}\eta^a_4+\spa{4}.{l_5} \eta^a_{4} \eta^a_{l_5}+ \spa{l_4}.{l_5} \eta^a_{l_4}\eta^a_{l_5}+ \spa{l_3}.{l_5} \eta^a_{l_3}\eta^a_{l_5}+ \spa{l_3}.{l_4} \eta^a_{l_3}\eta^a_{l_4} \right)
\end{split} \\
=\left( \spb3.{l_3} \spa{l_3}.{l_5} \spb{5}.{l_5} \right)^4  \delta^{(8)}\left(\sum_{i=1}^{5}{\lambda_i \eta_i} \right)  
\end{gather}
Putting this together with the denominator we find
\begin{gather}
I(q) =  \frac{ \delta^{(8)}\left(\sum_{i=1}^{5}{\lambda_i \eta_i} \right) \left( \spb3.{l_3} \spa{l_3}.{l_5} \spb{5}.{l_5} \right)^4 } { \spa1.2 \spa2.{l_2} \spa{l_2}.{l_5} \spa{l_5}.1 \spb{l_2}.3 \spb3.{l_3} \spb{l_3}.{l_2} \spa{l_3}.4 \spa4.{l_4} \spa{l_4}.{l_3} \spb{l_4}.5 \spb{5}.{l_5} \spb{l_5}.{l_4} } .
\end{gather} 
Using momentum conservation to cancel as many spinor products as possible we arrive at equation (\ref{boxcuteq})
\begin{gather}
I(q) =  - \delta^{(8)}\left(\sum_{i=1}^{5}{\lambda_i \eta_i} \right)\frac{ \spab3.\slashed{l_5}.5 \spb3.4} {\spa1.2 \spa2.3 \spa3.4 \spa5.1} .
\end{gather}

\section{Monomials in the Residues}
The monomials which parametrize the  residues of the pentabox, crossed pentabox and the double penta have been obtained in \cite{twoloopisp}. For convenience we will just state the main results here.

\subsection{Pentabox}
The pentabox in $\mathcal{N}$=4 sYM had four sevenfold-cut residues $\Delta^{PB}_{1235678}(1,2,3,4,5)$ , $\Delta^{PB}_{1234678}(1,2,34,5)$ , $\Delta^{PB}_{1234578}(1,2,3,45)$ and $\Delta^{PB}_{1234568}(1,2,3,4,5)$ and an eightfold-cut residue $\Delta^{PB}_{12345678}(1,2,3,4,5)$ . If we use the following general decomposition for the loop momentum
\begin{gather}
q^\mu=-p_0^\mu + \sum_{i=1}^4{x_i \tau_i^\mu} \ \ \ \ \  k^\mu=-r_0^\mu + \sum_{i=1}^4{y_i \epsilon_i^\mu}
\label{gendecom} 
\end{gather}
we can write down a basis for each cut
\bea
\label{def:ebasis8fold}
&&\begin{cases}
\begin{aligned}
 r_0^\mu  &= 0^\mu,   &  e^\mu_1      &= p_3^\mu, &   e^\mu_2      &= p_4^\mu, & e_3^\mu     &= \frac{\langle 3|\gamma^\mu |4 ]}{2} , & e_4^\mu &= \frac{\langle 4|\gamma^\mu | 3 ]}{2} ,  \quad  \\
 p_0^\mu  &=0^\mu,    &   \tau^\mu_1 &= p_2^\mu, & \tau^\mu_2 &= p_1^\mu, & \tau_3^\mu &= \frac{\langle 2|\gamma^\mu | 1 ]}{2} , & \tau_4^\mu &= \frac{\langle 1 |\gamma^\mu |2 ]}{2} , \quad
 \end{aligned} \\[5.0ex]
  x_1 = \frac{(q\cdot p_1)}{(p_1 \cdot p_2)}\ , \qquad
x_2 = \frac{(q\cdot p_2)}{(p_1 \cdot p_2)}\ , \qquad
y_1 = \frac{(k\cdot p_4)}{(p_3 \cdot p_4)}\ , \qquad
y_2 = \frac{(k\cdot p_3)}{(p_3 \cdot p_4)}\ ;
\end{cases}\pagebreak[1]  \\[2.0ex]
\label{def:ebasis7fold1}
&&\begin{cases}
\begin{aligned}
 r_0^\mu  &= 0^\mu, &  e^\mu_1    &= p_1^\mu, &   e^\mu_2    &= p_3^\mu, & e_{3,4}^\mu     &= \frac{
 \langle 3| 2 |1]
 \langle 1|\gamma^\mu | 3] \pm   \langle 1 |2 |3]
 \langle 3|\gamma^\mu | 1]
 }{4}    , \quad  \\
p_0^\mu  &=0^\mu,  &   \tau^\mu_1 &= p_1^\mu, &   \tau^\mu_2 &= p_3^\mu, &   \tau^\mu_{3,4} &= \frac{
 \langle  3| 2 | 1]
 \langle  1|\gamma^\mu | 3] \pm   \langle  1| 2 | 3]
 \langle  3|\gamma^\mu | 1]
 }{4}  ,  \quad
  \end{aligned} \\[5.0ex]
 x_1 =\frac{ (q\cdot p_3)}{(p_1\cdot p_3)}, \qquad
x_4 = \frac{(q\cdot \tau_4)}{\tau_4^2}, \qquad
y_1 = \frac{(k\cdot p_3)}{(p_1\cdot p_3)}, \qquad
y_4 = \frac{(k\cdot e_4)}{e_4^2} ;
\end{cases} \pagebreak[1]  \\[2.0ex]
&&\begin{cases}
\begin{aligned}
 r_0^\mu  &= 0^\mu, &  e^\mu_{1,4}      &= \frac{\langle 3|2 |1]
 \langle 1|\gamma^\mu |3] \mp   \langle 1|2 |3]
 \langle 3|\gamma^\mu |1]
 }{4} &  e^\mu_2 &=  p_3^\mu, \\
    e^\mu_3      &= p_1^\mu  ,  \\
p_0^\mu  &=-p_4^\mu,  &   \tau^\mu_{1,4} &=\frac{\langle 3|2 |1]
 \langle 1|\gamma^\mu |3] \mp   \langle 1|2 |3]
 \langle 3|\gamma^\mu |1]
 }{4}, &  \tau^\mu_2 &= p_3^\mu,\\
    \tau^\mu_{3} &= p_{1}^\mu ,
\end{aligned} \\[5.0ex]
 x_1 =\frac{ ((q-p_4) \cdot e_1)}{e_1^2}, \qquad
x_2 = \frac{((q-p_4) \cdot p_1)}{(p_1\cdot p_3)}, \qquad
y_1 = \frac{(k\cdot \tau_1)}{\tau_1^2}, \qquad
y_3 = \frac{(k\cdot p_3)}{(p_1\cdot p_3)} . \qquad
\end{cases}  \pagebreak[1]
\label{def:wbasis7fold:1235678}.
\eea
The monomials of the residues can be obtained through a multivariate polynomial division of a generic integrand. These monomials are shown in table \ref{Tab:Pentabox}.\footnote{The differences to the tables in \cite{twoloopisp} stem from a different labeling of the propagators.}

\begin{table}
\begin{tabular}[t]{l | l | l | l}
\hline \hline \trule  cut              & bases                                          & $z$ & Monomials in the residue  \\   \hline \trule
$(12345678)$     & \small Eq. (\ref{def:ebasis8fold})  & \small $(y_4,y_3,y_2,y_1,x_4,x_3,x_2,x_1)$               & \small $\mathcal{S}_{12345678} = \{ 1, x_1, y_1, y_2 \} $  \\ \hline \trule
$(1234568)$       &  \small Eq. (\ref{def:ebasis8fold})  & \small $(y_4,y_3,y_2,y_1,x_4,x_3,x_2,x_1)$                  & \small $\mathcal{S}_{1234568} =  \{  1,x_1,x_1^2,x_1^3,x_1^4,x_2,x_1 x_2,$ \\
&&&                    \small         $x_1^2 x_2,x_1^3 x_2, y_1,x_1 y_1,x_1^2 y_1,x_1^3 y_1, x_1^4 y_1,$ \\
&&&                    \small         $x_2  y_1,x_1 x_2 y_1, x_1^2 x_2 y_1,x_1^3 x_2 y_1,y_1^2,x_1y_1^2,$ \\
&&&                    \small         $x_2 y_1^2, y_1^3, x_1 y_1^3, x_2 y_1^3,y_1^4,x_1y_1^4,x_2 y_1^4,y_2,$ \\
&&&                   \small           $x_1 y_2,y_1 y_2,y_1^2 y_2,y_1^3 y_2  \}$  \\ \hline \trule
$(1235678)$       &\small Eq.  (\ref{def:ebasis8fold})  & \small$(y_4,y_3,y_2,y_1,x_4,x_3,x_2,x_1)$               & \small $\mathcal{S}_{1234678} =\mathcal{S}_{1234568}$ \\ \hline \trule
$(1234578)$       & \small Eq. (\ref{def:ebasis7fold1})  & \small $(y_4,y_3,y_2,y_1,x_4,x_3,x_2,x_1)$              & \small $\mathcal{S}_{1234578} = \{1,x_2,x_2^2,x_2^3,x_2^4,x_4,x_2 x_4,$ \\
&&& \small      $x_2^2 x_4,x_2^3x_4, y_1,x_2 y_1,x_2^2 y_1,x_2^3 y_1,x_2^4 y_1,x_4y_1,$ \\
&&& \small      $x_2 x_4 y_1,x_2^2 x_4 y_1,x_2^3 x_4 y_1,y_1^2,x_2y_1^2,x_4 y_1^2,y_1^3,x_2 y_1^3,$ \\
&&& \small      $x_4 y_1^3,y_1^4,x_2y_1^4,x_4 y_1^4,y_4,x_2 y_4,y_1 y_4,$\\
&&& \small      $y_1^2 y_4,y_1^3y_4 \}$ \\ \hline \trule
$(1234678)$       &\small Eq.  (\ref{def:wbasis7fold:1235678}) &  \small  $(y_4,y_3,y_2,y_1,x_4,x_3,x_2,x_1)$ &   \small $\mathcal{S}_{1235678} =\{ 1,x_1,x_1^2,x_1^3,x_1^4,x_2,$\\
&&&  \small $x_1 x_2,x_1^2 x_2,x_1^3x_2,y_1,x_1 y_1,x_1^2 y_1,$ \\
&&&  \small $x_1^3 y_1,x_1^4 y_1,x_2 y_1,  x_1 x_2 y_1,x_1^2 x_2 y_1,x_1^3 x_2y_1,$ \\
&&&  \small $y_1^2,x_1 y_1^2,x_2 y_1^2,y_1^3,x_1 y_1^3,x_2 y_1^3,$ \\
&&&  \small $y_1^4,  x_1 y_1^4,x_2 y_1^4,y_3,x_1 y_3,y_1 y_3,y_1^2 y_3,y_1^3 y_3 \}$ \\    \hline \hline
\end{tabular}
\label{Tab:Pentabox}
\caption{Set of monomials parametrizing the residues entering the decomposition of the five-point pentabox diagram.  They have all been found using degree lexicographic monomial ordering.  For each cut the bases and the chosen ordering for loop-variables are shown as well.
}
\end{table}

\subsection{Crossed Pentabox}
The crossed pentabox has also four sevenfold-cut residues $\Delta^{CPB}_{1235678}(1,2,3,4,5)$ , $\Delta^{CPB}_{1234678}(1,2,34,5)$ , $\Delta^{CPB}_{1234578}(1,2,3,45)$ and $\Delta^{CPB}_{1234568}(1,2,3,4,5)$ and an eightfold-cut residue $\Delta^{CPB}_{12345678}(1,2,3,4,5)$. With the same basis as for the pentabox we can find the monomials displayed in table \ref{Tab:Pentacross}.  

\begin{table}
\begin{tabular}[t]{l | l | l | l}
\hline \hline \trule  cut              & bases                                          & $z$ & Monomials in the residue  \\   \hline \trule
$(12345678)$    & \small Eq. (\ref{def:ebasis8fold})  & \small $(x_4,x_3,x_2,y_3,y_4,x_1,y_2,y_1)$                & \small $\mathcal{S}_{12345678} = \{ 1, x_1, y_1, y_2 \} $  \\ \hline \trule
$(1234568)$      &  \small Eq. (\ref{def:ebasis8fold})  & \small  $(y_4,y_3,y_2,y_1,x_4,x_3,x_2,x_1)$                 & \small $\mathcal{S}_{1234568} = \{1,x_1,x_1^2,x_1^3,x_1^4,x_1^5,x_1^6,x_2,$\\
&&&  \small $x_1 x_2,x_1^2  x_2,x_1^3 x_2,x_1^4 x_2,x_1^5 x_2,$\\
&&& \small $y_1,x_1 y_1,x_1^2   y_1,x_1^3 y_1,x_1^4 y_1,x_1^5 y_1,x_2 y_1,$\\
&&& \small $x_1 x_2   y_1,x_1^2 x_2 y_1,x_1^3 x_2 y_1,x_1^4 x_2 y_1,y_1^2,x_1y_1^2,$\\
&&& \small $x_2 y_1^2,y_1^3,x_1 y_1^3,x_2 y_1^3,y_1^4,x_1 y_1^4,x_2 y_1^4,y_2,$\\
&&& \small $x_1 y_2, y_1 y_2,y_1^2 y_2,y_1^3   y_2 \}$ \\  \hline \trule
$(1235678)$     &\small Eq.  (\ref{def:ebasis8fold})  & \small  $(y_4,y_3,y_2,y_1,x_4,x_3,x_2,x_1)$                 & \small $\mathcal{S}_{1234678} =\mathcal{S}_{1234568}$ \\ \hline \trule
$(1234578)$       & \small Eq. (\ref{def:ebasis7fold1})  & \small  $(y_4,y_3,y_2,y_1,x_4,x_3,x_2,x_1)$                 & \small $\mathcal{S}_{1234578} = \{1,x_2,x_2^2,x_2^3,x_2^4,x_2^5,x_2^6,x_4,$\\
&&&   \small  $x_2 x_4,x_2^2 x_4,x_2^3 x_4,x_2^4 x_4,x_2^5 x_4,y_1,x_2 y_1,$\\
&&&  \small   $x_2^2 y_1,x_2^3 y_1,x_2^4   y_1,x_2^5 y_1,x_4 y_1,x_2 x_4 y_1,x_2^2 x_4 y_1,$ \\
&&&   \small $x_2^3 x_4 y_1,x_2^4 x_4 y_1,y_1^2,x_2 y_1^2,x_4 y_1^2,y_1^3,x_2 y_1^3,$\\
&&&  \small  $x_4 y_1^3,y_1^4,x_2 y_1^4,x_4 y_1^4,y_4,x_2 y_4,y_1y_4,y_1^2 y_4,y_1^3 y_4 \}$ \\ \hline \trule
$(1234678)$       &\small Eq.  (\ref{def:wbasis7fold:1235678}) &  \small  $(y_4,y_2,y_3,y_1,x_4,x_3,x_2,x_1)$ &   \small $\mathcal{S}_{1235678} = \{1,x_1,x_1^2,x_1^3,x_1^4,x_1^5,x_1^6,x_2,$\\
&&& \small $x_1 x_2,x_1^2  x_2,x_1^3 x_2,x_1^4 x_2,x_1^5 x_2,y_1,$\\
&&& \small $x_1 y_1,x_1^2   y_1,x_2 y_1,x_1 x_2 y_1,y_1^2,x_1 y_1^2,x_1^2   y_1^2,$\\
&&& \small $x_2 y_1^2,x_1 x_2 y_1^2,y_1^3,x_1 y_1^3,x_1^2y_1^3,x_2 y_1^3,x_1 x_2 y_1^3,$\\
&&& \small $y_1^4,x_1 y_1^4,x_1^2   y_1^4,x_2 y_1^4,x_1 x_2 y_1^4,y_3,x_1 y_3,$\\
&&& \small $y_1   y_3,y_1^2 y_3,y_1^3 y_3\}$ \\ \hline \hline
\end{tabular}
\label{Tab:Pentacross}
\end{table}

\subsection{Double Penta}
The double penta has eight sevenfold-cut residues. But we also know that it is invariant under the transformation
\begin{gather}
p_1^\mu \leftrightarrow p_3^\mu \ \ \  p_4^\mu \leftrightarrow p_2^\mu \ \ \ q^\mu \leftrightarrow k^\mu 
\end{gather}
therefore it is enough to parametrize the residues $\Delta^{DP}_{12345678}$  , $\Delta^{DP}_{2345678}$ , $\Delta^{DP}_{1345678}$ , $\Delta^{DP}_{1245678}$ and $\Delta^{DP}_{1234567}$. The basis for each cut in terms of the general decomposition (\ref{gendecom}) are given by:
\bea
\label{def:ebasis7foldtwist0}
&&\begin{cases}
\begin{aligned}
  r_0^\mu  &=0^\mu,    &   e^\mu_1 &= p_4^\mu, & e^\mu_2 &= p_3^\mu, & e_3^\mu &= \frac{\langle 4|\gamma^\mu | 3 ]}{2} , & e_4^\mu &= \frac{\langle 3 |\gamma^\mu |4 ]}{2} , \quad \\
 p_0^\mu  &= 0^\mu,   &  \tau^\mu_1      &= p_2^\mu, &   \tau^\mu_2      &= p_1^\mu, & \tau_3^\mu     &= \frac{\langle 2|\gamma^\mu |1 ]}{2} , & \tau_4^\mu &= \frac{\langle 1|\gamma^\mu | 2 ]}{2} ,  \quad
 \end{aligned} \\[5.0ex]
x_1 = \frac{(q\cdot p_1)}{(p_2 \cdot p_1)}\ , \qquad
x_2 = \frac{(q\cdot p_2)}{(p_2 \cdot p_1)}\ , \qquad
  y_1 = \frac{(k\cdot p_3)}{(p_3 \cdot p_4)}\ , \qquad
y_2 = \frac{(k\cdot p_4)}{(p_3 \cdot p_4)}\ ;
\end{cases}\pagebreak[1]  \\[2.0ex]
&&\begin{cases}
\begin{aligned}
 r_0^\mu &=0^\mu,  &   e^\mu_1 &= p_1^\mu, & e^\mu_2 &= p_3^\mu, \\
   e^\mu_{3,4} &= \frac{
 \langle 1|4 |3]
 \langle 3|\gamma^\mu |1] \pm   \langle 3|4 |1]
 \langle 1|\gamma^\mu |3]
 }{4}   , \\
 p_0^\mu &= -p_3^\mu,  & \tau^\mu_1  &= p_1^\mu, &   \tau^\mu_2   &= p_3^\mu, \\
  \tau_{3,4}^\mu &=  \frac{
 \langle 1|4 |3]
 \langle 3|\gamma^\mu |1] \pm   \langle 3|4 |1]
 \langle 1|\gamma^\mu |3]
 }{4}   ,  \\
 \end{aligned} \\[5.0ex]
x_2 = \frac{((q+p_3)\cdot p_1)}{(p_1\cdot p_3)}, \quad
x_4 = \frac{((q+p_3)\cdot \tau_4)}{\tau_4^2}, \quad
 y_1 =\frac{ (k\cdot p_3)}{(p_1\cdot p_3)}, \quad
y_4 = \frac{(k\cdot e_4)}{e_4^2}.
\end{cases}
\label{def:ebasis7foldtwist1}
\eea 
 The monomials appearing in each residue are given in table \ref{Tab:Pentatwist} 
\begin{table}
\begin{tabular}[t]{l | l | l | l}
\hline \hline \trule  cut              & bases                                          & $z$ & Monomials in the residue  \\   \hline \trule
$(12345678)$    & \small Eq. (\ref{def:ebasis7foldtwist0}) & \small $(y_4,y_3,y_2,y_1,x_4,x_3,x_2,x_1)$ &  \small $\mathcal{S}_{12345678} = \{1,y_1,x_1,y_1 x_1,x_1^2,x_1^3,x_2,x_1 x_2 \}$  \\  \hline \trule
$(2345678)$      &  \small Eq. (\ref{def:ebasis7foldtwist0}) &  \small  $(x_4,x_3,x_2,x_1,y_4,y_3,y_2,y_1)$ &  \small $\mathcal{S}_{1345678} = \{1,y_1,y_1^2,y_1^3,y_1^4,y_1^5,y_1^6,y_2,y_1y_2,$ \\
&&& \small $y_1^2y_2,y_1^3 y_2,y_1^4 y_2,y_1^5 y_2,x_1,$\\
&&& \small $y_1 x_1,y_1^2   x_1,y_1^3 x_1,y_1^4 x_1,y_1^5 x_1,$ \\
&&& \small $y_2 x_1,y_1 y_2 x_1,y_1^2  y_2 x_1, y_1^3 y_2 x_1,$\\
&&& \small $y_1^4 y_2 x_1,x_1^2,y_1 x_1^2,y_2 x_1^2,x_1^3,y_1 x_1^3,$\\
&&& \small $y_2 x_1^3,x_1^4,y_1 x_1^4,y_2 x_1^4,$ \\
&&& \small $x_2,y_1 x_2,x_1 x_2,x_1^2 x_2,x_1^3 x_2\}$\\ \hline \trule
$(1245678)$     &\small Eq.  (\ref{def:ebasis7foldtwist0}) & \small  $(x_4,x_3,x_2,x_1,y_4,y_3,y_2,y_1)$ &   \small $\mathcal{S}_{1245678} =\mathcal{S}_{1345678}$ \\ \hline \trule
$(1345678)$       & \small Eq. (\ref{def:ebasis7foldtwist1}) &  \small  $(x_1,x_3,x_2,x_4,y_3,y_4,y_2,y_1)$   &   \small $\mathcal{S}_{2345678} =\{1,y_1,y_1^2,y_1^3,y_1^4,y_1^5,y_1^6,y_4,y_1y_4,$ \\
&&& \small $y_1^2 y_4,y_1^3y_4,y_1^4 y_4, y_1^5 y_4,x_2,y_1 x_2,$\\
&&& \small $x_4,y_1 x_4,y_1^2 x_4,y_1^3 x_4,y_1^4 x_4,y_1^5 x_4,$\\
&&& \small $y_4 x_4,y_1 y_4 x_4,y_1^2   y_4 x_4,y_1^3 y_4 x_4,$\\
&&& \small $y_1^4 y_4 x_4,x_2 x_4,x_4^2,y_1   x_4^2,y_4 x_4^2,$\\
&&& \small $x_2 x_4^2,x_4^3,y_1 x_4^3,y_4 x_4^3,x_2 x_4^3,x_4^4,y_1 x_4^4,y_4 x_4^4\Big\} \; , $\\ \hline \trule
$(1234568)$       &\small Eq.  (\ref{def:ebasis7foldtwist0}) &  \small  $(x_4,x_3,x_2,x_1,y_4,y_3,y_2,y_1)$   &   \small $\mathcal{S}_{123567} = \{1,y_1,y_1^2,y_1^3,y_1^4,y_2,y_1 y_2,y_1^2y_2,y_1^3y_2,$\\
&&& \small$x_1,y_1 x_1,y_1^2 x_1,y_1^3 x_1,y_1^4 x_1,y_2 x_1,y_1   y_2 x_1,$\\
&&& \small$y_1^2 y_2 x_1,y_1^3 y_2 x_1,x_1^2,y_1 x_1^2,y_2x_1^2,x_1^3,$\\
&&& \small$y_1 x_1^3,y_2 x_1^3,x_1^4,y_1 x_1^4,y_2 x_1^4,x_2,y_1 x_2,$\\
&&& \small$x_1 x_2,x_1^2 x_2,x_1^3 x_2\}$ \\ \hline \hline
\end{tabular}
\label{Tab:Pentatwist}
\end{table}

\end{document}